\documentclass[11pt, onesie]{article}   	
\usepackage{geometry}                		
\usepackage{graphicx}				
\usepackage{caption}
\usepackage{float}
\usepackage{amssymb}
\usepackage{amsmath}
\usepackage[title]{appendix}

\usepackage{footnote}
\usepackage{tablefootnote}

\usepackage{multirow}
\usepackage{array}

\newcommand\numberthis{\addtocounter{equation}{1}\tag{\theequation}}

\usepackage{draftwatermark}
\SetWatermarkScale{7}

\title{On the new and accurate (Goudsmit-Saunderson) model for describing e$^-$/e$^+$ multiple Coulomb scattering\newline\\A \texttt{Geant4} Technical Note}
\author{Mih{\'a}ly Nov{\'a}k}
\date{17 January 2018}							

\begin{document}
\maketitle

\begin{abstract}\noindent
A new model, for the accurate simulation of multiple Coulomb scattering (MSC) of e$^-$/e$^+$, has been implemented in \texttt{Geant4} 
recently and made available with version \texttt{Geant4-10.4}. The model is based on Goudsmit-Saunderson (GS) angular distributions 
computed by utilising the screen Rutherford (SR) DCS and follows very closely the formulation developed by Kawrakow 
\cite{kawrakow1998representation,kawrakow2000accurate} and utilised in the 
\texttt{EGSnrc} toolkit \cite{kawrakow2000egsnrc}. Corrections, for taking into account\textit{energy loss} \cite{kawrakow2000accurate} neglected by the GS theory, 
\textit{spin-relativistic effects} \cite{kawrakow2000egsnrc} not included in the SR but might be accounted on the basis of Mott DCS as well as the so-called 
\textit{scattering power correction} 
\cite{kawrakow1997improved}, i.e. appropriately incorporating deflections due to sub-threshold delta ray productions, are all included similarly to the \texttt{EGSnrc} 
model \cite{kawrakow2000egsnrc}. Furthermore, an accurate electron-step algorithm \cite{kawrakow1996electron,kawrakow1998condensed,kawrakow2000accurate} is 
utilised for path length correction, i.e. for calculating the post-step position in each condensed history simulation steps such that the corresponding single-scattering 
longitudinal and lateral (post step point) distributions are very well reproduced. An e$^-$/e$^+$ stepping algorithm, including the simulation step-limit due to the MSC 
and boundary crossing \cite{kawrakow2000accurate}), free from step-size artefacts, makes the model complete. Details on this new model, including all the 
above-mentioned components and corrections, are provided in this \texttt{Geant4} technical note.
\\\\
It must be noted, that a Goudsmit-Saunderson model for MSC was available before Geant4-10.4., documented in \cite{kadri2009incorporation}, that has been 
completely replaced by the model described in this technical note (keeping only the \texttt{G4GoudsmitSaundersonMscModel} name of the \texttt{C++} class 
from that previous version). 
\end{abstract}

\section{Introduction}
Elastic interaction of electrons(positrons) with free atoms are well understood and the corresponding differential cross 
sections (DCS) can be computed based on first principles. The electron transport problem in matter, including the effects of elastic 
scattering, can be solved exactly by using these DCSs (with corrections to account solid state effects) in a traditional event-by-event 
Monte Carlo simulation. However, event-by-event simulation is feasible only if the mean number of interactions per particle track is
below few hundred. This limits the applicability of the detailed simulation model only for electrons with relatively low kinetic energies 
(up to $E_{kin} \sim 100$ keV) or thin targets. A fast ($E_{kin} > 100$ keV) electron undergoes a high number of elastic collisions in the 
course of its slowing down in tick targets. Since detailed simulation becomes very inefficient, high energy particle transport simulation 
codes employ condensed history simulation models. Each particle track is simulated by allowing to make individual steps 
that are much longer than the average step length between two successive elastic interactions. The net effects of these high number 
of elastic interactions such as angular deflection and spacial displacement is accounted at each individual condensed history step
by using multiple scattering theories. The accuracy of modelling the cumulative effects of the many elastic scattering in one step, 
strongly depends on the capability of the employed multiple scattering theory and model to describe the angular distribution of 
electrons after travelling a given path length.

The MSC angular distribution employed by the model is described in Section \ref{sec:ANGULAR-DR}. The Goudsmit-Saunderson (GS) theory, that provides the 
basis for calculating such MSC angular distributions, is discussed first in Section \ref{sec:GS-angular} followed by the so-called hybrid simulation model explained 
in Section \ref{sec:hybrid-model}. The GS theory requires a single Coulomb scattering DCS as input. The selected screened Rutherford DCS, the computation of 
the corresponding GS angular distributions are discussed then in Section \ref{sec:USING-SR-DCS} in light of the hybrid simulation model. 
The details of the applied variable transformation, that results in a smooth and very compact representation of the pre-computed angular distributions eventually 
leading to a very efficient model, are also given. Section \ref{sec:ANGULAR-DR} on the screen Rutherford DCS based Goudsmit-Saunderson angular distributions 
is completed then in Section \ref{sec:summary-GS-with-SR} that provides technical details on the actual in-memory representation of these angular PDFs 
as well as on the corresponding very efficient algorithm for sampling of the corresponding angular deflections due to MSC at run-time.

Corrections, that are for taking into account the energy loss \cite{kawrakow2000accurate} neglected by the GS theory, 
spin-effects \cite{kawrakow2000egsnrc} not included in SR but accounted in the Mott DCS as well as the so-called scattering power correction 
\cite{kawrakow1997improved}, i.e. appropriately incorporating deflections due to sub-threshold delta ray productions, are all discussed in the 
dedicated sub-sections of Section \ref{sec:COR-to-DR}.

The note is completed by Section \ref{sec:electron-step-alg} describing the electron stepping algorithm, including the model employed
for the accurate final (longitudinal and lateral) positions in condensed history simulation steps as well as the corresponding simulation 
step-limit and boundary crossing that results in an accurate and artefact free electron tracking.

\section{Angular distribution in multiple scattering}
\label{sec:ANGULAR-DR}
The accuracy of modelling the cumulative effects of many elastic scattering in one step strongly depends on the capability of the 
employed multiple scattering theory to describe the angular distribution of electrons after travelling a given path length. Goudsmit and 
Saunderson~\cite{goudsmit1940multiple} derived an expression for this angular distribution in the form of an expansion in Legendre 
polynomials (neglecting energy loss along the step). One of the biggest advantage of the Goudsmit-Saunderson distribution is its 
formal independence of the form of the single scattering DCS. Therefore, any DCS can be combined with the GS angular distribution. 
First the GS angular distribution will be derived in this section followed by its form when using the screened Rutherford DCS including 
the variable transformation that makes the corresponding MSC angular PDF smooth, i.e. convenient for storing in memory and using in 
Monte Carlo simulation.

\subsection{Goudsmit-Saunderson angular distribution}
\label{sec:GS-angular}
The Goudsmit-Saunderson angular distribution will be derived in this section by following the notations in~\cite{fernandez1993theory}. 
Assuming that the particle was initially moving into the $\hat{\bar{z}}$ direction 
the probability density $F(s;\theta)$ of finding the particle moving into a direction falling into the solid 
angle element $\mathrm{d}\Omega$ around the direction defined by the polar angle $\theta$ after having travelled a
path length $s$ is
\begin{equation}
\label{eq:gs_angular_base}
 F(s;\theta) = \sum_{n=0}^{\infty} f_{n}(\theta)\mathcal{W}_n(s)
\end{equation} 
$f_{n}(\theta)$ is the angular distribution after $n$ elastic interactions and $\mathcal{W}_n(s)$ is the probability of having $n$ elastic 
interactions while travelling a path length $s$.
\\\\
The single scattering angular distribution $f_{n=1}(\theta)$ is given by  
\begin{equation}
\label{eq:gs_single_scattering_distribution_def}
f_{1}(\theta) = \frac{1}{\sigma}\frac{\mathrm{d}\sigma}{\mathrm{d}\Omega}(\theta)
\end{equation}
where  $\mathrm{d}\sigma/\mathrm{d}\Omega$ is the DCS for elastic scattering 
and $\sigma =  2\pi \int_{0}^{\pi} \mathrm{d}\sigma / \mathrm{d}\Omega \sin(\theta) \mathrm{d}\theta$ 
is the corresponding elastic scattering cross section (assuming spherical symmetry). 
It can be seen (with the variable change $\theta \to \cos(\theta)$) that
\begin{equation}
\label{eq:gs_single_scattering_pdf_cost}
 2\pi f_{1}(\cos(\theta)) = 2\pi\frac{1}{\sigma}\frac{\mathrm{d}\sigma}{\mathrm{d}\Omega}(\cos(\theta))=p(\cos(\theta))
\end{equation}
i.e. the probability density of having a direction in single elastic event that corresponds to $\cos(\theta)$.
The single scattering distribution can be expressed in terms of Legendre polynomials (see Appendix ~\ref{app:GS_series} for more details)
\begin{equation}
\label{eq:gs_single_scattering_distribution_withLeg}
 f_{1}(\cos(\theta)) = \sum_{\ell=0}^{\infty} \frac{2\ell+1}{4\pi} F_{\ell} P_{\ell}(\cos(\theta))
\end{equation}
where 
\begin{equation}
\label{eq:gs_fel_def}
 F_{\ell} = 2\pi \int_{-1}^{1} f_{1}(\cos(\theta)) P_{\ell}(\cos(\theta)) \mathrm{d}(\cos(\theta)) 
  = \langle P_{\ell}(\cos(\theta)) \rangle
\end{equation}
and $ P_{\ell}(x)$ is the $\ell$-th Legendre polynomial.
The quantity 
\begin{equation}
\label{eq:gs_transport_cefs_def}
 G_{\ell} \equiv 1- F_{\ell} = 1-\langle P_{\ell}(\cos(\theta))\rangle 
\end{equation}
is the transport coefficient and the $\ell$-th inverse transport mean free path is defined by
\begin{equation}
\label{eq:transport_mean_free_path_with_Gl}
 \lambda_{\ell}^{-1} \equiv \frac{G_{\ell}}{\lambda} = \frac{1- F_{\ell}}{\lambda} 
  = \frac{1-\langle P_{\ell}(\cos(\theta))\rangle}{\lambda} 
\end{equation}
where 
\begin{equation}
\label{eq:elastic_mfp_def}
\lambda \equiv \frac{1}{\mathcal{N}\sigma}
\end{equation}
is the elastic mean free path ($\mathcal{N}$ is the number of atoms per unit volume). The corresponding $l$-th transport cross section $\sigma_l$ can then be written as 
\begin{equation}
 \sigma_{\ell} = 2\pi \int_{-1}^{1} [1-P_{\ell}(\cos(\theta))]\frac{\mathrm{d}\sigma}{\mathrm{d}\Omega}\mathrm{d}(\cos(\theta))
\end{equation}     
with the first transport cross section of
\begin{equation}
\label{eq:first-trans-xsec}
 \sigma_{1} = 2\pi \int_{-1}^{1} [1-\cos(\theta)]\frac{\mathrm{d}\sigma}{\mathrm{d}\Omega}\mathrm{d}(\cos(\theta))
\end{equation}     
using $P_{\ell=1}(x)=x$.

Notice that $\lambda^{-1}$ is the mean number of elastic interaction per unit path length.
Furthermore, $F_0 =1$, $G_0=0$ and the value of $F_{\ell}$ decreases with $\ell$ due to the faster oscillation of $P_{\ell}(\cos(\theta))$~\cite{fernandez1993theory}. 
Therefore, $G_{\ell}$ goes to unity and hence $\lambda_{\ell}$ tends to $\lambda$ when $\ell$ goes to infinity. 

The angular distribution of particles after $n$ elastic interactions can be written as (see Appendix ~\ref{app:GS_series} for more details) 
\begin{equation}
\label{eq:gs_n_scattering_pdf_cost}
 f_{n}(\cos(\theta)) = \sum_{\ell=0}^{\infty} \frac{2\ell+1}{4\pi} (F_{\ell})^{n} P_{\ell}(\cos(\theta))
\end{equation}
The number of elastic interactions along a path length $s$ travelled by the particle follows Poisson 
distribution with parameter $s/\lambda$ (mean number of elastic interactions along $s$ )
\begin{equation}
\label{eq:prob_of_n_elastic_poission}
 \mathcal{W}_{n}(s)=\exp(-s/\lambda)\frac{(s/\lambda)^{n}}{n!} 
\end{equation}
Integrating Eq.(\ref{eq:gs_angular_base}) over $\phi$ (that gives a $2\pi$ factor due to cylindrical symmetry), changing the variable $\theta$ to $\cos(\theta)$
and inserting Eqs.(\ref{eq:gs_n_scattering_pdf_cost}, \ref{eq:prob_of_n_elastic_poission})
\begin{equation}
\label{eq:gs_angual_distribution_final_in_cost}
 \begin{split}
   F(\cos(\theta); s ) = & 2\pi \sum_{n=0}^{\infty} f_{n}(\cos(\theta))\mathcal{W}_n(s)
   = \sum_{n=0}^{\infty} \sum_{\ell=0}^{\infty} \frac{2\ell+1}{2} (F_{\ell})^{n} P_{\ell}(\cos(\theta)) \exp(-s/\lambda)\frac{(s/\lambda)^{n}}{n!}
\\ = & \sum_{\ell=0}^{\infty} \frac{2\ell+1}{2} \exp(-s/\lambda) \left[\sum_{n=0}^{\infty} \frac{(s/\lambda)^{n}}{n!} (F_{\ell})^{n} \right]  P_{\ell}(\cos(\theta))    
\\ = & \sum_{\ell=0}^{\infty} \frac{2\ell+1}{2} \exp(-s/\lambda)  
     \underbrace{ \left[\sum_{n=0}^{\infty} \frac{(s/\lambda)^{n}}{n!} \langle P_{\ell}(\cos(\theta) \rangle ^{n} \right] }_{\exp(s/\lambda \langle P_{\ell}(\cos(\theta) \rangle) = \exp(s/\lambda (1-\lambda/\lambda_{\ell}))}  
     P_{\ell}(\cos(\theta))    
\\ = & \sum_{\ell=0}^{\infty} \frac{2\ell+1}{2} \exp(-s/\lambda)  \exp(s/\lambda (1-\lambda/\lambda_{\ell}))  P_{\ell}(\cos(\theta))    
\\ = & \sum_{\ell=0}^{\infty} \frac{2\ell+1}{2} \exp(-s/\lambda_{\ell})  P_{\ell}(\cos(\theta))    
\end{split}
\end{equation}
that is the PDF of having a final direction that corresponds to $\cos(\theta)$ after travelling a 
path $s$ derived by Goudsmit and Saunderson~\cite{goudsmit1940multiple}. 

In order to compute the angular distribution for a given path length $s$ based on a given single scattering elastic DCS one needs to compute the transport mean free paths $\lambda_{\ell}$ that practically means the computation of the transport coefficients $G_{\ell}$ given by Eq.(\ref{eq:gs_transport_cefs_def}) that involves the computation of the integral Eq.(\ref{eq:gs_fel_def}). This integral needs to be computed numerically when the single scattering DCS is given in numerical form that can be challenging in case of large $\ell$ values due to the strong oscillation of the corresponding Legendre polynomials. 
The convergence of the Goudsmit-Saunderson series given by Eq.(\ref{eq:gs_angual_distribution_final_in_cost}) is determined by the exponential factor 
$\exp(-s/\lambda_{\ell})=\exp(-sG_{\ell}/\lambda)$. The number of terms needed to reach convergence of the series increases by decreasing path lengths and for short $s$ path lengths a large number of terms are needed to be computed. 
The convergence can be improved by separating the contribution of unscattered ($n=0$) and single-scattered ($n=1$) electrons in the series as it was suggested by Berger and Wang~\cite{berger1988multiple} and applied e.g. in~\cite{negreanu2005calculation}.  This from of the Goudsmit-Saunderson angular distribution was used by 
Bielajew~\cite{bielajew1996hybrid} to construct his hybrid simulation model that will be discussed in the next section.

\subsection{Hybrid simulation model}
\label{sec:hybrid-model}
According to Eq.(\ref{eq:prob_of_n_elastic_poission}) the probability of having no ($n=0$), single ($n=1$) or at least two ($n\geq 2$) elastic scattering along a path length $s$ travelled by the particle are 
\begin{equation}
\begin{split}
 \mathcal{W}_{n=0}(s)&=\exp(-s/\lambda)
\\ \mathcal{W}_{n=1}(s) & =\exp(-s/\lambda)(s/\lambda)\;
\\ \mathcal{W}_{n\geq 2}(s) & =1-\exp(-s/\lambda)-\exp(-s/\lambda)(s/\lambda) 
\end{split}
\end{equation}
Separating these three terms in Eq.(\ref{eq:gs_angual_distribution_final_in_cost}) gives 
\begin{equation}
 \begin{split}
   F(\mu;s) = & 2\pi \sum_{n=0}^{\infty} f_{n}(\mu)\mathcal{W}_n(s)
   = 2\pi f_{n=0}(\mu)\mathcal{W}_{n=0} + 2\pi f_{n=1}(\mu)\mathcal{W}_{n=1}
     + 2\pi \sum_{n=2}^{\infty}f_{n}(\mu)\mathcal{W}_n(s)      
\\ = & \underbrace{ \exp(-s/\lambda) \delta(1-\mu)+ \exp(-s/\lambda)(s/\lambda)2\pi f_{n=1}(\mu)  }_{\chi}
\\ & + \sum_{n=2}^{\infty} \sum_{\ell=0}^{\infty} \frac{2\ell+1}{2} (F_{\ell})^{n} P_{\ell}(\mu) \exp(-s/\lambda)\frac{(s/\lambda)^{n}}{n!} 
\\ = & \chi + \sum_{\ell=0}^{\infty} \frac{2\ell+1}{2}  P_{\ell}(\mu)  \exp(-s/\lambda) 
     \underbrace{  \sum_{n=2}^{\infty}  (F_{\ell})^{n} \frac{(s/\lambda)^{n}}{n!} }
               _{\sum_{n=0}^{\infty} x^{n}/n! -1 -x,\; x=F_{\ell} (s/\lambda) }
\\ = & \chi + \sum_{\ell=0}^{\infty} \frac{2\ell+1}{2}  P_{\ell}(\mu)  \exp(-s/\lambda) 
       \left[ \exp( F_{\ell} (s/\lambda) ) - 1- F_{\ell} (s/\lambda) \right]
\\ = & \chi + \sum_{\ell=0}^{\infty} \frac{2\ell+1}{2}  P_{\ell}(\mu)  \mathrm{e}^{-s/\lambda} 
       \left[ \mathrm{e}^{(1-G_{\ell}) (s/\lambda) } - 1- (1-G_{\ell}) (s/\lambda) \right]
\\ = & \chi + \sum_{\ell=0}^{\infty} \frac{2\ell+1}{2}  P_{\ell}(\mu)  \mathrm{e}^{-s/\lambda} 
       \left[ \mathrm{e}^{s/\lambda}\mathrm{e}^{-(s/\lambda)G_{\ell}} - 1- (1-G_{\ell}) (s/\lambda) \right]
\\ = & \mathrm{e}^{-s/\lambda}\delta(1-\mu) +(s/\lambda)\mathrm{e}^{-s/\lambda}2\pi f_{n=1}(\mu) + 
\\ &   \sum_{\ell=0}^{\infty} \frac{2\ell+1}{2}  P_{\ell}(\mu) 
       \left\{ 
              \mathrm{e}^{-(s/\lambda)G_{\ell}} -\mathrm{e}^{-(s/\lambda)}\left[1 +(s/\lambda)(1-G_{\ell}) \right]
       \right\}       
\end{split}
\end{equation}
with $\mu=\cos(\theta)$. This is the same probability density function as Eq.(\ref{eq:gs_angual_distribution_final_in_cost}) in the form of
suggested by Berger and Wang~\cite{berger1988multiple} and used by Bielajew~\cite{bielajew1996hybrid} to construct his hybrid simulation model.
Notice that the first term is the probability of having no elastic scattering along the path $s$ ($\mathrm{e}^{-s/\lambda}$) multiplied by the corresponding 
angular distribution (a delta function; $\int_{-1}^{+1} \delta(1-\mu) \mathrm{d}\mu = 1$).  The second term is the probability of having exactly one elastic scattering along the path $s$
($(s/\lambda)\mathrm{e}^{-s/\lambda}$) multiplied by the corresponding single scattering angular distribution ($2\pi f_{n=1}(\mu) = p(\mu)$ i.e. PDF of $\mu$ in single scattering;
 $\int_{-1}^{+1} 2\pi f_{n=1}(\mu)\mathrm{d}\mu = 1$). Since $F(\mu;s)$ is the PDF of $\mu$ for a given path length $s$ it is normalised to unity i.e. $\int_{-1}^{+1} F(\mu;s) \mathrm{d}\mu = 1$. Therefore, integrating the last term(s) over $\mu$ gives $1-\mathrm{e}^{-s/\lambda}-(s/\lambda)\mathrm{e}^{-s/\lambda}$ i.e. the probability of having 
 at least two elastic scattering along the path length $s$. In order to have this last term in the same form as the first and second (i.e. probability of event multiplied by the corresponding normalised distribution) one needs to normalise this term i.e.  
\begin{equation}
\label{eq:hybrid_base}
  F(\mu;s)=
  \mathrm{e}^{-s/\lambda}\delta(1-\mu) +(s/\lambda)\mathrm{e}^{-s/\lambda}2\pi f_{n=1}(\mu) + 
  (1-\mathrm{e}^{-s/\lambda}-(s/\lambda)\mathrm{e}^{-s/\lambda})F(\mu;s)^{2+}
\end{equation}
~\cite{bielajew1996hybrid,kawrakow1998condensed} where $F(\mu;s)^{2+}$ is the probability density function of having a 
direction given by $\mu=\cos(\theta)$ after at least two elastic interactions along the 
path $s$~\cite{bielajew1996hybrid,kawrakow1998condensed}
\begin{equation}
\label{eq:more_scattering_base}
F(\mu;s)^{2+} \equiv \sum_{\ell=0}^{\infty} (\ell+0.5)  P_{\ell}(\mu) 
       \frac{ 
              \mathrm{e}^{-(s/\lambda)G_{\ell}} -\mathrm{e}^{-(s/\lambda)}\left[1 +(s/\lambda)(1-G_{\ell}) \right]
            }{
              1-\mathrm{e}^{-s/\lambda}-(s/\lambda)\mathrm{e}^{-s/\lambda}
            }       
\end{equation}

When a model based on Eq.(\ref{eq:hybrid_base}) is used to describe the angular distribution of the particles after travelling a given path, the single and 
multiple scattering cases are combined into one model that suggests the adverb \textit{hybrid}. In case of short path lengths  (when $s \sim \lambda$)Eq.(\ref{eq:hybrid_base}) 
is almost entirely contributed by the second, single scattering term. In case of longer path lengths ($s \gg \lambda$) the multiple scattering part will be dominated. Geometry 
adaptive particle transport algorithms, that limits the allowed path length near the boundary, can exploit this property: by limiting the path length near the boundary the
the condensed history algorithm transforms to single scattering algorithm. Such a combination of single and multiple scattering 
algorithm into one model results in a \textit{hybrid} model for elastic scattering that can exploit both the computational efficiency of multiple scattering theories and the accuracy of single scattering approach~\cite{bielajew1996hybrid,kawrakow1998condensed} when needed.   

Kawrakow and Bielajew~\cite{kawrakow1998condensed}  developed an any-angle multiple scattering model based on the Goudsmit-Saunderson angular distribution and 
the screened Rutherford elastic scattering cross section.  Using the screened Rutherford elastic DCS in the Goudsmit-Saunderson series results in a compact numerical representation of the corresponding angular distributions that makes possible on-the-fly sampling of the angular deflections including any angles for any condensed history steps.  
~\cite{kawrakow1998condensed,kawrakow2000egsnrc,kawrakow2000accurate}. 
This model will be discussed in the next section.

\subsection{Using the screen Rutherford DCS for elastic scattering of electrons in the Goudsmit-Saunderson model}
\label{sec:USING-SR-DCS}

\subsubsection{The screened Rutherford DCS and some derived expressions}
The screened Rutherford DCS can be obtained by solving the scattering equation under the first Born approximation with 
a simple exponentially screened Coulomb potential as scattering potential (see Appendix~\ref{app:SRDCS} for all details and derivations)
\begin{equation}
\label{eq:exp_screened_pot_for_SR}
 V(r)=\frac{ZZ'e^{2}}{r} e^{-r/R}
\end{equation}
with a screening radius $R$ (target atomic number of $Z$ and projectile charge $Z'e$) that leads to the screened Rutherford DCS for elastic scattering
\begin{equation}
\label{eq:SRF-DCS}
  \frac{\mathrm{d}\sigma}{\mathrm{d}\Omega} ^{(SR)}= 
  \left( \frac{ZZ'e^{2}}{pc\beta} \right)^{2} \frac{ 1}{(1-\cos(\theta) + 2A)^{2}}    
\end{equation}
where $p$ is the momentum, $\beta$ is the velocity of the particle in units of speed of light $c$ and $A$ is the screening parameter
\begin{equation}
\label{eq:screening_param_SR}
  A \equiv \frac{1}{4}\left(\frac{\hbar}{p}\right)^2 R^{-2} 
\end{equation}
The corresponding total elastic scattering cross section  
\begin{equation}
\label{eq:total_cross_section_SR}
 \sigma^{(SR)} = \left( \frac{ZZ'e^{2}}{pc\beta} \right)^{2} \frac{ \pi }{ A(1+A)}
\end{equation}
and the single elastic scattering angular distribution 
\begin{equation}
\label{eq:sigle_scattering_distr_SR}
 f_1(\theta) ^{(SR)}= \frac{ 1 }{\pi } \frac{A(1+A)}{ (1-\cos(\theta) + 2A)^{2} }
\end{equation}
The corresponding $\ell$-th transport coefficient $G_\ell$ (see Appendix \ref{app:SRDCS})
\begin{equation}
\label{eq:transport_coefs_SR}
 G_\ell^{(SR)} (A) =  1-\ell[Q_{\ell-1}(1+2A)-(1+2A)Q_{\ell}(1+2A)]              
\end{equation}
where $Q_\ell(x)$ are Legendre functions of the second kind. Substituting $\ell=1$ with the required 
Legendre functions 
\begin{equation}
\label{eq:first_transport_coef_SR}
 G_{\ell=1}^{(SR)} (A)= 2A\left[ \ln \left(\frac{1+A}{A}\right) (A+1) -1 \right]   
\end{equation}
and for $\ell=2$
\begin{equation}
\label{eq:second_transport_coef_SR}
 G_{\ell=2}^{(SR)} (A)=  6A(1+A)\left[ (1+2A)\ln \left(\frac{1+A}{A}\right) -2 \right]   
\end{equation}
while all details can be found in Appendix \ref{app:SRDCS}. 
Note that Eq.(\ref{eq:transport_mean_free_path_with_Gl}) defines the corresponding transport mean free paths.

\subsubsection{Single scattering angular distribution}
\label{sec:single_scattering_SR}
According to Eqs.(\ref{eq:gs_single_scattering_pdf_cost},\ref{eq:sigle_scattering_distr_SR}) the PDF of $\mu=\cos(\theta)$ in single scattering 
can be written as
\begin{equation}
\label{eq:single_scattering_pdf_SR}
 2\pi f_{n=1}(\mu) = 2\pi \frac{1}{\pi}\frac{A(A+1)}{[1-\mu+2A]^{2}} =\frac{2A(A+1)}{[1-\mu+2A]^{2}}
\end{equation}
when the screened Rutherford DCS is used to describe elastic scattering. The corresponding cumulative 
density function (CDF) $\mathcal{P}(\mu;A)$ is
\begin{equation}
\label{eq:single_scattering_CDF_SR}
\begin{split}
 \mathcal{P}(\mu=\cos(\theta);A) &= \int_{0}^{\theta} \frac{2A(A+1)}{[1-\cos(\theta)+2A]^{2}} \sin(\theta)\mathrm{d}\theta
    = \int_{\mu}^{+1}   \frac{2A(A+1)}{[1-\mu+2A]^{2}} \mathrm{d}\mu
\\& = \left[\frac{2A(A+1)}{1-\mu+2A}\right]_{\mu}^{+1}
    = (A+1)\left[1-\frac{2A}{1-\mu+2A}\right] =\frac{(A+1)(1-\mu)}{1-\mu+2A} 
\end{split}
\end{equation}
Having $\eta \in \mathcal{U}(0,1)$ random variable one can derive the inverse function $\mathcal{P}^{-1}(\eta;A)$ by 
solving $\eta= \mathcal{P}(\mu;A)=(A+1)(1-\mu)/(1-\mu+2A)$ for $\mu$ which results in 
\begin{equation}
\label{eq:sampling-single-SR}
 \mu = \mathcal{P}^{-1}(\eta;A) = 1 - \frac{2A\eta}{1-\eta+A}
\end{equation}
Therefore, given the screening parameter $A$ and a random number $\eta \in \mathcal{U}(0,1)$,
deflection $\mu$ in single elastic scattering can be sampled analytically by using this expression.  

\subsubsection{More than one scattering case}
\label{sec:more-than-one-scattering}
$F(\mu;s)^{2+}$ as given by Eq.(\ref{eq:more_scattering_base}) is the conditional PDF of the angular deflection $\mu$ along the path length $s$ with 
the condition that at leat two elastic interaction happen along this path. The shape of this PDF shows high variation as a function of possible parameter values 
and can easily span sever order of magnitudes that makes the numerical treatment (sampling, interpolation, etc.) of these PDF delicate. In order to avoid numerical
problems one either needs to store a high amount of these precomputed PDF or come up with a proper variable transformation that makes these PDF as flat as possible. 
A special variable transformation for screened Rutherford DCS was suggested in~\cite{bielajew1994plural,bielajew1996hybrid}, generalised 
in~\cite{kawrakow1998condensed}. 

Let the transformation function $f$ and the transformed variable $u$ 
\begin{equation}
 u = f(a_1,...,a_n;\mu)
\end{equation}
where $u\in \mathcal{U}(0,1)$ and $a_1,...,a_n$ are parameters of the transformation that control 
the shape of the result of the transformation.
We want the transformed PDF of $q^{2+}(u)$ to satisfy
\begin{equation}
 q^{2+}(s;u)\mathrm{d}u = F(s;\mu)_{GS}^{2+}\mathrm{d}\mu
\end{equation}
i.e. the probability of having $u$ falling into the $\mathrm{d}u$ interval around $u$ according to the 
transformed PDF $q^{2+}(u)$ is equal to the probability of having $\mu$ falling into the $\mathrm{d}u$ 
interval around $\mu$ according to the original PDF $F(s;\mu)_{GS}^{2+}$.
From this~\cite{kawrakow1998condensed}
\begin{equation}
\label{eq:kaw_generalTransfom}
 q^{2+}(s;u) = F(s;\mu)_{GS}^{2+}\frac{\mathrm{d}\mu}{\mathrm{d}u}
\end{equation}
where 
\begin{equation}
\label{eq:kaw_generalTransform_plus}
 \frac{\mathrm{d}\mu}{\mathrm{d}u} = \left( \frac{\mathrm{d}u}{\mathrm{d}\mu} \right)^{-1}
 =\left( \frac{\partial f(a_1,...,a_n;\mu)}{\partial \mu} \right)^{-1}
\end{equation}
The values of $a_1,...,a_n$ parameters (that gives a transformation that results in a 
"as flat as possible" transformed distribution) can be determined by minimizing 
\begin{equation}
 \int_{0}^{1} \left[ q^{2+}(s;u) -1\right]^{2} \mathrm{d}u  
\end{equation}
with respect to $a_1,...,a_n$ (i.e. try to find $a_1,...,a_n$ that results in a transformation 
$f(a_1,...,a_n;\mu)$ that makes the transformed PDF $q^{2+}(s;u)$ as close to the uniform 
distribution on $[0,1]$ as possible). The optimal values of $a_1,...,a_n$ can be found by 
solving the set of equations~\cite{kawrakow1998condensed} 
\begin{equation}
\label{eq:kaw_general_optimalaity}
 \begin{split}
   0 & = \frac{\partial}{\partial a_i} \left[ \int_{0}^{1} \left[ q^{2+}(s;u) -1\right]^{2} \mathrm{d}u  \right]
\\& = \frac{\partial}{\partial a_i} \left\{
            \int_{-1}^{+1} \left[ F(s;\mu)_{GS}^{2+} 
                        \left( \frac{\partial f(a_1,...,a_n;\mu)}{\partial \mu} \right)^{-1}    
                 \right]^{2}  \left( \frac{\partial f(a_1,...,a_n;\mu)}{\partial \mu} \right)
        \mathrm{d}\mu \right\}
\\&\;   -
         \frac{\partial}{\partial a_i} \left\{
            \int_{-1}^{+1} \left[ 2 F(s;\mu)_{GS}^{2+} 
                        \left( \frac{\partial f(a_1,...,a_n;\mu)}{\partial \mu} \right)^{-1}    
                 \right]  \left( \frac{\partial f(a_1,...,a_n;\mu)}{\partial \mu} \right)
        \mathrm{d}\mu \right\}
\\& =   \int_{-1}^{+1} \frac{\partial}{\partial a_i} 
            \left\{ \left[ F(s;\mu)_{GS}^{2+} \right]^{2}  
                        \left( \frac{\partial f(a_1,...,a_n;\mu)}{\partial \mu} \right)^{-1} 
            \right\}            
        \mathrm{d}\mu 
\\&\;    -2 \int_{-1}^{+1} \underbrace{
                  \frac{\partial}{\partial a_i}
                   2 F(s;\mu)_{GS}^{2+}
                  }_{0} 
        \mathrm{d}\mu
\\& =   \int_{-1}^{+1}  
             \left[ F(s;\mu)_{GS}^{2+} 
                            \left( \frac{\partial f(a_1,...,a_n;\mu)}{\partial \mu} \right)^{-1} 
             \right]^{2}  
             \left[ \frac{\partial^{2} f(a_1,...,a_n;\mu)}{ \partial \mu \partial a_i} 
             \right] 
        \mathrm{d}\mu, \quad i=1,...,n 
 \end{split}
\end{equation}
Notice that this formalism is independent of the form of the adopted DCS. 

When the single elastic scattering is described by the screened Rutherford DCS (i.e. DCS for elastic scattering computed under the 
first Born approximation by using a simple exponentially screened Coulomb potential; see Appendix~\ref{app:SRDCS}) one can take~\cite{kawrakow1998condensed} 
\begin{equation}
\label{eq:kaw_transform}
 u = f(\mu;a) = \frac{(a+1)(1-\mu)}{1-\mu+2a}
\end{equation}
that is the single scattering CDF corresponding to the screened Rutherford DCS with a scaled screening 
parameter $a=w^{2}A$ (see Eq.(\ref{eq:single_scattering_CDF_SR})) where the scaling factor $w$ is arbitrary at the moment. Choosing 
this form of transformation function is motivated by the fact, that according to Eq.(\ref{eq:kaw_generalTransfom}) the transformed distribution 
becomes $q^{2+}(s;u) \equiv \mathcal{U}[0,1]$ i.e. the uniform distribution on the $[0,1]$ interval when the transformation function is chosen 
to be the CDF. The corresponding inverse transformation can be derived by solving $u=f(a;\mu)$ for $\mu$ that gives
\begin{equation}
\label{eq:kaw_inverstrans}
  \mu = 1- \frac{2au}{1-u+a}
\end{equation}
Taking the partial derivatives
\begin{equation}
\label{eq:kaw_firstderiv}
 \frac{\partial f(a;\mu)}{\partial \mu} = -\frac{2a(1+a)}{[1-\mu+2a]^{2}}
\end{equation}
\begin{equation}
\label{eq:kaw_secondderiv}
 \frac{\partial^{2}  f(a;\mu)}{\partial \mu \partial a} = -2\frac{1-\mu(1+2a)}{[1-\mu+2a]^{3}}
\end{equation}
plugging them into Eq.(\ref{eq:kaw_general_optimalaity}) and solving for $a$ gives the optimal value of the parameter $a$ of the chosen 
transformation function given by Eq.(\ref{eq:kaw_transform}) (where optimality means that the corresponding transformed $q^{2+}(s;u)$ PDF is as close to the uniform distribution as possible)~\cite{kawrakow1998condensed} (see Appendix~\ref{app:kaw_optimal_value_of_a})
\begin{equation}
\label{eq:kaw_optimala}
 a=\frac{\alpha}{4\beta}+\sqrt{\left(\frac{\alpha}{4\beta}\right)^{2}+\frac{\alpha}{4\beta}}
\end{equation}
where 
\begin{equation}
\label{eq:kaw_opt_alpha}
\begin{split}
 \alpha = & \sum_{\ell=0}^{\infty} \xi_{\ell}(s,\lambda,A) \left\{ 
   \left( 1.5\ell +\frac{0.065}{\ell+1.5}+\frac{0.065}{\ell-0.5}+0.75 \right) \xi_{\ell}(s,\lambda,A)
        \right.
\\&\;   \left.
   - 2(\ell+1)  \xi_{\ell+1}(s,\lambda,A)
   + \frac{(\ell+1)(\ell+2)}{(2\ell+3)} \xi_{\ell+2}(s,\lambda,A)  
   \right\}
\end{split}   
\end{equation}
\begin{equation}
\label{eq:kaw_opt_beta}
 \beta =  \sum_{\ell=0}^{\infty} (\ell+1) \xi_{\ell}(s,\lambda,A)  \xi_{\ell+1}(s,\lambda,A)  
\end{equation}
with 
\begin{equation}
\label{eq:kaw_termxi}
\xi_{i}(s,\lambda,A) = 
       \frac{ 
              \mathrm{e}^{-(s/\lambda)G_{i}(A)} -\mathrm{e}^{-(s/\lambda)}\left[1 +(s/\lambda)(1-G_{i}(A)) \right]
            }{
              1-\mathrm{e}^{-s/\lambda}-(s/\lambda)\mathrm{e}^{-s/\lambda}
            }       
\end{equation}
i.e. the last part of $F(s;\mu)_{GS}^{2+}$ given by Eq.(\ref{eq:more_scattering_base}).

\begin{figure}
\centering
\begin{minipage}[b]{.45\linewidth}
\centering
	  \includegraphics[width = 1.0\linewidth]{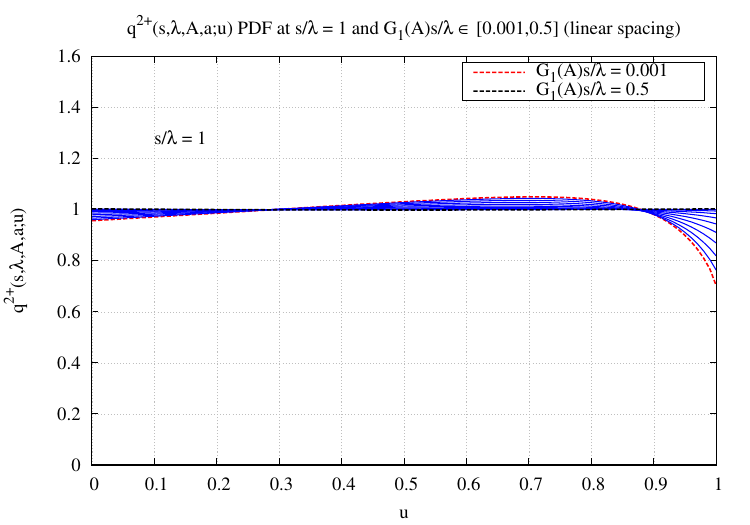}
\end{minipage}
\begin{minipage}[b]{.45\linewidth}
\centering
	  \includegraphics[width = 1.0\linewidth]{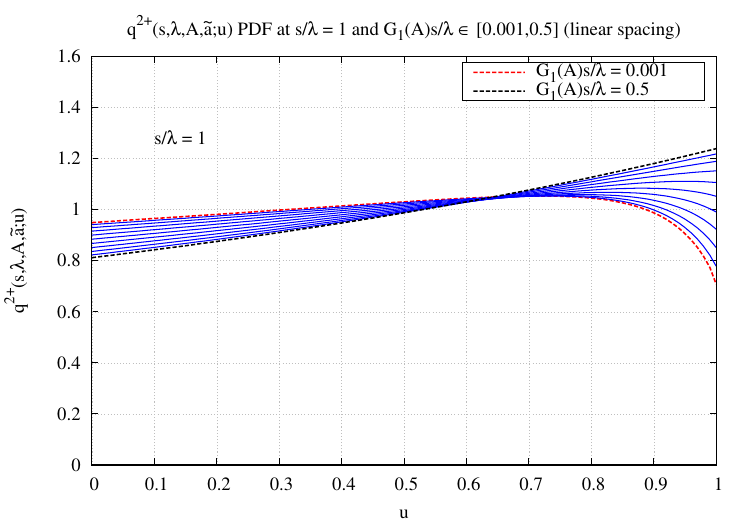}
\end{minipage}
\\
\begin{minipage}[b]{.45\linewidth}
\centering
	  \includegraphics[width = 1.0\linewidth]{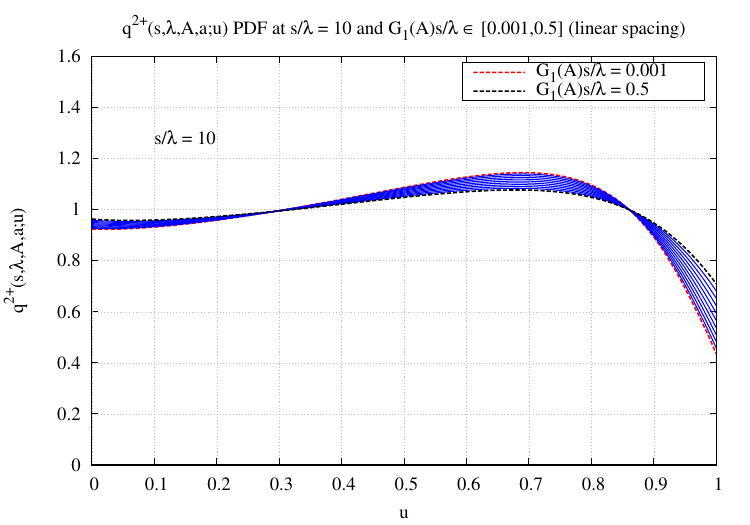}
\end{minipage}
\begin{minipage}[b]{.45\linewidth}
\centering
	  \includegraphics[width = 1.0\linewidth]{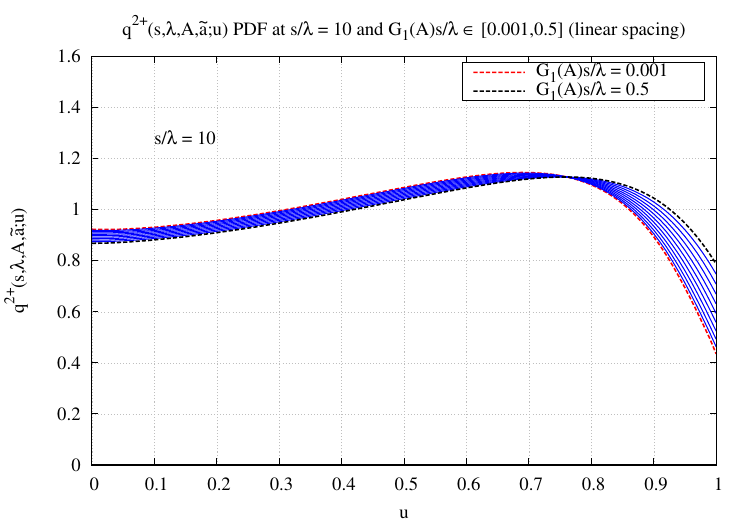}
\end{minipage}
\\
\begin{minipage}[b]{.45\linewidth}
\centering
	  \includegraphics[width = 1.0\linewidth]{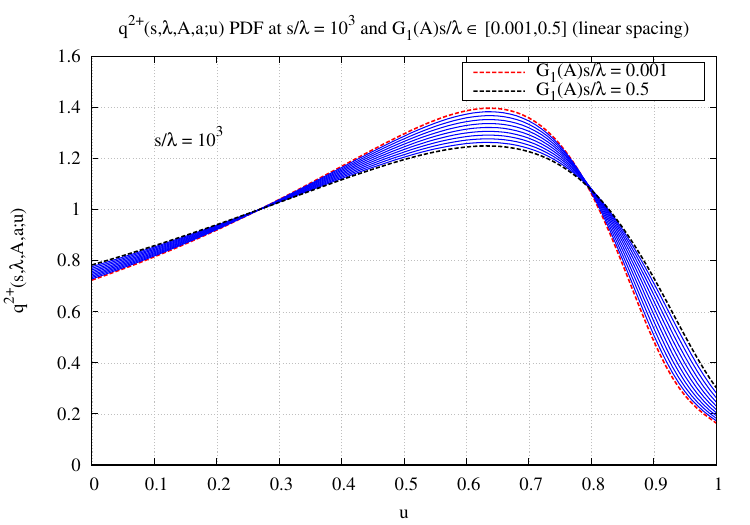}
\end{minipage}
\begin{minipage}[b]{.45\linewidth}
\centering
	  \includegraphics[width = 1.0\linewidth]{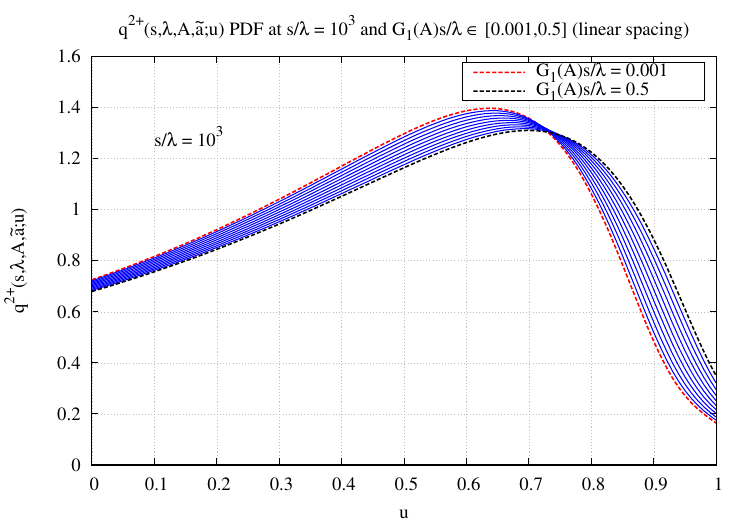}
\end{minipage}
\\
\begin{minipage}[b]{.45\linewidth}
\centering
	  \includegraphics[width = 1.0\linewidth]{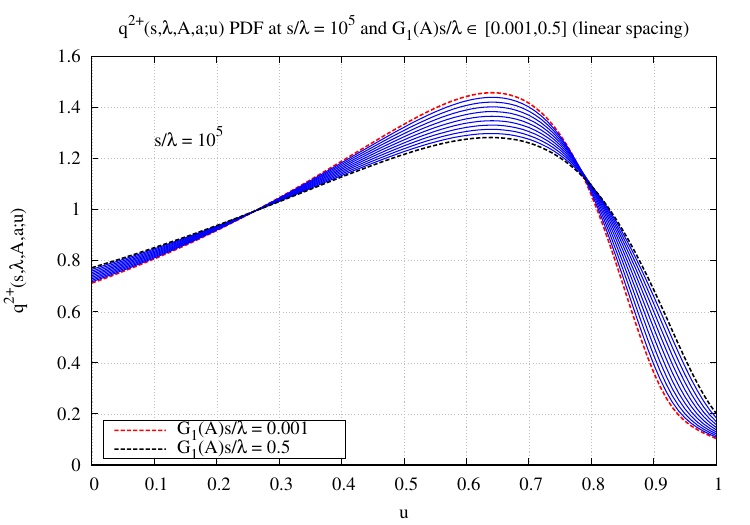}
\end{minipage}
\begin{minipage}[b]{.45\linewidth}
\centering
	  \includegraphics[width = 1.0\linewidth]{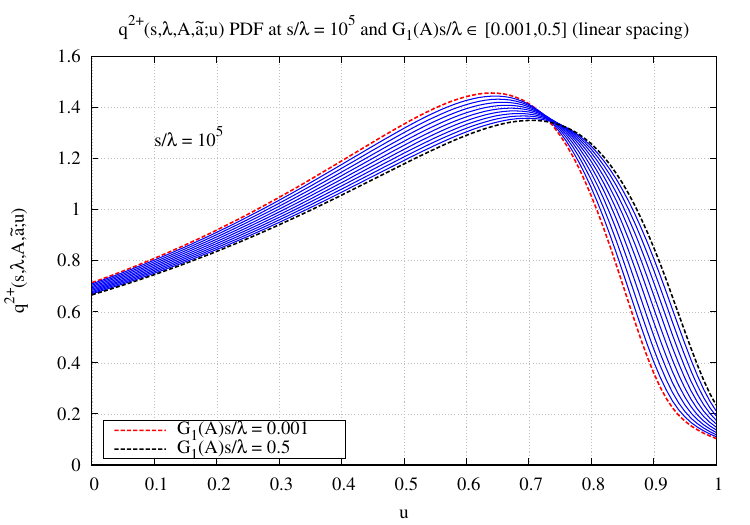}
\end{minipage}
\caption{$q^{2+}(s,\lambda,a,A;u)$ (left) and $q^{2+}(s,\lambda,\tilde{a},A;u)$  transformed PDFs that correspond to the parameter values shown in  
Table~\ref{tb:kaw_table1}-\ref{tb:kaw_table4}. }
\label{fig:kaw_transformed_PDFs}
\end{figure}

When the screened Rutherford DCS is adopted with the transformation given by Eq.(\ref{eq:kaw_transform}), the transformed PDF 
can be given by using Eq.(\ref{eq:kaw_generalTransfom}), substituting Eq.(\ref{eq:more_scattering_base}), Eq.(\ref{eq:kaw_firstderiv}) according to 
Eq.(\ref{eq:kaw_generalTransform_plus}) and using Eq.(\ref{eq:kaw_inverstrans}) to replace $\mu$ in the Legendre polynomials that yield 
\begin{equation}
\label{eq:kaw_transformed_PDF}
 q^{2+}(s,\lambda,a,A;u) = \frac{2a(1+a)}{[1-u+a]^{2}} \sum_{\ell}^{\infty} (\ell+0.5) P_{\ell}
 \left[  1- \frac{2au}{1-u+a} \right] \xi_{\ell}(s,\lambda,A)
\end{equation}
In order to compute this transformed PDF as a function of the transformed variable $u \in [0,1]$ the parameters $s,\lambda,a,A$ need to be given (in a form but see below 
for more details). The computation then have some key points that will be discussed in the remaining: (\textit{i}) stable computation of the required $Q_{\ell}(x)$ 
Legendre functions of second kind;  (\textit{ii}) efficient computation of the transformation parameter $a$.   
\\
\\
The computation involves Eq.(\ref{eq:kaw_termxi}) which means that one needs to be able to compute $G_{\ell}(A)$ transport coefficients up to a high enough 
value of $\ell$ in order to achieve convergence even in the case of short path lengths (i.e. small number of elastic interactions along the given $s$ path length). 
When the screened Rutherford DCS is adopted these $G_{\ell}(A)$ transport 
coefficients can be computed by using Eq.(\ref{eq:Wentzel_transport_coefs}) that involves the computation of Legendre functions of the second kind  $Q_{\ell}(x)$ with 
$x=1+2A>1$. Special care needs to be taken to evaluate the Legendre functions of second kind $Q_{\ell}(x)$ based on their recurrence relation in order to avoid numerical
inaccuracies. One solution is given in Appendix A. of~\cite{fernandez1993theory} that ensures numerical stability of the computed $Q_{\ell}(x)$ up to high values of $\ell$ even in the case of $x>1$. The other possibility is to use the approximate expression for $G_{\ell}(A)$ given in~\cite{kawrakow1998condensed}
\begin{equation}
\label{eq:kaw_aprxtranscoef}
 G_{\ell}(A) = 1-y_{\ell}(A)K_{1}(y_{\ell}(A))\left\{1+0.5y_{\ell}^{2}(A)[\Phi(\ell) - 0.5\ln(\ell(\ell+1)) - \gamma] \right\}
\end{equation}
where $y_{\ell}(A)=2\sqrt{\ell(\ell+1)A}$, $K_1(x)$ is the modified Bessel function of the second kind,
$\Phi(\ell) = \sum_{m=1}^{\ell}1/m$ is the $\ell$-th harmonic number and $\gamma$ is the Euler constant. In the present work the $G_{\ell}(A)$ transport coefficients 
were computed up to $\ell_{max}=10 000$ by using Eq.(\ref{eq:kaw_aprxtranscoef}) that ensures convergence in Eq.(\ref{eq:kaw_transformed_PDF}) even for small number of 
elastic interactions and/or small value of $A$.
\newline\newline\noindent
The second point regarding the transformed PDF Eq.(\ref{eq:kaw_transformed_PDF}) is the computation of the parameter $a$ (i.e. parameter of the variable transformation 
Eq.(\ref{eq:kaw_transform})). 
Computation of the optimal value of $a$ by Eq.(\ref{eq:kaw_optimala}) is a numerically intensive task that not an issue regarding the precomputation phase. However, 
one needs to compute the proper values of $a$ during the simulation after each sampling of $u$ from $q^{2+}(s,\lambda,a,A;u)$ in order to be able to apply the proper inverse 
transformation Eq.(\ref{eq:kaw_inverstrans}) that delivers the corresponding sampled $\mu=\cos(\theta)$ according to $F(s;\mu)_{GS}^{2+}$ given by 
Eq.(\ref{eq:more_scattering_base}).  
One can replace the computation of the optimal $a$ (Eq.(\ref{eq:kaw_optimala})) by an approximation of the corresponding $w^{2}$ values in the $A\to0$ case
~\cite{kawrakow2000egsnrc} 
\begin{equation}
\label{eq:kaw_aprxw2}
 \frac{\tilde{w}^{2}}{0.5(s/\lambda)+2} = 
  \left\{
   \begin{array}{ll}
      1.347+ t(0.209364-t(0.45525-t(0.50142-t0.081234))) & \quad \mathrm{if}\; \lambda < 10\\
      & \\
      -2.77164+t(2.94874-t(0.1535754-t0.00552888))       & \quad \mathrm{otherwise} \\
   \end{array}
  \right. 
\end{equation}
where $t=\ln(s/\lambda)$.
Then corresponding approximate value of $a$ can be computed by $\tilde{a}=\tilde{w}^{2}A$ both for the pre-computation of $q^{2+}(s,\lambda,\tilde{a},A;u)$ and run time.
The parameter $\tilde{a}$ won't be as optimal as the one obtained by the exact computation but will be very close to that especially in case of small values of $A$ i.e. at high 
energies when the non-transformed distributions $F(s;\mu)_{GS}^{2+}$ strongly peaked to the forward direction. Some example $q^{2+}(s,\lambda,\tilde{a},A;u)$ 
distributions, that clearly demonstrate their smoothness and low variation over a wide range of practical values of the input parameters, will be shown below using 
both $a$ and $\tilde{a}$. However, some important points regarding the actual input parameters are discussed briefly before. 
\newline\newline\noindent
It was mentioned after Eq.(\ref{eq:kaw_transformed_PDF}) that $s, \lambda, A$ and $a$ or $\tilde{a}$ are the parameters required for the calculation of 
the corresponding $q^{2+}(u)$ transformed distribution. However, providing the values of $s/\lambda$ and $G_{1}s/\lambda$ are already sufficient when using the 
screened Rutherford DCS to describe the single scattering. This is because the value of $G_{1}(A)$ can be obtained easily as their ratio and the corresponding 
screening parameter can be computed then by solving Eq.(\ref{eq:first_transport_coef_SR}) for $A$. These are already sufficient for calculating the transformation 
parameter using either its optimal value $a$ (Eq.(\ref{eq:kaw_optimala}) and Eq.(\ref{eq:kaw_termxi})) or its approximation $\tilde{a}$ (Eq.(\ref{eq:kaw_aprxw2})). 
On the same time, $s/\lambda$ is the mean number of elastic scattering along the $s$ travelled path affecting strongly the corresponding multiple scattering angular 
distribution while $G_{1}s/\lambda$ characterises the mean $\cos(\theta)$ scattering angle of the distribution 
$\langle \cos(\theta) \rangle_{GS} = \exp(-sG_{\ell=1}/\lambda)$ as given by Eq.(\ref{eq:first-gs-mom}). 
Therefore, the $q^{2+}(u)$ transformed angular distributions might be pre-calculated over an appropriate 2D $s/\lambda$, $G_{1}s/\lambda$ grid leading to 
a moderate size data base thanks to the variable transformation while allowing to utilise very accurate angular distributions at run time through interpolation. 
See Section \ref{sec:summary-GS-with-SR} for all details on the pre-computation, the actual form and run time usage of the screened Rutherford DCS based GS angular distributions.     
\newline\newline\noindent
Some example $q^{2+}(u)$ distributions are shown in Fig.~\ref{fig:kaw_transformed_PDFs} in order to demonstrate how well-behaved they are irrespectively of using  
the exact $a$ or the more practical approximate $\tilde{a}$ values of the transformation parameter. It is clear that the distributions are smooth, stay close to uniform 
above the entire $u \in [0,1]$ transformed variable interval showing a small variation over a wide range of practical values of the $s/\lambda$ 
and $G_{1}s/\lambda$ input parameter (the $s/\lambda \in[1,10^5]$ and $G_{1}s/\lambda \in [0.001,0.5]$ limits will be discussed in Section \ref{sec:summary-GS-with-SR}). 

\subsubsection{Screening parameter}
\label{sec:screening-par}
When elastic scattering is described by the screened Rutherford DCS, the screening parameter $A$ given by Eq.(\ref{eq:screening_param_SR}) plays an essential role. 
The screening radius $R$ can be estimated from the Thomas-Fermi model of atom that gives
\begin{equation}
\label{eq:screening_radius_TF}
R=C_{TF}\frac{a_{0}}{Z^{1/3}}
\end{equation}
where $a_0$ is the Bohr radius 
\begin{equation}
a_0=\hbar/(m_e c \alpha)
\end{equation}
with $\hbar$, $m_e$, $c,$ $\alpha$ reduced Planck constant, electron rest mass, speed of light and fine structure constant and $C_{TF}$ is the Thomas-Fermi constant 
\begin{equation}
\label{eq:thomas_fermi_const}
C_{TF}\equiv \frac{(3\pi)^{\frac{2}{3}}}{2^{\frac{7}{3}}} \approx 0.88534
\end{equation}
Substituting Eq.(\ref{eq:screening_radius_TF}) into Eq.(\ref{eq:screening_param_SR}) and using the above expression for the Bohr radius yields 
\begin{equation}
\label{eq:screening_TF}
A^{(TF)} = \frac{m_{e}^{2} c^{2} \alpha^{2} Z^{2/3}}{4 p C_{TF}^{2}} =  \frac{ (m_{e} c^{2})^{2} \alpha^{2} Z^{2/3}}{4 (pc)^{2} C_{TF}^{2}} 
\end{equation}

Moliere derived the DCS for elastic scattering based on a Thomas-Fermi potential and applying WKB-expansion to the radial equation without 
making use of the Born approximation~\cite{moliere1947theorieI}. Then he derived a correction to $A^{(TF)}$ by fitting the average scattering angle squared computed from the
screened Rutherford DCS to that resulted from his DCS and he finally obtained
\begin{equation}
\label{eq:screening_Moliere}
A^{(M)} = A^{(TF)} \left[ 1.13 + 3.76 \left(\frac{\alpha Z}{\beta}\right)^2\right]   
\end{equation}
The corresponding inverse elastic mean free path for electron or positron incident can be given by Eq.(\ref{eq:elastic_mfp_def}) using Eq.(\ref{eq:total_cross_section_SR}) 
with $A=A^{(M)}$ that results in
\begin{equation}
\label{eq:el_mfp}
\lambda^{-1} = \mathcal{N}  \sigma^{(SR)} = \mathcal{N} \left( \frac{Ze^{2}}{pc\beta} \right)^{2} \frac{ \pi }{ A^{(M)}(1+A^{(M)})} 
 =  \mathcal{N} \frac{Z^{2} r_{0}^{2}(m_e c^2)^2  \pi}{(pc)^2 \beta ^2 A^{(M)}(1+A^{(M)})} 
\end{equation}
where $\mathcal{N}=\rho\mathcal{N}_{\mathcal{A}}/M = \rho/(u A_{M})$ is the number of atoms per unit volume ($\rho$ is the density of the material, $\mathcal{N}_{\mathcal{A}}$ is 
the Avogadro number, $M$ is the molar mass, $u=M_{u}/\mathcal{N}_{\mathcal{A}}$ is the atomic mass unit, $M_{u}=1 $ [g/mol] is the molar mass constant and 
$A_M$ is the relative atomic mass) and $e^2=r_0 m_e c^2$ was used with $r_0$ being the classical electron radius. Replacing $Z^2$ by $Z(Z+1)$ in order to take into account 
scattering by atomic electrons~\cite{bethe1953moliere} one can write 
\begin{equation}
\label{eq:inverse_elastic_Molier_atom}
\lambda^{-1} (1+A^{(M)}) = \frac{ \rho Z(Z+1) r_{0}^{2}(m_e c^2)^2  \pi}{(pc)^2 \beta ^2 u A_M A^{(M)} } = 
\frac{1}{\beta ^2}   \frac{  4\pi r_{0}^{2} C_{TF}^{2} }{ \alpha^2 u 1.13 }  \frac{ \rho }{ A_M}  \frac{Z^{1/3}(Z+1) }{ 1 + 3.33 \left(\frac{\alpha Z}{\beta}\right)^{2}} 
\approx \frac{b_{c}^{*}}{\beta^2}
\end{equation}
where the material dependent parameter
\begin{equation}
\label{eq:bc_Molier_atom}
b_{c}^{*}= \frac{  4\pi r_{0}^{2} C_{TF}^{2} }{ \alpha^2 u 1.13 }  \frac{ \rho }{ A_M}  \frac{Z^{1/3}(Z+1) }{ 1 + 3.33 (\alpha Z)^{2}} 
= 7827.68 \left[ \frac{ \textrm{cm}^2 }{\textrm{g}} \right] \frac{ \rho }{ A_M}  \frac{Z^{1/3}(Z+1) }{ 1 + 3.33 (\alpha Z)^{2}} 
\end{equation}
with $1/\alpha=137.0356$, $r_0=2.81794\times10^{-13}$ [cm], $u=1.660538\times10^{-24}$ [g] and $C_{TF}$ as given by Eq.(\ref{eq:thomas_fermi_const})
was used and $\beta^2$ was neglected. According to Eq.(\ref{eq:screening_Moliere}), $A^{(M)}$ can be expressed by using this material dependent $b_{c}^{*}$ parameter as 
\begin{equation}
\label{eq:screening_Molier_with_material_params_atom}
A^{(M)} = \frac{1}{4(pc)^2} \frac{1}{b_{c}^{*}} \chi_{cc}^{*2}
\end{equation}
where 
\begin{equation}
\label{eq:xcc2_Molier_atom}
\chi_{cc}^{*2} = \frac{4\pi r_{0}^{2} (m_e c^2)^2}{u} \frac{\rho}{A_M} Z(Z+1) = 0.156914 \left[ \frac{\textrm{cm}^2 \textrm{MeV}^2}{\textrm{g}} \right] \frac{\rho}{A_M} Z(Z+1) 
\end{equation}
is also a material dependent parameter ($m_e c^2=0.510998$ [MeV]).

Moliere's screening angle $\chi_\alpha$~\cite{moliere1947theorieI} for mixtures was derived in~\cite{nelson1985egs4} based on the corresponding expression given by  
Scott~\cite{scott1963theory}. Moliere's screening parameter for mixtures that consist of $N_e$ elements with having $n_i$ the proportion by number of element $Z_i$
can be written as~\cite{nelson1985egs4}  
\\ \newline
\begin{scriptsize}
Our starting point is Eq. (2.14.43) from~\cite{nelson1985egs4}
\begin{equation}
\label{eq:egs4_compound_screeningAngle}
\ln{\chi_{\alpha}^2} = \frac{\sum_{i}^{N_e} n_i Z_i(Z_i+\xi_e)\ln\chi_{\alpha_i}^2}{\sum_{i}^{N_e} n_i Z_i(Z_i+\xi_e)}
\end{equation} 
where $\chi_{\alpha}$ is Moliere's screening angle (the screening parameter is $A^{(M)}=\chi_{\alpha}^2/4$) of the compound that builds up from $N_e$ elements, $\chi_{\alpha_i}$ are the 
screening angles characterising the $i-th$ components with corresponding atomic number $Z_i$ and $n_i$ is the proportion by number of the $i-th$ atom ($n_i=\rho\mathcal{N}_{\mathcal{A}}/M_i = \rho/(u A_{M_i})$ (Note, that $Z^2$ is replaced by $Z(Z+\xi_e)$ 
with  $\xi_e \in [0,1]$ to account scattering from sub-threshold inelastic events). The individual Moliere's screening angles are expressed in terms of the corresponding Thomas-Fermi 
screening angle $\chi_{0_{i}}$ as 
\begin{equation}
\label{eq:egs4_perComponent_screeningAngle}
 \ln \chi_{\alpha_i}^{2} = \ln \chi_{0_i}^2 + \ln \left[  1.13 + 3.76 (\alpha Z_i)^2 \right]
\end{equation}
(where the $1/\beta^2$ part was dropped from the last term similarly to~Eq.(\ref{eq:bc_Molier_atom})). The Thomas-Fermi screening angle for the $i-th$ components with atomic number $Z_i$ is
\begin{equation}
  \ln \chi_{0_i}^{2} = \ln \left[ \frac{(m_e c^2)^2\alpha^2}{(pc)^2 C_{TF}^2 Z_i^{-2/3}}  \right] = \ln \left[ \frac{(m_e c^2)^2\alpha^2}{(pc)^2 C_{TF}^2 }  \right] - \ln Z_i^{-2/3} 
\end{equation} 
and the second term can be written as 
\begin{equation}
 \ln \left[  1.13 + 3.76 (\alpha Z_i)^2 \right] = \ln 1.13\left[  1 + 3.34 (\alpha Z_i)^2 \right]  = \ln 1.13 + \ln \left[  1 + 3.34 (\alpha Z_i)^2 \right]
\end{equation}
Substituting these two equations back into Eq.(\ref{eq:egs4_perComponent_screeningAngle}) one can get
\begin{equation}
 \ln \chi_{\alpha_i}^{2} =  \ln \left[ \frac{(m_e c^2)^2\alpha^2}{(pc)^2 C_{TF}^2 }  \right] - \ln Z_i^{-2/3} + \ln 1.13 + \ln \left[  1 + 3.34 (\alpha Z_i)^2 \right]
   = \ln \left[ \frac{(m_e c^2)^2\alpha^2 1.13}{(pc)^2 C_{TF}^2 }  \right] - \ln Z_i^{-2/3}  + \ln \left[  1 + 3.34 (\alpha Z_i)^2 \right]
\end{equation}
Inserting this expression to the component wise screening angles into  Eq.(\ref{eq:egs4_compound_screeningAngle}) one can get

\begin{equation}
\begin{split}
 \ln{\chi_{\alpha}^2} & = \frac{\sum_{i}^{N_e} n_i Z_i(Z_i+\xi_e) \left\{ 
   \ln \left[ \frac{(m_e c^2)^2\alpha^2 1.13}{(pc)^2 C_{TF}^2 }  \right] - \ln Z_i^{-2/3}  + \ln \left[  1 + 3.34 (\alpha Z_i)^2 \right]  \right\}} {\sum_{i}^{N_e} n_i Z_i(Z_i+\xi_e)}
 \\ & = \ln \left[ \frac{(m_e c^2)^2\alpha^2 1.13}{(pc)^2 C_{TF}^2 }  \right]  + 
          \frac{\sum_{i}^{N_e} n_i Z_i(Z_i+\xi_e) \left\{ - \ln Z_i^{-2/3}  + \ln \left[  1 + 3.34 (\alpha Z_i)^2 \right]  \right\} } {\sum_{i}^{N_e} n_i Z_i(Z_i+\xi_e)} 
 \\ & = \ln \left[ \frac{(m_e c^2)^2\alpha^2 1.13}{(pc)^2 C_{TF}^2 }  \right]  + 
           \frac{\sum_{i}^{N_e} n_i Z_i(Z_i+\xi_e) \ln \left[  1 + 3.34 (\alpha Z_i)^2 \right]  } {\sum_{i}^{N_e} n_i Z_i(Z_i+\xi_e)} -
           \frac{\sum_{i}^{N_e} n_i Z_i(Z_i+\xi_e)  \ln Z_i^{-2/3}  } {\sum_{i}^{N_e} n_i Z_i(Z_i+\xi_e)}          
 \end{split}         
 \end{equation}
Then using the relation $A^{(M)}=\chi_{\alpha}^2/4$ between the screening parameter  and screening angle, Moliere's screening parameter for compounds can be written as 
\begin{equation}
\begin{split}
 A^{(M)} = & \frac{ \chi_{\alpha}^2}{4} 
 \\ &  = \left[ \frac{(m_e c^2)^2\alpha^2 1.13}{4(pc)^2 C_{TF}^2 }  \right] \exp \left[ 
      \frac{\sum_{i}^{N_e} n_i Z_i(Z_i+\xi_e) \left\{ - \ln Z_i^{-2/3}  + \ln \left[  1 + 3.34 (\alpha Z_i)^2 \right]  \right\} }{\sum_{i}^{N_e} n_i Z_i(Z_i+\xi_e)}
      \right]
 \\ & =   \left[ \frac{(m_e c^2)^2\alpha^2 1.13}{4(pc)^2 C_{TF}^2 }  \right]  \frac{ 
       \exp \left[  \frac{\sum_{i}^{N_e} n_i Z_i(Z_i+\xi_e) \ln \left[  1 + 3.34 (\alpha Z_i)^2 \right]  } {\sum_{i}^{N_e} n_i Z_i(Z_i+\xi_e)} \right] 
       } {
       \exp \left[   
       \frac{\sum_{i}^{N_e} n_i Z_i(Z_i+\xi_e)  \ln Z_i^{-2/3}  } {\sum_{i}^{N_e} n_i Z_i(Z_i+\xi_e)}   \right] 
       }  
 \\ & =   \left[ \frac{(m_e c^2)^2\alpha^2 1.13}{4(pc)^2 C_{TF}^2 }  \right]  \frac{ 
       \exp \left[  \frac{Z_X} {Z_S}\right] 
       } {
       \exp \left[  \frac{Z_E} {Z_S}\right] 
       }           
 \end{split}      
 \end{equation}
by using the notations given in Eqs.(\ref{eqs:Moliere_mixture_shorts})
\end{scriptsize}

\begin{equation}
\label{eq:screening_Moliere_mixtures}
A^{(M)} = \frac{\chi_{\alpha}^{2}}{4}  = \left[ \frac{(m_e c^2)^2\alpha^2 1.13}{4(pc)^2 C_{TF}^2 }  \right] \exp \left[ 
      \frac{\sum_{i}^{N_e} n_i Z_i(Z_i+\xi_e) \left\{ - \ln Z_i^{-2/3}  + \ln \left[  1 + 3.34 (\alpha Z_i)^2 \right]  \right\} }{\sum_{i}^{N_e} n_i Z_i(Z_i+\xi_e)}
      \right]
\end{equation}
according to the $A^{(M)}=\chi_{\alpha}^{2}/4$ relation between the screening parameter and the corresponding Moliere's screening angle. Notice that the 
$1/\beta^2$ in Eq.(\ref{eq:screening_Moliere}) is also neglected in Eq.(\ref{eq:screening_Moliere_mixtures}). Furthermore, $Z(Z+1)$ has been changed to $Z(Z+\xi_e)$
where $\xi_e \in [0,1]$ which will be discussed at the end of this Section. 

Eq.(\ref{eq:screening_Moliere_mixtures}) can be written in the more compact form
\begin{equation}
\label{eq:screening_Moliere_mixtures_with}
A^{(M)} = \frac{\chi_{\alpha}^{2}}{4}  = \frac{ (m_{e} c^{2})^{2} \alpha^{2} 1.13}{4 (pc)^{2} C_{TF}^{2}} 
\frac{\exp (Z_X/Z_S)}{\exp(Z_E/Z_S)}
\end{equation}
by introducing~\cite{nelson1985egs4} 
\begin{equation}
\label{eqs:Moliere_mixture_shorts}
\begin{split}
 Z_S   &= \sum_{i=1}^{N_{e}}  n_i Z_i(Z_i+\xi_{e})
\\ Z_E &= \sum_{i=1}^{N_{e}}  n_i Z_i(Z_i+\xi_{e}) \ln Z_{i}^{-2/3} 
\\ Z_X &= \sum_{i=1}^{N_{e}}  n_i Z_i(Z_i+\xi_{e}) \ln \left[ 1+3.33 \alpha^{2} Z_{i}^{2}) \right] 
\\ A     &= \sum_{i=1}^{N_{e}}  n_i A_i
\end{split}
\end{equation} 
where $n_i$ is the fraction of the $i-th$ atom in the compound by number $n_i=$ number-of-atoms-per-unit-volume$_i$ per total number of atoms per volume. (Note, that only the ratios are important since 
a possible constant term will be cancelled later: (i) $Z_X/Z_S$ and $Z_E/Z_S$ will appear in Z-s; (ii) $b_c, \chi_{cc}^2$ will contain a $Z_S/A$ term where $A=\sum_i n_i A_i$ relative molecular mass 
with $A_i$ being the molar mass of the $i-th$ atom).

Similarly to Eq.(\ref{eq:inverse_elastic_Molier_atom}), in case of mixtures
\begin{equation}
\label{eq:inverse_elastic_Molier_mix}
\lambda^{-1} (1+A^{(M)}) \approx \frac{b_{c}}{\beta^2}
\end{equation}
where the material dependent parameter now
\begin{equation}
\label{eq:bc_Molier_mix}
b_{c}= 7827.68 \left[ \frac{ \textrm{cm}^2 }{\textrm{g}} \right]  \frac{\rho Z_S}{A}\frac{\exp (Z_E/Z_S)}{\exp(Z_X/Z_S)}
\end{equation}
According to Eq.(\ref{eq:screening_Moliere_mixtures_with}), $A^{(M)}$ for mixtures can be expressed by using this material dependent $b_{c}$ parameter as 
\begin{equation}
\label{eq:screening_Molier_with_material_params_mix}
A^{(M)} = \frac{1}{4(pc)^2} \frac{1}{b_{c}} \chi_{cc}^{2}
\end{equation}
where now
\begin{equation}
\label{eq:xcc2_Molier_mix}
\chi_{cc}^{2}  = 0.156914 \left[ \frac{\textrm{cm}^2 \textrm{MeV}^2}{\textrm{g}} \right] \frac{\rho Z_S}{A} 
\end{equation}
\newline\newline\noindent
Two things must be noted here before moving further. First, an element wise, i.e. Z (as well as projectile kinetic energy and projectile type) dependent, correction 
to Moliere's screening parameter will be introduced in Section \ref{sec::spin-effect} such that for the $i$-th element $A_i^{(M_{corr})}=\kappa_i A_i^{(M)}$. One 
can use Eq.(\ref{eq:egs4_compound_screeningAngle}) then to obtain the expression for the corresponding corrected screening parameter of the compound as
\begin{equation}
\label{eq:screening-corrected}
A^{(M_{corr})} = \frac{(m_e c^2)^2\alpha^2}{4 (pc)^2 C_{TF}^2}  \exp \left[  \frac{\sum_{i}^{N_e} n_i Z_i(Z_i+\xi_e)\ln [\kappa_iZ_i^{2/3}(1.13+3.76(\alpha Z_i)^2)] }{\sum_{i}^{N_e} n_i Z_i(Z_i+\xi_e)} \right] 
\end{equation}
\newline\noindent
Second, the replacement of $Z(Z+1)$ with $Z(Z+\xi_e)\;\xi_e \in [0,1]$  mentioned after Eq.(\ref{eq:screening_Moliere_mixtures}). Note that $Z^2$ was previously 
change to $Z(Z+1)$ in order to take into account scattering from atomic electrons in delta ray production (i.e. ionisation and not Coulomb scattering). 
However, in a usual \texttt{Geant4} simulation some of the delta electrons, i.e. those having an initial kinetic energy above a user defined threshold or secondary production cut, 
are produced explicitly. Therefore, the deflections of the primary electron/position due to the corresponding interactions are already accounted. As a consequence, keeping 
$\xi_e=1$ would lead to double-counting of these deflections related to the corresponding discrete secondary electron generations. $\xi_e \in [0,1]$ must be chosen 
such that it accounts only the angular deflections from sub-threshold delta ray productions that are not simulated explicitly. The $\xi_e = 0$ and $\xi_e = 1$ limits correspond 
to a pure discrete (i.e. zero cut) and pure continuous (i.e. infinite cut) description of the electron/position ionisation. The appropriate value will be given in Section 
\ref{sec::scattering-power-cor} where the corresponding so called \textit{scattering power correction} will be introduced based on \cite{kawrakow1997improved}.

\subsection{Summary on the screened Rutherford (SR) DCS based GS angular distributions}
\label{sec:summary-GS-with-SR}
The $F(\cos(\theta);s)$ Goudsmit-Saunderson multiple scattering angular distribution, i.e. PDF of having a final direction that corresponds to 
$\mu=\cos(\theta)$ after travelling a path $s$,
was derived in Section \ref{sec:GS-angular} (given by Eq.(\ref{eq:gs_angual_distribution_final_in_cost})). The zero, single and at least two scattering contributions to this PDF
were separated then in Section \ref{sec:hybrid-model} (given by Eq.(\ref{eq:hybrid_base})) that makes possible to handle separately these different cases during the simulation.
The zero and single scattering cases are not only straightforward to model but removing their contributions from the GS angular distribution helps to concentrate more on  
the challenging contributions coming from at least two elastic scattering. The corresponding $F(\mu;s)^{2+}$ PDF is given by 
Eq.(\ref{eq:more_scattering_base}) and the ultimate goal is 
to provide a representation that allows efficient and accurate sampling from these PDF-s at run time based on a limited amount of pre-computed data. This is rather 
challenging as the $F(\mu;s)^{2+}$ distributions are often strongly peaked to the forward directions especially at higher energies and/or small number of scatterings. 

The screened Rutherford (SR) DCS was considered as the model for describing single Coulomb scattering in Section \ref{sec:USING-SR-DCS}. 
While the SF DCS is a reasonable 
approximation its simple analytical form allows to express the variables, required for computing the above $F(\mu;s)$ multiple scattering angular distributions, 
in closed form. Moreover, this made possible to introduce a variable transformation in Section \ref{sec:more-than-one-scattering} which moves the original, often strongly peaked  
$F(\mu;s)^{2+}$ to the smooth transformed $q^{2+}(u;s)$ distributions (given by Eq.(\ref{eq:kaw_transformed_PDF})). Unlike the original, the transformed PDF 
are not only smooth but compact and stay close to the uniform distribution over a wide range of practical values of the input parameters. Finally, the appropriate input 
parameters for calculating the $q^{2+}(u;s)$ transformed distributions, which are the $s/\lambda$ and the corresponding $G_{1}s/\lambda$ values when relying on the SR DCS, 
were discussed at the end of Section \ref{sec:more-than-one-scattering}. Therefore, the GS angular distributions can be pre-computed and stored over an appropriate 2D 
$s/\lambda$, $G_1s/\lambda$ grid then utilised at \textit{run-time} for generating samples from these multiple scattering angular distributions. 
The amount of pre-computed data, required for accurate run time interpolation of the distributions, can be kept very low thanks to the above variable transformation.
Some details on the actual representation of the multiple scattering angular distributions, their pre-computation and run-time utilisation are given below.

\subsubsection{On the actual representation of the $q^{2+}(u)$ transformed PDF} 
In case of \texttt{EGSnrc} \cite{kawrakow1998representation,kawrakow2000egsnrc}, 
the $q^{2+}(s/\lambda,G_1s/\lambda,u)$ angular PDF-s were pre-computed at each point of a 2D $s/\lambda$, $G_1s/\lambda$ grid  
using $N_u=101$ values on the $u\in [0,1]$ transformed variable interval. Then an Alias table \cite{walker1977efficient} was generated for each of these PDF-s 
that allows optimally fast sampling from a discrete distribution which is the probability of having $u \in [u_i, u_{i+1}], i \in {0,...,N_{u}-1}$ in this case. 
After selecting an $u$-bin this way, the actual $u$-value is generated by using a linear approximation of the corresponding $q^{2+}(u)$ PDF within each $u$-bin. 
The variable transformation and the $N_u=101$ discrete values of the transformed variable ensures the validity of this linear approximation.   

This solution provides the possibility of generating samples from the corresponding PDF $q^{2+}(s/\lambda,G_1s/\lambda,u)$ without any grid search or rejection
that makes it very efficient. On the other hand, at each $s/\lambda,G_1s/\lambda$ point of the 2D grid, it requires to store the PDF, i.e the arrays of 
$u$ and $q^{2+}(s/\lambda,G_1s/\lambda,u)$ with sizes of $N_u$, and the two additional arrays of the corresponding Alias probabilities 
and alias indices both with a size of $N_u-1$. Furthermore, it requires two uniform random numbers to generate a sample from the corresponding PDF: one 
consumed during the Alias sampling when selection of the $u$-bin while the other is used to sample from the linear distribution within that bin.

After investigating this method, it has been decided to utilise a different solution in the present work. The solution outlined below relies on a \textit{Rational 
interpolation} based \textit{Inverse Transform} (RIT) of the CDF \cite{salvat2006penelope} to generate samples from the corresponding PDF while exploits more 
that the transformed PDF-s are rather close to the uniform distribution. The former helps to reduce the data size required to generate samples form a given 
transformed PDF while the latter ensures that the sampling is efficient as free from any grid searches or rejections. Moreover, generating a random samples 
requires only a single uniform random number. 

Given the continuous $q^{2+}(u)$ PDF calculated at the discrete $u_i, i=0,...,N$ points, the corresponding $\eta_i, i=0,...,N$ CDF values can be calculated.
Then $u$ samples, distributed according to the PDF, can be provided by generating $\eta \in \mathcal{U}(0,1)$, determining the $\eta_i \leq \eta < \eta_{i+1}$ CDF 
bin and calculating the corresponding $u_i \leq u < u_{i+1}$ value by the numerical inversion of the CDF. The accuracy of this sampling method is determined 
by the accuracy of the applied numerical inverse transform. Using RIT makes possible to perform the numerical inversion accurately based on a small set of 
discrete values especially when the PDF is so smooth and not far from the uniform distribution like in our case. In the case, when 
$\eta \in \mathcal{U}(0,1)$ such that $\eta_i \leq \eta < \eta_{i+1}$, the RIT based sampling formula is \cite{salvat2006penelope} 
\begin{equation}
\label{eq:rit-sampling}
  u = u_i + \frac{(1+a_i+b_i)\xi}{1+a_i\xi+b_i\xi^2}(u_{i+1}-u_i)
\end{equation}
where 
\begin{equation}
  a_i  =  \frac{\delta}{p(u_i)}-b_i-1,\;
  b_i  =  1- \frac{\delta^2}{p(u_i)p(u_{i+1})}\;\text{with }
  \delta \equiv \frac{\eta_{i+1}-\eta_i}{u_{i+1}-u_i},\; \xi\equiv\frac{\eta-\eta_i}{\eta_{i+1}-\eta_i}
\end{equation}
and $p(u)\equiv q^{2+}(u)$, i.e. the continuous PDF. The parameters $a_i, b_i, i=0,...,N-1$ can be precomputed for the given $u_i, i=0,...,N$ discrete transformed 
variable and corresponding $\eta_i, i=0,...,N$ CDF values then $u$ samples can be generated by Eq.(\ref{eq:rit-sampling}) according to the $q^{2+}(u)$ PDF 
without requiring the PDF itself.

The above method is rather efficient as free from any rejections and the RIT consist of only few arithmetic operations. However, finding the index $i$ such that 
$\eta_i < \eta \leq \eta_{i+1}$ for $\eta \in \mathcal{U}(0,1)$ requires either a bin search (time consuming) or an Alias table (memory consuming). This is because 
the $\eta_{i+1}-\eta_{i}, i=0,...,N-1$ CDF bins are not equal sized which would allow a fast and simple computation of the above bin index. 
However, when the PDF is so smooth and not far from the uniform distribution as in our case, the corresponding CDF is steadily increasing, distributed 
nicely and relatively evenly on the $\eta \in [0,1]$ interval. Therefore, one might go the other way around by defining the discrete $\eta_i, i=0,...,N$ CDF values 
first with equal $\eta_{i+1}-\eta_{i}, i=0,...,N-1$ bin sizes then determines the corresponding discrete $u_i, i=0,...,N$ variable values that yield them. 
This can also be done by relying on RIT, built over a dense $u$-grid, i.e. the PDF and CDF computed at large $N$, that allows very accurate inversion.
Furthermore, the above mentioned procedure can be turned into an iterative refinement of the discretisation of the $\eta \in [0,1]$ CDF interval by starting with 
a small $N$ division and increasing this number till the corresponding RIT based approximate PDF reaches the required accuracy. Having an actual $N$ point 
based $\eta_i, i=0,...,N$ discretisation of the CDF interval with the corresponding $u_i, i=0,...,N$ variable and $a_i, b_i, i=0,...,N-1$ RIT parameters determined, the 
corresponding approximate PDF for a given $u_i \leq u < u_{i+1}$ random variable value is \cite{salvat2006penelope}  
\begin{equation}
\label{eq:rit-aprx-pdf}
 \tilde{p}(u) = \frac{(1+a_i\xi+b_i\xi^2)^2}{(1+a_i+b_i)(1-b_i\xi^2)}\frac{\eta_{i+1}-\eta_i}{u_{i+1}-u_i}
\end{equation}
 with   
\begin{equation}
 \xi = \frac{1+a_i+b_i-a_i\tau}{2b_i\tau}\left[ 1- \sqrt{1-\frac{4b_i\tau^2}{(1+a_i+b_i-a_i\tau)^2}}\right],\; \tau \equiv \frac{u-u_i}{u_{i+1}-u_i}
\end{equation}
that can be used to calculate the actual approximation error during the iterative refinement.

The above procedure was applied in the present work to generate the final RIT based sampling tables. At each $s/\lambda,G_1s/\lambda$ point of the 2D grid, 
the $q^{2+}(u)$ PDF, the corresponding CDF and RIT parameters were calculated over a dense $u \in[0,1]$ interval division with 1001 discrete points that 
allowed very accurate inversion of the CDF. Then the above iterative procedure was applied to find the division of the $\eta \in [0,1]$ CDF into $N$ equal size bins, 
starting with $N=11$, that yields with an RIT based approximate PDF with the required accuracy. At each round of the iteration, the $u_i,i=0,...,N$ discrete values 
of the transformed variable, that yield the actual $\eta_i, i=0,...,N$ CDF division, were determined by using the dense $u$-grid based RIT. Then the CDF and 
RIT, that correspond to these actual $u_i,i=0,...,N$ values were computed. At the end, the integrated error of the corresponding PDF approximation 
was computed by using Eq.(\ref{eq:rit-aprx-pdf}). The procedure was either repeated by an 
increased $N=N+1$ number of CDF interval division or the current $N$ point based approximation was accepted if the integrated error was less than 1E-5
within each bin. At termination, the corresponding $u_i, i=0,...,N$ discrete values of the transformed variable and $a_i,b_i, i=0,...,N-1$ RIT parameters 
were stored and formed the sampling table associated to the given pint of the 2D $s/\lambda,G_1s/\lambda$ grid. 

The followings must be noted here. RIT of the CDF can provide a good approximation of the PDF based on a small number of discrete values thanks to the variable 
transformation that ensures a smooth and well-behaved $q^{2+}(u)$ PDF. The iterative procedure helps to find an optimally small number $N$ of the discrete points
which was as small as the initial $N=11$ value in many case (i.e. when the transformed PDF was very close to the uniform distribution). As these discrete $u_i,i=0,...,N$ 
transformed variable values were determined to correspond an equally 
spaced $\eta_i, i=0,...,N, \eta_i \in [0,1]$ CDF interval, the index $i$ such that $\eta_i \leq \eta < \eta_{i+1}$ can be calculated easily for a given $\eta \in \mathcal{U}(0,1)$ 
without a grid search. Moreover, its equal bin division makes unnecessary to store the CDF grid as knowing the single number $N$ is sufficient. Therefore, a sampling table, 
that allows to generate samples from a corresponding $q^{2+}(u)$ PDF, requires to store the $u_i,i=0,...,N$ discrete values of the transformed variable and 
the two arrays of the RIT parameters $a_i, b_i, i=0,...,N-1$. While this is only one array less than required by the above mentioned solution utilised in \texttt{EGSnrc}
(Alias table for sampling the PDF interval and linear PDF approximation within each interval), using RIT makes $N$, the size of these arrays significantly 
smaller than $N=101$ (even the initial $N=11$ is already sufficient in many cases) utilised by that alternative method. Therefore, this solution requires significantly 
smaller size data to be stored. On the same time, the sampling procedure is free from any grid searches or rejections, similarly to the alternative, while requires 
only a single uniform random number to generate an $u$ sample from the corresponding $q^{2+}(u)$ PDF. Therefore, the adaptive refinement based RIT 
of the CDF, utilised in this work, provides a very time and memory efficient solution for generating samples from the $q^{2+}(u)$ transformed angular distributions. 

\subsubsection{Pre-computation of the $q^{2+}(u)$ transformed PDF} 
As mentioned above, sampling tables are generated over an appropriate 2D grid of discrete $s/\lambda, G_1s/\lambda$ values. 
Practical considerations in \cite{kawrakow1998representation} suggested the $s/\lambda \in [1,10^5]$ and $G_1s/\lambda \in [0.001, 0.99]$
limits with $64$ linearly spaced points on $\ln{(s/\lambda)}$ scale and $15$ $G_1s/\lambda$ values spaced linearly.  

Note that a smaller $G_1s/\lambda = 0.5$ upper limit was chosen in \cite{kawrakow1998representation,kawrakow2000egsnrc} which corresponds 
to an average $\theta \approx 0.92$ radian ($\approx 53^{\circ}$) deflection due to multiple scattering along a single condensed-history step 
that already approaches the upper limit of the validity of this technique. This implies a maximally allowed step size 
$s_{\text{max}} = 0.5\lambda/G_1$ (without any loss) used in \texttt{EGSnrc} \cite{kawrakow2000egsnrc} (accounting also energy loss) to maintain a high 
accuracy. Furthermore, the Longitudinal and Lateral Correlation Algorithm (LLCA) \cite{kawrakow1996electron,kawrakow1998condensed,kawrakow2000accurate}, 
that provides the accurate final (longitudinal and lateral) position of a multiple scattering step and utilised in \texttt{EGSnrc} \cite{kawrakow2000egsnrc}, also has the 
same upper limit of validity (see Section X.). However, the corresponding step limit might be relaxed in case of some applications (e.g. in High Energy Physics detector 
simulations) that can tolerate the corresponding precision loss in exchange for the computing performance gain due to the longer condensed-history simulation steps. 
This is why the higher $G_1s/\lambda = 0.99$ upper limit was chosen in the present work as it allows to sample angular deflections in case of longer condensed-history 
simulation steps. It must be noted though, that nothing prevents to impose an $s_{\text{max}} = 0.5\lambda/G_1$ limit on the condensed-history simulation steps 
in \texttt{Geant4} to ensure the accuracy which is the case when the above LLCA electron stepping algorithm is used (see more details in Section X).

As mentioned above, for generating the sampling table at a given $s/\lambda,G_1s/\lambda$ value pair of the 2D grid, the corresponding 
$q^{2+}(s/\lambda,G_1s/\lambda,u)$ PDF is computed over a given set of discrete transformed variable values $u_i, i=0,...,N,\in [0,1]$. 
As mentioned at the end of Section \ref{sec:more-than-one-scattering}, this is done by first determining all derived parameter values required for the transformed PDF 
computation. Namely, the value of $G_{1}(A)$ is obtained as the ratio of the actual input parameter values then Eq.(\ref{eq:first_transport_coef_SR}) 
is solved for determining the corresponding screening parameter value $A$. These are already sufficient for calculating the required $\xi_\ell, \ell=0...\ell_{\text{max}}$ 
values using Eq.(\ref{eq:kaw_termxi}) and the approximated value of the optimal transformation parameter $\tilde{a}$ with Eq.(\ref{eq:kaw_aprxw2}). The 
$q^{2+}(s/\lambda,G_1s/\lambda,A, \tilde{a}; u)$ transformed PDF can then be computed by using Eq.(\ref{eq:kaw_transformed_PDF}) at any $u$ values. 
Note that $\ell_{\text{max}}$ must be chosen such that it ensures convergence of the summation in Eq.(\ref{eq:kaw_transformed_PDF}). Also note that $\xi_\ell$ 
does not depend on the actual $u$ value so $\xi_\ell, \ell=0...\ell_{\text{max}}$ values can be computed only once for a given $s/\lambda,G_1s/\lambda$ parameter 
combination and reused in Eq.(\ref{eq:kaw_transformed_PDF}) for the different $u$ values.   

\subsubsection{Run-time: sampling MSC deflections from GS angular distributions}
\label{sec:run_time_sampling}
The hybrid-simulation model, discussed in Section \ref{sec:hybrid-model}, was employed in the present work for modelling angular deflections in condensed-history 
simulation steps due to multiple scattering. While the complete model for multiple scattering, using angular distributions based on the Goudsmit-Saunderson angular 
distributions, is implemented in \texttt{G4GoudsmitSaundersonMscModel}, it is the \texttt{G4GoudsmitSaundersonTable} that is responsible to handle the 
precomputed angular distributions and to provide angular deflections for the model. The corresponding algorithm is outlined below\footnote{this algorithm will be altered 
by the corrections discussed in the next section}. 

Given the actual (initial) kinetic energy and material information with the current step length $s$, first the required $A, \lambda, G_1$ parameters are computed 
using Eq.(\ref{eq:screening_Molier_with_material_params_mix}) for $A\equiv A^{M}$, Eq.(\ref{eq:inverse_elastic_Molier_mix}) for $\lambda$ with the material dependent 
$b_c$ and $\chi_{cc}^2$ values precomputed at initialisation time for each materials using Eq.(\ref{eq:bc_Molier_mix}) and Eq.(\ref{eq:xcc2_Molier_mix})
respectively. Eq.(\ref{eq:first_transport_coef_SR}) can be used then to compute the corresponding value of $G_1(A)$.

The next step is then to determine the actual case of scattering type handled separately in the hybrid model, i.e. one out of the three contributions in  Eq.(\ref{eq:hybrid_base}). 
A uniform random number $\eta \in \mathcal{U}(0,1)$ and the probabilities of zero $\exp(-s/\lambda)$ and single $\exp(-s/\lambda)s/\lambda$ scattering are used 
to determine the actual scattering type then provide a $ \mu \equiv \cos(\theta)$ sample accordingly as:

\begin{equation}
\label{eq:algo}
\begin{array}{lllcl}
    \text{if} &  \eta < \exp(-s/\lambda)                       & \text{no scattering}               & \to & \mu = 1 \\
    \text{else if} & \eta < \exp(-s/\lambda) [1+s/\lambda] & \text{single scattering}   & \to & \mu \leftarrow Eq.(\ref{eq:sampling-single-SR}) \\
                &  \text{otherwise:}                                 & \text{at least two scattering}  & \to & \mu \sim  F^{2+}(\mu; s,\lambda, G_1)\; Eq.(\ref{eq:more_scattering_base})\\
\end{array}
\end{equation}
\noindent
In the third, i.e. at least two scattering, case the pre-computed sampling tables can be utilised to generate samples of the transformed variable $u$ according to the 
corresponding $q^{2+}(u; s,\lambda, G_1, \tilde{a})$ transformed PDF. The final $\mu(u)$ sample can then be computed by the inverse transform given by 
Eq.(\ref{eq:kaw_inverstrans}). 

It must be noted, that while the run-time $s/\lambda$ and $G_1$ values are continuous, the sampling stables are generated 
only at some discrete $s/\lambda,\; G_1s/\lambda$ value pairs. Therefore, an appropriate interpolation needs to be utilised at run 
time. As mentioned in the previous section, the $64$ discrete $s/\lambda \in [1,10^5]$ values (equally spaced on $\ln{(s/\lambda)}$ scale) 
and the $15$ discrete $G_1s/\lambda \in [0.001, 0.99]$ values (equally spaced on linear scale) were determined to provide an accurate 
interpolation when using linear interpolation in $\ln(s/\lambda)$ and linear interpolation in $G_1s/\lambda$. Furthermore, the 
algorithm, utilised in this work to interpolate PDF-s and described in general in \ref{app:pdf-interpol}, requires samples only from the 
PDF-s computed at the discrete parameter values. Therefore, the pre-generated sampling tables (over the appropriate 2D grid of 
$s/\lambda,\; G_1s/\lambda$ value pairs) in combination with the employed interpolation algorithm make possible a very fast and accurate 
run time sampling of the angular deflection along the simulation steps corresponding to any continuous values of $s/\lambda \in [1,10^5]$ 
and $G_1s/\lambda \in [0.001, 0.99]$.

The lower limit of the discrete $s/\lambda$ grid, i.e. the smallest value at which sampling table is still available to generate angular deflections according to 
more than one elastic scattering, is $s/\lambda = 1$. On the same time, the probability of having more than one scattering when $s/\lambda < 1$ 
is $1-\exp(-s/\lambda)[1+s/\lambda]$ which is not zero having a maximum value of $\sim 0.26$ (i.e. at $s/\lambda=1$).
However, as the number of elastic scattering along the path $s$ follows a Poisson distribution with the mean number of elastic scattering $s/\lambda$, the probability of having two, three, four or more scattering drops very quickly when $s/\lambda < 1$ (reaching their maximum at $s/\lambda=1$ with $\sim 0.184, 0.061, 0.015$ 
and $\sim 0.00366$ values). In other words, the actual number of elastic scattering is so small when $s/\lambda < 1$ than sampling those individually as successive single 
scatterings becomes more efficient than using any sampling tables. Therefore, the third case of the algorithm, given in Eq.(\ref{eq:algo}), is actually split depending 
on the $s/\lambda$ value. Sampling tables, with the above mentioned procedure, are utilised to generate $\mu(u)$ samples when $s/\lambda \geq 1$ while 
angular deflection is computed as the net effect of successive single scatterings otherwise. In this latter case, first the actual number of elastic scattering along 
the given step is sampled from Poisson distribution with the $s/\lambda < 1$ parameter then Eq.(\ref{eq:sampling-single-SR}) is utilised to generate angular 
deflection in single scatterings successively.

The procedure, described above and utilised in the \texttt{G4GoudsmitSaundersonMscModel} to provide angular deflections,
is implemented in the \texttt{G4GoudsmitSaundersonTable::Sampling} method.

\section{Corrections to the screened Rutherford DCS based GS angular distributions}
\label{sec:COR-to-DR}

\subsection{Energy loss correction to the GS angular distributions}
\label{sec:eloss_cor_gs}
The GS theory provides angular distributions after the particle, with initial energy of $E_0$, travels a given $s$ path length by assuming that the underlying elastic 
DCS is constant along the step. The DCS for elastic scattering of electron depends on the electron energy (beyond the material) that is continuously changing 
along the step from $E_0$ to $E_0-\Delta E$ due to the sub-threshold energy losses of $\Delta E$ (when using Class II condensed history simulation scheme 
that is the case in \texttt{Geant4} except in some special cases such as \texttt{Geant4-DNA} or \texttt{microelectronics}). As the effect of this energy loss, i.e. the path 
dependence of the underlying DCS, is neglected the computed GS angular distribution requires an appropriate energy loss correction. 

It has been shown in \cite{kawrakow2000accurate} (see all details in APPENDIX A. there), that accurate energy loss corrected angular distributions can be 
obtained by using the original, i.e. non-corrected, GS angular distributions but evaluating at an $E_{\text{eff}}$ effective energy and $s_{\text{eff}}$ effective 
step length instead of $E_0$ and $s$. The expressions for these effective quantities 
\begin{equation}
\begin{split}
E_\text{eff} & = E_0 \left[ 
   1 - \frac{\epsilon}{2}  - \frac{\epsilon^2}{12(2-\epsilon)} 
      \left( 
          \frac{5\tilde{\tau}^2+10\tilde{\tau}+6}{(\tilde{\tau}+1)(\tilde{\tau}+2)} + 2b(\tilde{E}) 
      \right)      + \mathcal{O}(\epsilon^3)
 \right] \\
s_\text{eff} & = s \left[
  1 - \frac{\epsilon^2}{3(2-\epsilon)} 
      \left( 
          \frac{\tilde{\tau}^4+4\tilde{\tau}^3+7\tilde{\tau}^2  + 6\tilde{\tau} +4} {(\tilde{\tau}+1)^2(\tilde{\tau}+2)^2} 
      \right)      + \mathcal{O}(\epsilon^4)
 \right]
\end{split}
\end{equation}
were derived in \cite{kawrakow2000accurate} for the case of utilising the screened Rutherford DCS for computing the GS angular distributions (as our case) with 
$\tilde{E}=E_0 - \Delta E/2$ mid-point energy along the step, $\epsilon=\Delta E/E_0$ fractional energy loss along the step and $\tilde{\tau} = \tilde{E}/m_e$ is 
the mid-point energy in electron rest-mass $m_e$ units. Interestingly, only $E_\text{eff}$ has dependence on the stopping power $L(E)$ through $b(\tilde{E})$ but 
even that can be neglected as $b(\tilde{E})$ is proportional to $\mathrm{d}C(E)/\mathrm{d}E$ with $C(E)=L(E)\beta^2$ being only a slowly varying function 
with $E$. Introducing then $\tilde{\epsilon} = \Delta E/\tilde{E}$ one can write the above as (keeping the same $\mathcal{O}(\epsilon^2)$ precision as above 
also when moving to $\tilde{\epsilon}$)
\footnote{there seems to be a typo in the \texttt{EGSnrc} implementation ($E_\text{eff}$ having $(\tilde{\tau}^2+2\tilde{\tau}+3)$ instead of $(\tilde{\tau}^2+3\tilde{\tau}+2)$) 
though not causing significant difference}
\begin{equation}
\begin{split}
E_\text{eff} & = \tilde{E} \left[ 
   1 - \frac{\tilde{\epsilon}^2}{24}  
          \frac{(5\tilde{\tau}^2+10\tilde{\tau}+6)}{(\tilde{\tau}^2+3\tilde{\tau}+2)}
 \right] \\
s_\text{eff} & = s \left[
  1 - \frac{\tilde{\epsilon}^2}{6} 
          \frac{(\tilde{\tau}^4+4\tilde{\tau}^3+7\tilde{\tau}^2  + 6\tilde{\tau} +4)} {(\tilde{\tau}+1)^2(\tilde{\tau}+2)^2} 
 \right]
\end{split}
\end{equation}
Therefore, the effects of energy loss to the angular distributions along a step can be accounted by replacing $E_0 \to E_\text{eff}, s \to s_\text{eff}$ 
then using the default, non-corrected pre-computed GS angular distributions at the corresponding $s_\text{eff}/\lambda(E_\text{eff})$ and $G_{l=1}(A(E_\text{eff}))$.

\subsection{Accounting spin-effect}
\label{sec::spin-effect}

\begin{figure}
\centering
 \includegraphics[width = 0.9\linewidth]{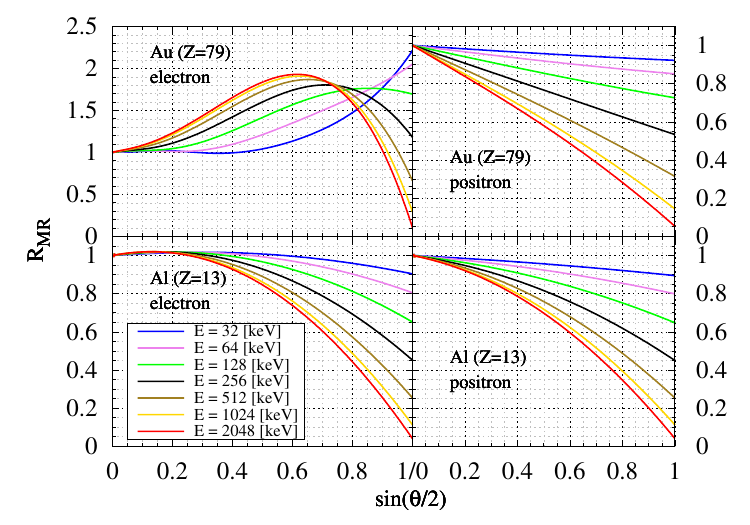}
\caption{The $R_{\text{MR}}$ Mott-to-Rutherford elastic DCS ratio, that shows the effect of the spin to the DCS, for $Au$ and $Al$ target atoms,  electron and position projectiles 
at various projectile kinetic energy. See APPENDIX \ref{app:mott-dcs} for the details of the computation.}
\label{fig:Rmott} 
\end{figure}

The screened Rutherford DCS has been considered so far as the electron-atom Coulomb scattering DCS input to the GS angular distribution 
computation. The screened Rutherford DCS is the result of describing scattering of spinless (unpolarised) 
electrons in the exponentially screened, point like Coulomb potential of the atom and solving the related Schr\"{o}dinger equation (that describes 
spinless particles) under the first Born approximation (see APPENDIX \ref{app:SRDCS}). The relativistic Dirac theory needs to be considered instead 
for accounting properly the electron spin. Very accurate electron/positron-atom Coulomb scattering DCS can be computed by \texttt{ELSEPA} \cite{salvat2005elsepa} using 
numerical Dirac partial-wave analysis, describing the target atom by the static-filed approximation (i.e. as a frozen charge distribution) with the possibility 
of including polarisation (by the field of the projectile) and exchange (for electron projectile) effects, the finite size of the target nucleus (i.e. charge distribution 
instead of a point) and accurate screening computed from numerical Dirac-Fock electron densities (instead of the simple exponential). The resulted numerical  
DCS, referred as ELSEPA-DCS hereafter, can also be used as input for the GS theory for calculating the multiple scattering angular distributions (accurately 
including spin, screening and all other affects). While these are the most accurate angular distributions unfortunately they are not very suitable for general-purpose 
simulations such as \texttt{Geant4} as either the amount of the pre-computed data required for the simulation is far too large and/or the numerical effort required 
during each simulation step is not affordable \cite{kawrakow2000egsnrc}. Therefore, simpler alternative DCS needs to be considered that accounts only some of 
the above mentioned effects, especially those that might affect significantly the multiple scattering distributions, while keeps the amount of the corresponding 
pre-computed data reasonable. The solution, utilised in \texttt{EGSnrc} \cite{kawrakow2000egsnrc} and followed in the present work, is described in the followings 
giving some insights into the underlying motivations with some details on the calculations performed.
\newline\newline \noindent
The simplest DCS, that already includes the electron/positron spin, can be obtained by solving the Dirac equation of elastic scattering in a point like, 
\textit{unscreened} Coulomb potential. The resulted DCS is know as the Mott DCS (M-DCS) and while it includes the spin, totally neglects the screening of 
the potential of the nucleus by the atomic electrons, i.e. describes elastic scattering of Dirac particles on the point-like Coulomb potential of the bare nucleus. 
The corresponding spinless DCS can be obtained by solving the same scattering problem (i.e. using the same point like, 
\textit{unscreened} Coulomb potential) but described by the (relativistic) Schr\"{o}dinger equation instead. This later results in the \textit{unscreened} Rutherford 
DCS (R-DCS) which is practically the same as the \textit{screened} Rutherford DCS derived in APPENDIX \ref{app:SRDCS} but having now $R \to \infty$ (i.e.
no screening of the potential) in Eq.(\ref{eq:SRF-DCS-R}) leading to 
\begin{equation}
\label{eq:RF-DCS}
  \frac{\mathrm{d}\sigma^{(R)}}{\mathrm{d}\Omega} = 
  \left( \frac{ZZ'e^{2}}{pc\beta} \right)^{2} \frac{ 1}{(1-\cos(\theta) )^{2}}    
\end{equation}
The most convenient form to express the Mott DCS is then (see e.g. \cite{fernandez1993cross,salvat2005elsepa} 
\begin{equation}
\label{eq:MOTT-DCS}
  \frac{\mathrm{d}\sigma^{(M)}}{\mathrm{d}\Omega} =  \frac{\mathrm{d}\sigma^{(R)}}{\mathrm{d}\Omega} R_{MR} (\theta)
\end{equation}
with $R_{MR}(\theta) \equiv \mathrm{d}\sigma^{(M)}(\theta)  / \mathrm{d}\sigma^{(R)}(\theta)$ being the Mott-to-Rutherford DCS ratio which is a smooth function 
of $\theta$ going to one when $\theta \to 0$ as shown in Fig. \ref{fig:Rmott}. 
As both DCS-s correspond to scattering on the same \textit{unscreened}, point like Coulomb potential but the Mott includes the spin while the Rutherford 
not, $R_{MR}(\theta)$ captures the spin effect to the DCS. Note, that the computation of the Mott DCS and $R_{MR}(\theta)$ adopted in the present work is summarised in  APPENDIX \ref{app:mott-dcs}. 

Combining and accounting then both the spin and screening effects might be done by replacing the \textit{unscreened} Rutherford DCS with its \textit{screend} version 
in the above equation leading to (see e.g. \cite{zeitler1964screening,fernandez1993cross,kawrakow2000egsnrc})
\begin{equation}
\label{eq:spin-cor-dcs}
  \frac{\mathrm{d}\sigma^{(SR\times R_{MR})}}{\mathrm{d}\Omega} =  \frac{\mathrm{d}\sigma^{(SR)}}{\mathrm{d}\Omega} R_{MR} (\theta)
\end{equation}
This assumes that spin and screening effects are decoupled which is motivated by the fact that 
$R_{MR}(\theta)$ goes to unity (i.e. no spin effects) at small scattering angles (especially with decreasing projectile energies), while screening affects most significantly 
these small angle scatterings (especially at lower projectile energies) \cite{zeitler1964screening,fernandez1993cross}. Furthermore, spin effects are more significant 
at larger target atomic numbers in case of electron projectile as demonstrated in Fig. \ref{fig:Rmott}. Therefore, Eq.(\ref{eq:spin-cor-dcs}), that was also adopted in 
\texttt{EGSnrc} to account the combined spin and screening effects, might be a reasonable approximation at projectile energies $> ~100$ [keV] in case of low-Z while 
requires $> ~1$ [MeV] for high-Z materials \cite{kawrakow2000egsnrc}.

\begin{table}
\caption{Summary of the DCS-s shown in Fig.\ref{fig:dcs-spin}}
\centering
  \begin{tabular}{| c | c | c | c | c |  }\hline
    DCS name         &           notation       & expression                    & screening & spin\\\hline\hline
    Rutherford          & DCS$_{\text{R}}$  & Eq.(\ref{eq:RF-DCS})    & x               & x    \\\hline
    \textit{screened} Rutherford  & DCS$_{\text{SR}}$ & Eq.(\ref{eq:SRF-DCS}) & exponential & x \\\hline
     Mott\tablefootnote{computed as described in APPENDIX \ref{app:mott-dcs}} & DCS$_{\text{M}}$ & Eq.(\ref{eq:MOTT-DCS}) & x & \checkmark \\\hline
     spin-corrected \textit{screened} Rutherford & \multirow{2}{*}{DCS$_{\text{SR}\times\text{R}_{\text{MR}}}$} & \multirow{2}{*}{Eq.(\ref{eq:spin-cor-dcs})} & \multirow{2}{*}{exponential} & \multirow{2}{*}{\checkmark} \\
     or \textit{screened} Mott DCS & & & & \\\hline
     numerical Dirac PWA & DCS$_{\text{ELSEPA}}$ & \texttt{ELSEPA}\tablefootnote{computed by numerical Dirac partial wave analysis using 
     \texttt{ELSEPA}\cite{salvat2005elsepa} including spin,
     polarisation, exchange effects (for e$^-$ projectile) as well as accurate screening of the potential of the (finite-size) nucleus by the atomic electrons computed from 
     numerical Dirac-Fock electron densities of the atom} & \checkmark & \checkmark \\\hline
  \end{tabular}
  \label{tb:DCS-PLOT}
\end{table}

\begin{figure}[H]
\centering
\begin{minipage}[b]{.45\linewidth}
\centering
	  \includegraphics[width = 1.0\linewidth]{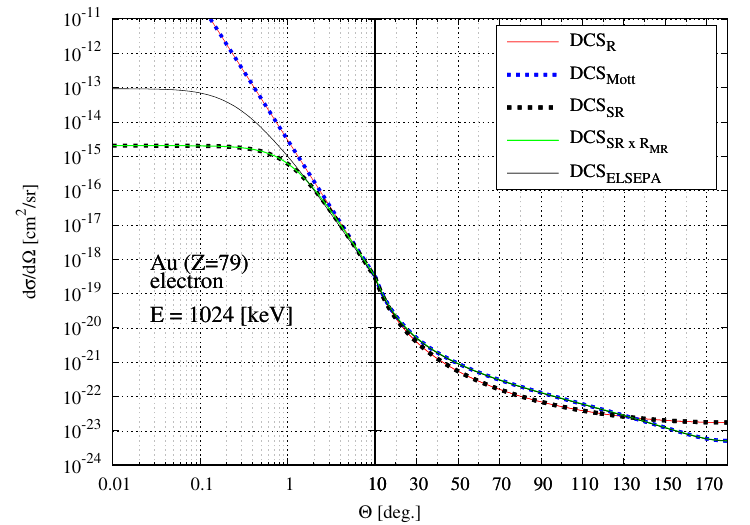}
\end{minipage}
\begin{minipage}[b]{.45\linewidth}
\centering
	  \includegraphics[width = 1.0\linewidth]{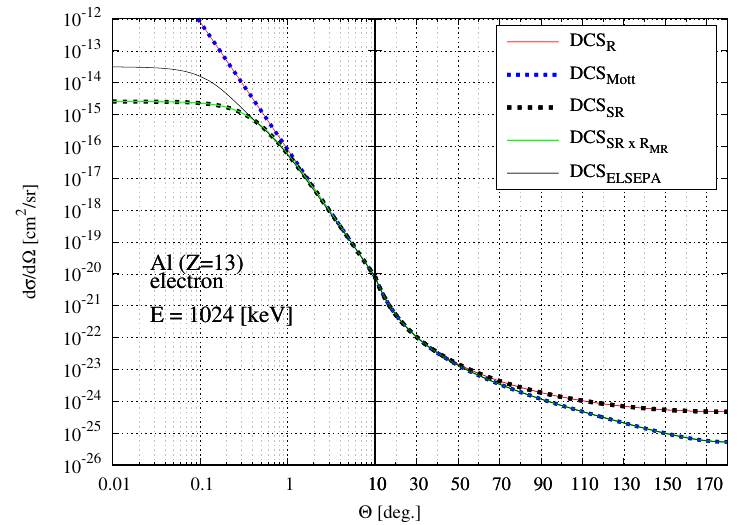}
\end{minipage}
\\
\begin{minipage}[b]{.45\linewidth}
\centering
	  \includegraphics[width = 1.0\linewidth]{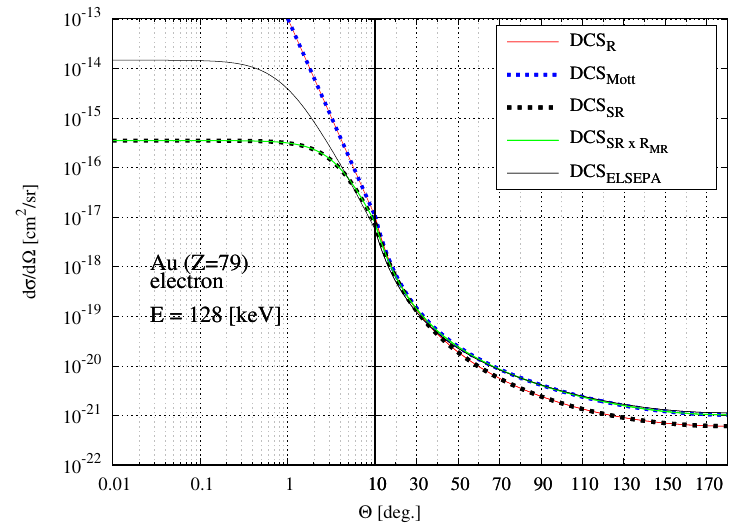}
\end{minipage}
\begin{minipage}[b]{.45\linewidth}
\centering
	  \includegraphics[width = 1.0\linewidth]{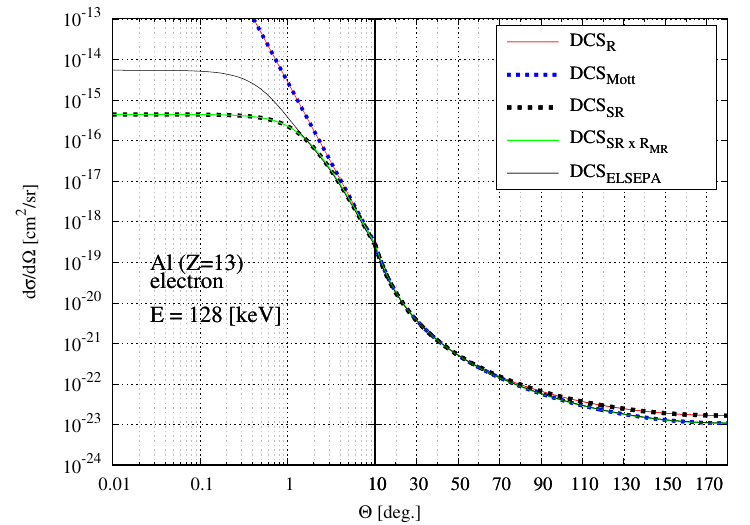}
\end{minipage}
\\
\begin{minipage}[b]{.45\linewidth}
\centering
	  \includegraphics[width = 1.0\linewidth]{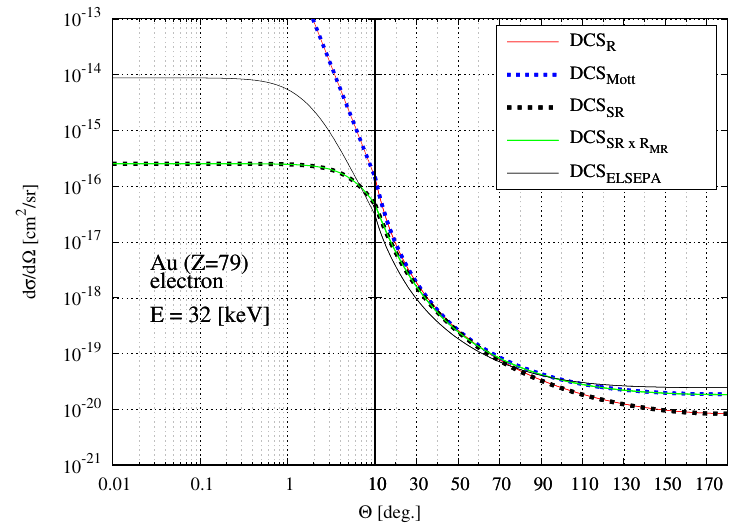}
\end{minipage}
\begin{minipage}[b]{.45\linewidth}
\centering
	  \includegraphics[width = 1.0\linewidth]{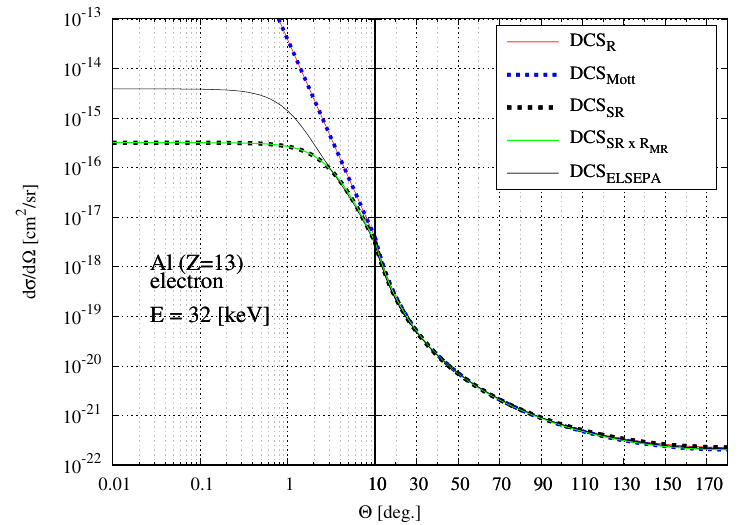}
\end{minipage}
\\
\begin{minipage}[b]{.45\linewidth}
\centering
	  \includegraphics[width = 1.0\linewidth]{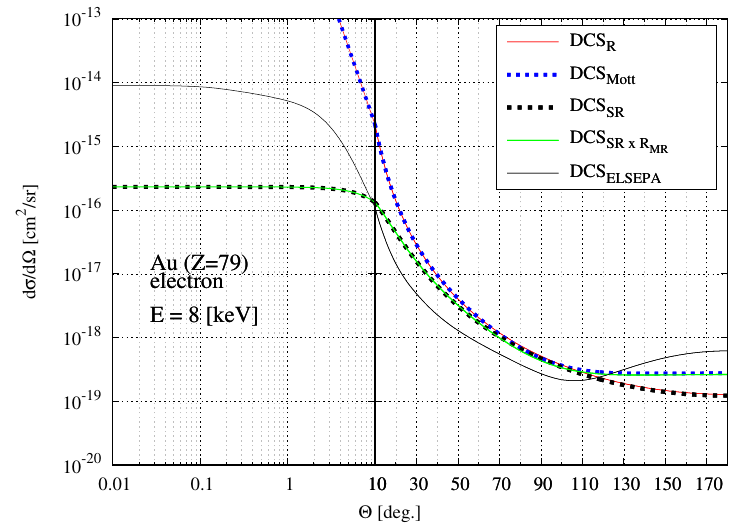}
\end{minipage}
\begin{minipage}[b]{.45\linewidth}
\centering
	  \includegraphics[width = 1.0\linewidth]{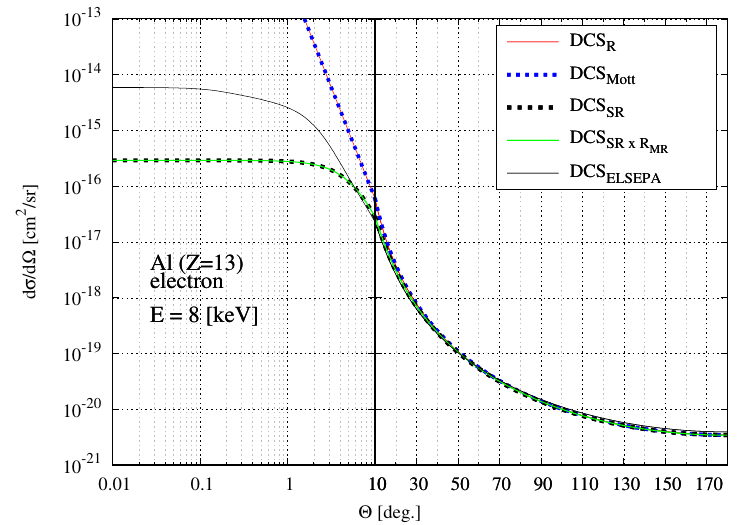}
\end{minipage}
\caption{Differential Cross Sections (DCS-s) for Coulomb scattering of electrons on $Au$ and $Al$ target atoms at different projectile kinetic energies. See Table 
\ref{tb:DCS-PLOT} and the text.}
\label{fig:dcs-spin}
\end{figure}

More might be seen regarding the applicability of the above factorisation by comparing the DCS-s computed with the different approximations 
summarised in Table \ref{tb:DCS-PLOT} and shown in Fig. \ref{fig:dcs-spin} for low- ($Al$) and high-Z ($Au$) target atoms at varying projectile kinetic energies. 
In general, one can see that the simple Rutherford (R) and screened Rutherford (SR) DCS-s are the same at large scattering angles while they differ only at small 
$\theta$ values where the effect of the screening is clearly visible. The Rutherford (R) and Mott (M) DCS are the same at these small scattering angles (and 
diverge in the same way due to the lack of screening) while they differ at larger $\theta$ values where spin effect can be important as expected from Fig. \ref{fig:Rmott}.
The spin-corrected screened Rutherford (or screening corrected Mott) DCS (SR$\times\text{R}_{\text{MR}}$) agrees with the Mott (M) DCS at large scattering angles while 
follows the screened Rutherford (SR) one at smaller $\theta$ values accounting both the spin and and screening effects respectively. 

Even more details on the validity of this factorisation can be revealed by investigating the relation of this relatively simple, spin-corrected, screened Rutherford DCS
to the most accurate (\texttt{ELSEPA}) DCS, especially the target atomic number (Z) and projectile kinetic energy dependence of their relation. The large and small 
scattering angle part of the DCS will be investigated separately in the followings as the projectile spin might affect the large scattering angle behaviour of the DCS 
while screening is more important at lower $\theta$ values.

\subsubsection*{Large scattering angles}
 
All the four DCS-s agree well in case of \textit{low-Z} ($Al$) target at \textit{smaller projectile energies} as the spin effect to the DCS tends to vanish for low-Z targets 
with decreasing kinetic energies (see Fig. \ref{fig:Rmott} right). At \textit{larger projectile energies}, when spin effects starts to be more pronounced even for \textit{low-Z targets},
the SR$\times\text{R}_{\text{MR}}$ DCS keeps following very well the \texttt{ELSEPA} one thanks to the spin correction. In contrast to the low-Z targets, spin effects 
might be important in case of \textit{high-Z} targets also at \textit{lower projectile energies} (see Fig. \ref{fig:Rmott} left) leading to large differences between those not including 
(e.g. R, SR) spin effects and the others. Furthermore, a relatively \textit{large projectile kinetic energy} $\sim128-1024$ [keV] is needed for reaching a good agreement between the 
SR$\times\text{R}_{\text{MR}}$ (same for M) and \texttt{ELSEPA} DCS-s in case of \textit{high-Z} targets even at these large scattering angles.

\subsubsection*{Small scattering angles}

A simple exponential screening is used to account the screening of the potential of the nucleus by the presence of the atomic electrons when deriving the screened 
Rutherford (SR) DCS. The limitation of this approximation is responsible for the relatively large differences observable at small scattering angles between the 
SR$\times\text{R}_{\text{MR}}$ and \texttt{ELSEPA} DCS-s that later includes accurate screening computed from the Dirac-Fock model of the atom. This difference 
reduces more rapidly with \textit{increasing projectile energies} in case of \textit{low-Z} targets and practically vanishes at $\sim 128$ [keV] for scattering angles 
$\theta > \sim 1^{\circ}$, while this requires larger $\sim 1024$ [keV] projectile energy for \textit{high-Z} targets. 
\newline\newline \noindent
Therefore, the SR$\times\text{R}_{\text{MR}}$ DCS given by Eq.(\ref{eq:spin-cor-dcs}), requires larger projectile kinetic energies $\sim 1$ [MeV] to be 
a reasonable approximation for high-Z target materials while somewhat lower $\sim 100$ [keV] energies are already sufficient for low-Z targets. 

\subsection{Screening parameter correction}
\label{sec:screening_parameter_correction}

\begin{figure}
\begin{minipage}[b]{.45\linewidth}
\centering
	  \includegraphics[width = 1.1\linewidth]{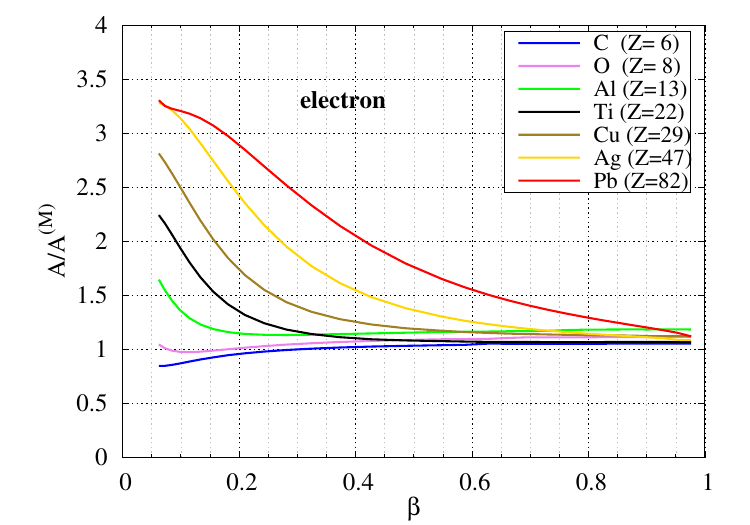}
\end{minipage}
\begin{minipage}[b]{.45\linewidth}
\centering
	  \includegraphics[width = 1.1\linewidth]{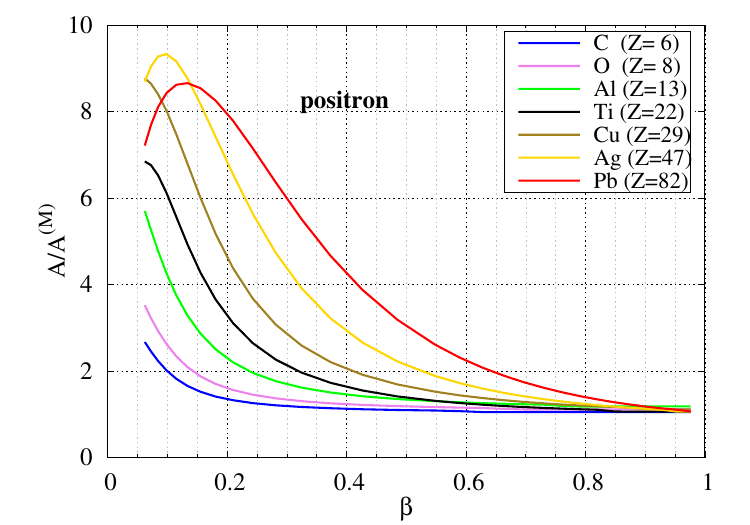}
\end{minipage}
\caption{Some $\kappa=A/A^{(M)}$ screening correction factors calculated by solving Eq.(\ref{eq:first-trans-correction}) numerically for the given projectile (left $e^-$; right $e^+$), target atom and kinetic energy to obtain the corrected $\kappa A^{(M)}$ screening parameter that satisfies Eq.(\ref{eq:first-trans-correction}) when used in the SR part of the 
SR$\times\text{R}_{\text{MR}}$ DCS. Similar correction is used in \texttt{EGSnrc} (see Figure 13. in \cite{kawrakow2000egsnrc} or Figure 1. in \cite{kawrakow2000cross}) and the reported correction factors are in a good agreement with the presented values (though different \texttt{PWA} DCS were used in 
\cite{kawrakow2000egsnrc,kawrakow2000cross} as reference which explains the small observable differences).}
\label{fig:screening-corr}
\end{figure}

The accuracy of the corresponding multiple scattering angular distributions can be increased even further though. To this end, it is important to recognise that the multiple scattering 
angular distributions are strongly determined by the first transport cross section
\footnote{\label{fn:on-first-GS-moment}according to Eq.(\ref{eq:first-gs-mom}) the first moment of the GS angular distribution, 
i.e. the mean scattering angle along an $s$ path is $<\cos(\theta)>_{GS}=\exp(-sG_{\ell=1}/\lambda) = \exp(-s/\lambda_1)$ where $\lambda_1$ is the first transport mean free path 
that is $1/\lambda_1=\mathcal{N}\sigma_1$ with the number of scattering centres (atoms) per unit volume $\mathcal{N}$ and 
first transport cross section $\sigma_1$ as given by Eq.(\ref{eq:first-trans-xsec})}
thus the large angle behaviour of the DCS ~\cite{fernandez1993theory}. It is 
obvious from Eq.(\ref{eq:first-trans-xsec}), that the small scattering angle part of the DCS do not contribute significantly to the first transport cross section due to the 
$[1-\cos(\theta)]$ factor. Therefore, the differences between the SR$\times\text{R}_{\text{MR}}$ and the \texttt{ELSEPA} DCS-s, observable at small $\theta < \sim 1^\circ$ 
scattering angles and discussed above, have only mild effect to the corresponding multiple scattering angular distributions. Furthermore, the $\theta > \sim 1^\circ$ 
accuracy of the SR$\times\text{R}_{\text{MR}}$ DCS, therefore the accuracy of the corresponding first transport cross section and the related multiple scattering angular 
distributions, can be increased further by applying a $\kappa$ correction to Moliere's screening parameter $A^{(M)}$ (used so far in $\mathrm{d}\sigma^{(SR)}/\mathrm{d}\Omega$
and given by Eq.(\ref{eq:screening_Molier_with_material_params_mix})) such that 
\begin{equation}
\label{eq:first-trans-correction}
 2\pi \int_{-1}^{1} [1-\cos(\theta)]  \frac{\mathrm{d}\sigma^{(SR\times R_{MR})}}{\mathrm{d}\Omega}  \mathrm{d}(\cos(\theta)) =
 2\pi \int_{-1}^{1} [1-\cos(\theta)]  \frac{\mathrm{d}\sigma^{\texttt{ELSEPA}}}{\mathrm{d}\Omega}\mathrm{d}(\cos(\theta))
\end{equation}
holds for the SR$\times\text{R}_{\text{MR}}$ DCS with the corrected $A=\kappa A^{(M)}$ screening parameter. In other words, 
the SR$\times\text{R}_{\text{MR}}$ DCS with the corrected screening parameter $A=\kappa A^{(M)}$ reproduces the first transport cross section 
computed from the corresponding accurate \texttt{ELSEPA} DCS. A data base of the $\kappa$ correction factor has been generated both for $e^-$ 
and $e^+$ projectiles, target atomic numbers Z = $1 - 98$ over an appropriate kinetic energy grid as suggested in \cite{kawrakow2000egsnrc} 
\footnote{an $E_{\text{kin}}$ kinetic energy grid on $1 - 100$ [keV] with 16 log-spacing points where linear interpolation on $\ln(E_{\text{kin}})$ is used 
and a second $\beta^2$ grid above from $E_{\text{kin}} = 100$ [keV] ($\beta^2 = 0.300546$) to $\beta^2 = 0.9999$ ($ E_{\text{kin}} = 50.5889$ [MeV]) 
with 16 linear-spacing points where linear interpolation on $\beta^2$ is used}. 
The numerical \texttt{ELSEPA} DCS and $R_{MR}$ Mott-to-Rutherford ratio were computed for $e^-/e^+$ at each target atomic number,
projectile energy point and the corresponding $\kappa(Z, E_\text{kin}, e^-/e^+)$ correction factors were determined by solving the above Eq.(\ref{eq:first-trans-correction}) 
numerically. Some examples of the $\kappa=A/A^{(M)}$ corrections factors are shown in Fig.\ref{fig:screening-corr} for different target atoms and projectile energies. 
The correction factors deviates significantly from unity only at \textit{high-Z} targets and lower projectile energies where the larger angle behaviour of the spin-corrected 
SR$\times\text{R}_{\text{MR}}$ DCS deviates from the corresponding \texttt{ELSEPA} DCS as discussed in the previous section.  These precomputed, element-wise 
correction factors $\kappa(Z, E_\text{kin}, e^-/e^+)$ are stored in separate files for $e^-/e^+$ over the pre-defined kinetic 
energy grid and used at initialisation for computing the material (compound) dependent screening correction factors for all materials present in the geometry using 
Eq.(\ref{eq:screening-corrected}) (separate factors for $e^-/e^+$ over the pre-defined kinetic energy grid). 

It must be noted, that it is the SR$\times\text{R}_{\text{MR}}$ DCS that gives the same first transport cross section with the corrected screening parameter as the 
corresponding \texttt{ELSEPA} DCS and not the pure SR DCS. Therefore, the pure SR DCS based (analytical) expressions for the 
elastic mean free path ($\lambda$ as given by Eq.(\ref{eq:inverse_elastic_Molier_mix})) and first transport coefficients ($G_{\ell=1}$ as given by 
Eq.(\ref{eq:wentzel_first_transport_coef}), used to compute the first transport mean free path as defined by Eq.(\ref{eq:inverse_elastic_Molier_mix}), do not provide 
a result that would agree with the corresponding more accurate \texttt{ELSEPA} DCS based value even when using the corrected screening parameter. 
On the same time, the accuracy of the first (and second) transport mean free path is important as it determines the first (and second) GS moment 
(see Eqs.(\ref{eq:first-gs-mom},\ref{eq:second-gs-mom})). These are especially important in the accurate electron-step algorithm discussed in Section \ref{sec:electron-step-alg}
as the first- and second-order spatial moments of the step are strongly determined by $Q_1\equiv sG_{\ell=1}/\lambda$ and $\gamma\equiv G_{\ell=2}/G_{\ell=1}$ 
(without energy loss along the $s$ step but see more at Section \ref{sec:electron-step-alg}).

Therefore, element-wise correction factors to the first transport cross sections (and to the second-to-first) have also been pre-computed based on the relation 
between the above mentioned analytical SR (using the corrected screening) and the more accurate numerical \texttt{ELSEPA} DCS based values. The corresponding 
element-wise correction factors are also stored in separate files for $e^-/e^+$ over the pre-defined kinetic energy grid, loaded at initialisation 
and utilised to prepare multiplicative correction factors to the SR model based analytical values such that the more accurate \texttt{ELSEPA} DCS related 
results are reproduced. Therefore, this material, projectile kinetic energy and type (i.e. $e^-$ or $e^+$) dependent multiplicative correction factors can be utilised at 
run time (in combination with the corrected screening parameter) to obtain more accurate $Q_1\equiv sG_{\ell=1}/\lambda$ and $\gamma\equiv G_{\ell=2}/G_{\ell=1}$
values for the electron-step computation similarly to \texttt{EGSnrc} \cite{kawrakow2000egsnrc} .

\subsection{Scattering power correction}
\label{sec::scattering-power-cor}
It has been discussed in Section \ref{sec:screening-par}, that $Z^2 \to Z(Z+\xi)$ was introduced with the usual $\xi=1$value to incorporate the angular 
deflections due to interactions with atomic electrons or more precisely the angular deflections due to the sub-threshold ionisations along the simulation 
step that are not modelled explicitly. However, secondary electrons with initial energy above the production threshold are generated and the corresponding 
angular deflections of the primary particle are accounted. Therefore, the corresponding contribution should not be included in the description of elastic 
scattering to avoid double counting and only the angular deflections due to sub-threshold ionisations should be accounted. It means that an appropriate, production 
threshold dependent $0\leq\xi\leq1$ value should be used. It will be assumed in the following that using $\xi=1$ is a valid approximation under the continuous 
slowing down case, i.e. when there is no delta ray production during the simulation\footnote{though the validity of this usual approximation is not clear}.

Kawrakow investigated this issue and derived an appropriate correction in \cite{kawrakow1997improved}. Exactly that correction is implemented thus 
summarised below (all details can be found in section IV of \cite{kawrakow1997improved}). First he obtained expressions for the 
mass scattering power (the rate of increase of the mean square scattering angle with traversed mass thickness) due to multiple Coulomb 
and Moller scatterings, i.e. due to scattering on the atomic potential and on the atomic electrons.
He obtained \cite{kawrakow1997improved}
\[
\frac{T_R(\zeta)}{\rho} = \frac{4 \pi N_A r_e^2}{\beta^2 \tau (\tau + 2)} \, \frac{Z(Z + \zeta)}{A} \, g_R(\chi_a)
\]
with 
\[
g_R(A^{(M)}) = \left( 1 + 2A^{(M)} \right) \ln \left( 1 + \frac{1}{A^{(M)}} \right) - 2.
\]
for the Coulomb scattering part of the mass scattering power, using the SR DCS with $A^{(M)}$ screening parameter (given in Section \ref{sec:screening-par}).
Depending on the value of $\xi$ it might include some, all or no contributions from electron-electron scattering.   
He obtained for the electron-electron part, that corresponds to the explicitly produced secondary electrons (i.e. above threshold), \cite{kawrakow1997improved}
\[
\frac{T_M(\tau_c)}{\rho} = \frac{4 \pi N_A r_e^2}{\beta^2 \tau (\tau + 2)} \, \frac{Z}{A} \, g_M(\tau, \tau_c),
\]
with 
\[
g_M(\tau, \tau_c) =
\ln \frac{\tau}{2 \tau_c}
+ \left[ \frac{(\tau + 2)}{(\tau + 1)^2} \right]^2 \ln \frac{2(\tau - \tau_c + 2)}{\tau + 4}
- \left[ \frac{(\tau + 2)^2}{4} + \frac{(\tau + 2)(\tau + 1/2)}{(\tau + 1)^2} \right]
\]

\[
\times \ln \frac{(\tau + 4)(\tau - \tau_c)}{\tau(\tau - \tau_c + 2)}
+ \frac{(\tau - 2 \tau_c)(\tau + 2)}{2} \left[ \frac{1}{\tau - \tau_c} - \frac{1}{(\tau + 1)^2} \right].
\]
using the M\o{}ller scattering and its free electron at rest approximation. $\tau, \tau_c, N_A, r_e, Z,  A$ are the primary electron kinetic energy in rest mass units, the 
secondary electron production kinetic energy threshold in rest mass units, Avogadro's constant, electron classical radius, target atomic number and relative atomic mass.

Assuming then that an elastic mass scattering power with the original $\xi=\xi_0=1$ parameter $T_R(\xi_0)/\rho$ correctly describes the total contributions to the 
scattering power (i.e. both the real Coulomb and the total (above and below) electron-electron contributions), the elastic scattering part should have a parameter 
$\xi_e$ that satisfies the following equation
\[
 \frac{T_R(\xi_0)}{\rho} = \frac{T_M(\tau,\tau_c)}{\rho} + \frac{T_R(\xi_e)}{\rho}
\]
This states that the real Coulomb plus the total electron-electron contribution ($T_R(\xi_0)/\rho$) should be equal to the above threshold electron-electron 
($T_M(\tau,\tau_c)/\rho$)  plus an elastic scattering contribution that includes both the real Coulomb and the sub-threshold electron-electron contributions 
($T_R(\xi_e)/\rho$). As we would like to have exactly this latter, we can solve the above equation for $\xi_e$ that gives
\[
  \xi_e = \xi_0 - \frac{g_M(\tau, \tau_c)}{g_R(A^{(M)})}
\] 
As $g_M(\tau, \tau_c)$ is 0 whenever $\tau$ is below the secondary production threshold (i.e. $\tau\leq \tau_c$) then $\xi_e=\xi_0$ with the usual $\xi=1$ value covering all 
electron-electron interaction contribution while otherwise $g_M(\tau, \tau_c)/g_R(A^{(M)}) \in (0,1]$ leading to $\xi_e<\xi_0$ removing the above threshold contributions.

This $\xi$ parameter is used in $Z_S$ defined in Eq.(\ref{eqs:Moliere_mixture_shorts}) that transforms now with this scattering power correction(\textit{spc}) to 
\[
Z_S^{\text{spc}}  = \sum_{i=1}^{N_{e}}  n_i Z_i(Z_i+\xi_{e}) = \sum_{i=1}^{N_{e}}  n_i Z_i \left[Z_i+\xi_{0} - \frac{g_M(\tau, \tau_c)}{g_R(A^{(M)})}\right] 
= Z_S \left[1 - \frac{\sum_{i=1}^{N_{e}}  n_i Z_i }{Z_S}  \frac{g_M(\tau, \tau_c)}{g_R(A^{(M)})} \right]
\]   
with $Z_S = \sum_{i=1}^{N_{e}}  n_i Z_i (Z_i + \xi_0)$. This multiplicative, kinetic energy and production threshold dependent scattering power correction parameter to $Z_S$ is
computed at initialisation for each material-cuts couple above an appropriate discrete kinetic energy threshold (from $2E_c$ for electron and $E_c$ for positron) and interpolated 
at run time. Note, that only the elastic mean free path depends on $Z_S$ through Moliere's $b_c$ parameter as given by Eq.(\ref{eq:inverse_elastic_Molier_mix})
and Eq.(\ref{eq:bc_Molier_mix}) ($Z_S$ from $\chi_{cc}^2$ and $b_c$ cancel each other in the case of $A^{(M)}$; see Eq.(\ref{eq:screening_Molier_with_material_params_mix})).
Therefore, we apply this scattering power correction only to the elastic mean free path at run time after calculated by using Eq.(\ref{eq:inverse_elastic_Molier_mix}).
\newline\newline
It has been mentioned above (and also discussed by Kawrakow in \cite{kawrakow1997improved}) that the full validity of the $Z^2\to Z(Z+\xi)$ replacement to 
incorporate angular deflections from electron-electron interactions to the elastic scattering is not clear. However, \textit{the main advantage of the above scattering 
power correction} to avoid double counting \textit{is that it makes the simulation of angular deflections independent from the secondary electron production threshold} 
used in the ionisation. 
This can be especially important in case of light (low Z) materials as the contribution from sub-threshold electron-electron interactions to the total angular 
deflection can be significant compared to the Coulomb scattering. Moreover, it must be noted that neglecting those angular deflections can 
lead to inaccurate modelling results independently from the applied, i.e. multiple or single, Coulomb scattering model. 



\section{Electron-step algorithm}
\label{sec:electron-step-alg}
The algorithm implemented for calculating the post-step point in a condensed history (CH) simulation step is described in this section. The implemented 
algorithm is the one developed by Kawrakow in \cite{kawrakow1998condensed} including the energy loss correction from \cite{kawrakow2000accurate}. 
Only the essentials required for the calculations are summarised in this section below while all details can be found in \cite{kawrakow1998condensed,kawrakow2000accurate}.
\newline\newline
Kawrakow \& Bielajew analysed the accuracy of the existed CH algorithms  in \cite{kawrakow1998condensed} by the direct comparison of the (low, up to 2nd order) 
moments of the spatial distributions and spatial-angular correlations to the exact results of Lewis' theory (neglecting energy loss!). Using the expansion in terms 
of $Q_1\equiv s/\lambda_1$\footnote{$Q_1 \equiv s/\lambda_1 = G_{1}s/\lambda$ which characterises the mean $\cos(\theta)$ scattering angle along the $s$ step in 
the GS distribution $\langle \cos(\theta) \rangle_{GS} = \exp(-sG_{\ell=1}/\lambda)$ as given by Eq.(\ref{eq:first-gs-mom}). The coefficients also depends 
on $\gamma\equiv G_{\ell=2}/G_{\ell=1}$ as will be discussed shorty.}, a new algorithm for computing the post-step 
positions was proposed. It has been shown, that the new algorithm well reproduces Lewis' results for $<z>, <zv_z>, <xv_x + yv_y>, <z^2>, <x^2+y^2>$ 
(with local error of $\mathcal{O}(Q_1^4)$) when energy loss along the step is neglected. Since $Q_1 = s/\lambda_1$, this imposes a step limit 
such that $Q_1 = s/\lambda_1 \leq 0.5$ to remain within the validity range of the series expansion. An energy loss correction was derived later in \cite{kawrakow2000accurate}
that completes the model. The algorithm, implemented within the model, will be described below staring with the formulas assuming constant energy along 
the step (derived in \cite{kawrakow1998condensed} section 4.4) then introducing the energy loss corrections (derived in Appendix B. of \cite{kawrakow2000accurate}).
\newline\newline
In order to reduce the approximation error, the CH step is divided to two equal sub-steps in the algorithm proposed in \cite{kawrakow1998condensed}. 
Polar and azimuthal scattering angles are sampled for the first $\theta_1, \phi_1$ and second $\theta_2, \phi_2$ sub-steps using the pre-computed 
GS angular distributions for  $\theta_1$ and $\theta_2$ as described in section \ref{sec:run_time_sampling}.
Having a coordinate system with $z$-axis pointing to the particle momentum direction, i.e. an initial momentum direction $\mathbf{\hat{d}_0}=(0,0,1)$ in this system, 
the direction after the first scattering
\begin{equation}
\mathbf {\hat{d}_1}
= R_z(\phi_1)\,R_y(\theta_1)\,\mathbf {\hat{d}_0}
=
\begin{pmatrix}
\sin\theta_1\cos\phi_1\\[2pt]
\sin\theta_1\sin\phi_1\\[2pt]
\cos\theta_1
\end{pmatrix}
\equiv
\begin{pmatrix}
u_1\\[2pt]
v_1\\[2pt]
w_1
\end{pmatrix}.
\end{equation}
where $R_z, R_y$ are the matrices for rotation around the $z$ and $y$ axes. The final direction, i.e. after the second scattering, then
\begin{align}
\mathbf {\hat{d}_f}
&=
\begin{pmatrix}
\sin\theta_1\cos\phi_1\,\cos\theta_2\cos\phi_2
-\sin\theta_1\sin\phi_1\,\sin\phi_2
+\cos\theta_1\,\sin\theta_2\cos\phi_2
\\[6pt]
\sin\theta_1\cos\phi_1\,\cos\theta_2\sin\phi_2
+\sin\theta_1\sin\phi_1\,\cos\phi_2
+\cos\theta_1\,\sin\theta_2\sin\phi_2
\\[6pt]
-\sin\theta_1\cos\phi_1\,\sin\theta_2
+\cos\theta_1\,\cos\theta_2
\end{pmatrix}
\\[8pt]
&=
\begin{pmatrix}
u_1\,\cos\theta_2\cos\phi_2
- v_1\,\sin\phi_2
+  \cos\theta_1\,\sin\theta_2\cos\phi_2
\\[6pt]
u_1\,\cos\theta_2\sin\phi_2
+ v_1\,\cos\phi_2
+  \cos\theta_1\,\sin\theta_2\sin\phi_2
\\[6pt]
- u_1\,\sin\theta_2
+ \cos\theta_1\,\cos\theta_2
\end{pmatrix}
\\[8pt]
&=
\begin{pmatrix}
\cos\phi_2\,t_m - v_1\,\sin\phi_2
\\[6pt]
\sin\phi_2\,t_m + v_1\,\cos\phi_2
\\[6pt]
- u_1\,\sin\theta_2 + \cos\theta_1\,\cos\theta_2
\end{pmatrix}
\;\equiv\;
\begin{pmatrix}
\sin\theta\,\cos\phi
\\[6pt]
\sin\theta\,\sin\phi
\\[6pt]
\cos\theta
\end{pmatrix}
\equiv
\begin{pmatrix}
u_f\\[2pt]
v_f\\[2pt]
w_f
\end{pmatrix}.
\end{align}
\noindent
where $t_m \;=\; \sin\theta_1\cos\phi_1\,\cos\theta_2 + \cos\theta_1\,\sin\theta_2$ while $\theta,\phi$ are the combined polar and azimuthal scattering angles.
Note that $\theta_i$ and $\phi_i$ are samples from the same distributions therefore the indices $1\to2$ are interchangeable. 
Having the pre-step point in the origin of the coordinate system, the post-step point coordinates of the CH step might be given \cite{kawrakow1998condensed} as
\begin{equation}
\label{eq:final_dir}
\begin{aligned}
x_f &= s\Big[\, \eta_1\sin\theta_1\cos\phi_1 
  &+& \eta_2\sin\theta_2\big(\cos\phi_1\cos\phi_2 - \cos\theta_1\sin\phi_1\sin\phi_2\big)
  &+& (1-\eta_0-\eta_1-\eta_2)\sin\theta\cos\phi \,\Big],\\[4pt]
y_f &= s\Big[\, \eta_1\sin\theta_1\sin\phi_1 
  &+& \eta_2\sin\theta_2\big(\sin\phi_1\cos\phi_2 + \cos\theta_1\cos\phi_1\sin\phi_2\big)
  &+& (1-\eta_0-\eta_1-\eta_2)\sin\theta\sin\phi \,\Big],\\[4pt]
z_f &= s\Big[\, \eta_0 + \eta_1\cos\theta_1
  &+& \eta_2\cos\theta_2 
  &+& (1-\eta_0-\eta_1-\eta_2)\cos\theta \,\Big].
\end{aligned}
\end{equation}
The coefficients $\eta_0, \eta_1, \eta_2$ were then determined such that the moments considered are well reproduced leading to the following equations
\begin{equation}
\label{eq:etas}
\begin{aligned}
\eta_0 & = (1 - \eta)/2,\\[4pt]
\eta_1 & =  \eta\delta,\\[4pt]
\eta_2 & =  \eta(1-\delta)
\end{aligned}
\end{equation}
with $\eta \in [0,1]$ sampled from the $2\eta\mathrm{d}\eta$ distribution and 
\begin{equation}
\label{eq:delta}
\delta = \frac{1}{2} + \frac{\sqrt{6}}{6} -\left(\frac{1}{4\sqrt{6}} - \gamma\frac{4-\sqrt{6}}{24\sqrt{6}} \right) Q_1 
\end{equation}
with the 
\begin{equation}
\label{eq:gamma}
\begin{aligned}
Q_1       & \equiv sG_{\ell=1}/\lambda    &\;\text{with}\; &G_{\ell=1}^{SR} =  2A  \left[ \ln \left(1+\frac{1}{A}\right) (A+1) -1 \right],\\[4pt]
\gamma & \equiv  G_{\ell=2}/G_{\ell=1}  &\;\text{with}\; &G_{\ell=2}^{SR} =  6A(1+A)\left[ (1+2A)\ln \left(1+\frac{1}{A}\right) -2 \right] 
\end{aligned}
\end{equation}
notation introduced in section \ref{sec:screening_parameter_correction} for these quantities that determines the first and second moments of the 
GS angular distribution along a CH step with a path-length of $s$. The expression for $G_{\ell=1}$ and $G_{\ell=2}$ are given by 
Eq.(\ref{eq:wentzel_first_transport_coef}) and Eq.(\ref{eq:wentzel_second_transport_coef}) as derived in APPENDIX \ref{app:SRDCS} 
under the Screened Rutherford DCS approximation with a screening parameter of $A$. Introducing $a_1 = (1-\eta_0-\eta_1-\eta_2)$ 
and $u_2=\sin\theta_2\cos\phi_2,\;v_2=\sin\theta_2\sin\phi_2$ the post-step coordinates can be written in  the compact form
\begin{equation}
\label{eq:final_pos}
\begin{aligned}
x_f &= s\Big[\, \eta_1u_1
  &+\;& \eta_2\big(\cos\phi_1u_2 - \cos\theta_1\sin\phi_1v_2\big)
  &+\;& a_1u_f \,\Big],\\[4pt]
y_f &= s\Big[\, \eta_1v_1
  &+\;& \eta_2\big(\sin\phi_1u_2 + \cos\theta_1\cos\phi_1v_2\big)
  &+\;& a_1v_f\,\Big],\\[4pt]
z_f &= s\Big[\, \eta_0 + \eta_1\cos\theta_1
  &+\;& \eta_2\cos\theta_2
  &+\;& a_1w_f \,\Big].
\end{aligned}
\end{equation}
With the above expressions for $\eta_0, \eta_1, \eta_2$, now $a_1 = (1-\eta_0-\eta_1-\eta_2) = \eta_0$. 

The above algorithm for computing the CH post-step point position was obtained to reproduce the moments of the spatial distributions 
and spatial-angular correlations of the Lewis' theory when neglecting the energy loss along the step.
Corrections for accounting energy loss were derived in \cite{kawrakow2000accurate} (see Appendix B there)
by comparing the corresponding first and second spatial moments, i.e. the mean longitudinal $<z>$ and squared lateral $<r^2> = <x^2+y^2>$ displacements, 
to the results of Lewis' theory including the energy loss. The first order correction (in $\Delta E$) corresponds to changing the parameters $\eta_0 \to eta_0(1+\alpha)$
and $\delta \to \delta + \alpha_2$ with the following expressions for $\alpha_1, \alpha_2$ resulting from the above-mentioned comparison \cite{kawrakow2000accurate}
\begin{equation}
\begin{aligned}
\alpha_1 &= - \frac{\kappa_1'(\tilde{E})}{\kappa_1(\tilde{E})} \frac{\Delta E}{2} + \mathcal{O}(\Delta E^2) ,\\[4pt]
\alpha_2 &=  \left[ \frac{\kappa_2'(\tilde{E})}{\kappa_2(\tilde{E})} -  \frac{\kappa_1'(\tilde{E})}{\kappa_1(\tilde{E})} \right] \frac{\Delta E}{2\sqrt{6}} + \mathcal{O}(\Delta E^2) 
\end{aligned}
\end{equation}
with $\kappa_\ell \equiv G_\ell/\lambda = \lambda_\ell^{-1} = \Sigma_\ell$ being the $\ell$-th macroscopic transport cross section (or inverse transport mean free path) 
as given by the definition Eq.(\ref{eq:transport_mean_free_path_with_Gl}), $\tilde{E}=E_0 - \Delta E/2$ is the mid-point energy along the step as used in section 
\ref{sec:eloss_cor_gs} while $(\cdot)'$ denotes their derivatives with respect to $E$. 
The expressions for $\Sigma_{\ell=1}$ and $\Sigma_{\ell=2}$ as given in \cite{kawrakow2000accurate} Eqs.(A5) (denoted by $\kappa_{\ell=1}, \kappa_{\ell=2}$ there) 
can be obtained by starting the general definition $\Sigma_\ell \equiv \lambda_{\ell}^{-1} = G_{\ell}/\lambda$ as given by Eq.(\ref{eq:transport_mean_free_path_with_Gl}) 
then substituting the SR DCS forms of $G_{\ell=1}$ and $G_{ell=2}$ as given by Eqs.(\ref{eq:wentzel_first_transport_coef},\ref{eq:wentzel_second_transport_coef}) with the expression 
for $\lambda$ as given by Eq.(\ref{eq:inverse_elastic_Molier_mix}) (with the usual approximation of dropping the $1/(1+A)$ factor from this latter)
\begin{equation}
\begin{aligned}
   \Sigma_{\ell=1}^{(SR)}  & = 2A\frac{b_c}{\beta^2}\left[ \ln \left(1+\frac{1}{A} \right)(1+A) - 1 \right],\\
   \Sigma_{\ell=2}^{(SR)}  & = 6A\frac{b_c}{\beta^2}(1+A)\left[ (1+2A)\ln \left(1+\frac{1}{A}\right) -2 \right]
\end{aligned}
\end{equation}
which are the same as Eqs.(A5) in \cite{kawrakow2000accurate} when substituting Eq.(\ref{eq:screening_Molier_with_material_params_mix}) for $A$.  Substituting these into 
the above expressions for $\alpha_1$ and $\alpha_2$ yields \cite{kawrakow2000accurate}\footnote{the $-\alpha_1^2/(4\sqrt{6})$ leading 2nd order contribution is also added 
to $\alpha_2$ in the EGSnrc implementation}
\begin{equation}
\label{eq:eloss_alphas}
\begin{aligned}
\alpha_1 & \approx
\left[
\frac{2+2\tilde{\tau}+\tilde{\tau}^2}{1+\tilde{\tau}}
-\frac{1+\tilde{\tau}}{\ln\!\big(1+1/\tilde{A}\big)\,(1+\tilde{A})-1}
\right]\,
\frac{\tilde{\epsilon}}{2+\tilde{\tau}}\,,\\
\alpha_2 & \approx
\frac{\tilde{\epsilon}}{
\sqrt{6}\,(1+\tilde{\tau})(2+\tilde{\tau})\,
\big[\ln\!\big(1+1/\tilde{A}\big)\,(1+\tilde{A})-1\big]\,
\big[\ln\!\big(1+1/\tilde{A}\big)\,(1+2\tilde{A})-2\big]
}\,.
\end{aligned}
\end{equation}
where $\tilde{\epsilon} = \Delta E/\tilde{E}$, $\tilde{\tau} = \tilde{E}/m_e$ is the mid-point energy in electron rest-mass $m_e$ units just as in Section \ref{sec:eloss_cor_gs} and 
$\tilde{A}$ is the screening parameter evaluated at the mid-point energy of the step (i.e. taking $(pc)^2=\tilde{E}(\tilde{E}+2m_e)$ in 
Eq.(\ref{eq:screening_Molier_with_material_params_mix})). 
\newline\newline
When accounting energy loss along the CH step, $\eta$ is sampled from the $2\eta \mathrm{d}\eta$ distribution, $\gamma$ and $\delta$ are computed 
by Eqs.(\ref{eq:gamma},\ref{eq:delta}), $\eta_0, \eta_1, \eta_2$ can be calculated by Eq.(\ref{eq:etas}) then $\eta_0 = \eta_0(1+\alpha_1)$ and $\delta = \delta+\alpha_2$  
can be obtained after evaluating the $\alpha_1$ and $\alpha_2$ energy loss correction parameters given by Eqs.(\ref{eq:eloss_alphas}). With these forms of $\eta_0$ and 
$\delta$, the parameter $a_1$, introduced at Eq.(\ref{eq:final_pos}), have the $a_1 \equiv 1-\eta_0-\eta_1-\eta_2 = \eta_0(1-\alpha_1)$ form now. The post-step position 
of the CH step can be calculated then by using these parameters in Eq.(\ref{eq:final_pos}) after dividing the step into two equal sub-steps, generating $\theta_1, \phi_1$ 
and $\theta_2, \phi_2$ for the sub-steps and calculating the corresponding final direction components as given  by Eq.(\ref{eq:final_dir}).

As described in Section \ref{sec:screening_parameter_correction}, corrections to the $A$ screening parameter of the SR DCS have been derived such that the 
SR$\times\text{R}_\text{Mott}$ DCS (i.e. the Mott corrected SR DCS) reproduces the first transport cross section computed from the accurate numerical ELSEPA DCS.  
It was also discussed that the expressions for $Q_1$ and $\gamma$ as given by Eqs.(\ref{eq:gamma}) correspond to the pure SR DCS and correction factors
to these quantities have also been computed such that the corrected parameters agree with the accurate numerical ELSPA DCS based values. When the Mott 
correction is active, these corrections to $A$, $Q_1$ and $\gamma$ are also used beyond sampling the angular deflections along the CH step from distributions 
that corresponds to the Mott corrected SR DCS. The additional scattering power correction, that is described in Section \ref{sec::scattering-power-cor}, is also active 
in this case that completes the most accurate setting of the model.


\section{Summary and final notes}
\label{sec:summary} 
\textcolor{red}{\textbf{TO BE COMPLETED...}}



\newpage

\begin{appendices}
\numberwithin{equation}{section}

\section{Angular distribution} 
\subsection{Goudsmit-Saunderson series}
\label{app:GS_series}
First the single scattering distribution, given by Eq.(\ref{eq:gs_single_scattering_distribution_def}), will be expressed in terms of Legendre polynomials.
Legendre polynomials form a \textit{complete orthogonal system} on $\{-1,+1\}$  where orthogonality is
\begin{equation}
\label{eq:Pell-otho}
 \int_{-1}^{1} P_{\ell}(x)P_{k}(x) \mathrm{d}x = 
 \left\{
  \begin{array}{l l}
    C_{\ell} = \frac{2}{2\ell+1} \quad & \text{if $\ell=k$ } \\
    0       \quad & \text{if $\ell \neq k$ } \\
  \end{array} 
  \right.
\end{equation}
and completeness means that for any $f(x)$ continuous functions on $\{-1,+1\}$  
\begin{equation}
\lim _{\ell \to \infty} E_{\ell}(a_0,\cdots,a_{\ell}) \equiv \lim _{\ell \to \infty} || f(x) - \sum_{\ell} a_{\ell}P_{\ell}(x)||^2 = 0 
\end{equation}
i.e. the least square error $E_{\ell}$ converges to zero.   
So any $f(x)$ function on 
$\{-1,+1\}$ can be written as
\begin{equation}
 f(x)=\sum_{\ell=0}^{\infty}a_{\ell}P_{\ell}(x)
\end{equation}
where the $a_{\ell}$ coefficients can be derived by multiplying both side by $P_{k}(x)$, integrating both side over $x$ while using 
the orthogonality property of the Legendre polynomials that yields 
\begin{equation}
 a_{\ell}=\frac{2\ell+1}{2}\int_{-1}^{+1}P_{\ell}(x)f(x)\mathrm{d}x
\end{equation}
Therefore, the single scattering distribution $f_1(\theta)$ can be written as 
\begin{equation}
f_{1}(\cos(\theta)) = \sum_{\ell=0}^{\infty}a_{\ell}P_{\ell}(\cos(\theta))
\end{equation}
where the coefficients are 
\begin{equation}
\begin{split}
 a_{\ell} = & \frac{2\ell+1}{2} \int_{-1}^{1} f_{1}(\cos(\theta)) P_{\ell}(\cos(\theta)) \mathrm{d}\cos(\theta)
  = \frac{2\ell+1}{4\pi} 
  \overbrace{  
    \int_{-1}^{1}  
       \underbrace{2\pi f_{1}(\cos(\theta))}_{p(\cos(\theta)) \gets Eq.(\ref{eq:gs_single_scattering_pdf_cost})} 
    P_{\ell}(\cos(\theta)) \mathrm{d}\cos(\theta)}^{\langle P_{\ell}(\cos(\theta))\rangle = F_{\ell}} 
\\ = &\frac{2\ell+1}{4\pi} F_{\ell} = \frac{2\ell+1}{4\pi} \langle P_{\ell}(\cos(\theta))\rangle
\end{split}
\end{equation}
and using them
\begin{equation}
\label{eq:app_gs_single_scattering_distribution_withLeg}
 f_{1}(\cos(\theta)) =  \sum_{\ell=0}^{\infty}  \frac{2\ell+1}{4\pi}  F_{\ell} P_{\ell}(\cos(\theta))
\end{equation}
\\
\\
The second step is to obtain the expression for $f_{n>1}(\cos(\theta))$ i.e. angular distribution of particles after $n>1$ elastic interactions. 
The angular distribution after $n=2$ elastic interaction will be derived first. 
Suppose that the particle was originally moving into the $\hat{\bar{z}}$ direction and the first interaction 
resulted $\theta_1,\phi_1$ with corresponding direction of $\hat{\bar{d}}_1$ while the second 
$\theta_2,\phi_2$. In order to get a direction $\hat{\bar{d}}$ defined by $\theta,\phi$ after the second 
interaction 
\begin{equation}
\label{eq:app_cost2_req}
\begin{split}
 \cos(\theta_2) = & {\hat{\bar{d}}_1}^{T} \hat{\bar{d}}
\\ = & [\sin(\theta_1)\cos(\phi_1),\sin(\theta_1)\sin(\phi_1),\cos(\theta_1)]
 \left[ 
   \begin{array}{c}
     \sin(\theta)\cos(\phi) \\ \sin(\theta)\sin(\phi) \\ \cos(\theta) 
   \end{array}
 \right]
\\ = & \cos(\theta)\cos(\theta_1)+\sin(\theta)\sin(\theta_1)
     \underbrace{[\cos(\phi_1)\cos(\phi)+\sin(\phi_1)\sin(\phi)]}_{\cos(\phi_1-\phi)}
\\ = & \cos(\theta)\cos(\theta_1)+\sin(\theta)\sin(\theta_1)\cos(\phi_1-\phi)     
\end{split}
\end{equation}
must be fulfilled. 
The probability of finding the 
particle after two interactions moving into the solid angle element $\mathrm{d}\omega$ 
around the direction $\hat{\bar{d}}$ defined by $\theta$ $\phi$
\begin{equation}
f_{n=2}(\theta) \mathrm{d}\omega 
= \int_{0}^{\pi}  \int_{0}^{2\pi}   f_{1}(\theta_1)f_{1}(\theta_2)
\sin(\theta_1) \mathrm{d}{\phi_1} \mathrm{d}{\theta_1}  \underbrace{ \sin(\theta) \mathrm{d}{\theta} \mathrm{d}{\phi} }_{\mathrm{d}\omega}
\end{equation}
and
\begin{equation}
\label{eq:app_second_scattering_pdf0}
\begin{split}
f_{n=2}(\cos(\theta)) = &
    \int_{-1}^{+1}\int_{0}^{2\pi} f_1(\cos(\theta_1))f_{1}(\cos(\theta_2))  \mathrm{d}\phi_1 \mathrm{d}(\cos(\theta_1))
\\  & \hspace*{-2cm} =
\int_{-1}^{+1}\int_{0}^{2\pi}
 \left[
 \sum_{\ell=0}^{\infty} \frac{2\ell+1}{4\pi} F_{\ell} P_{\ell}(\cos(\theta_1))
 \right] 
 \left[
 \sum_{\ell'=0}^{\infty} \frac{2\ell'+1}{4\pi} F_{\ell'} P_{\ell'}(\cos(\theta_2))
 \right]
 \mathrm{d}\phi_1  \mathrm{d}(\cos(\theta_1)) 
\end{split}
\end{equation}
(note that $F_k$ is $\langle P_{k}(\cos(\chi)) \rangle$ in a single interaction). To make the next step one need to use the requirement
for $\cos(\theta_2)$ given by Eq.(\ref{eq:app_cost2_req}) and the corresponding result from the addition theorem for spherical harmonics i.e.
\begin{equation}
\label{eq:app_from_addrule_spherical}
 P_{\ell}(\cos(\theta_2))
 = P_{\ell}(\cos(\theta))P_{\ell}(\cos(\theta_1))
   + 2\sum_{m=1}^{\ell}\frac{(\ell-m)!}{(\ell+m)!} P_{\ell}^{m}(\theta,\phi)P_{\ell}^{m}(\theta_1,\phi_1)
   \cos[m(\phi-\phi_1)]
\end{equation}

\begin{scriptsize}
\noindent
The addition theorem of spherical harmonics says that if 
\begin{equation}
 \cos(\theta_2) \equiv \cos(\theta)\cos(\theta_1)+\sin(\theta)\sin(\theta_1)\cos(\phi_1-\phi)     
\end{equation}
then 
\begin{equation}
\label{eq:app_addrule_sphericalh0}
 P_{\ell}(\cos(\theta_2))
 =  \frac{4\pi}{2\ell+1} \sum_{m=-\ell}^{\ell} Y_{\ell}^{m}(\theta,\phi)Y_{\ell}^{^{*}m}(\theta_1,\phi_1) 
\end{equation}
One can use:\\
the associated Legendre polynomials:
\begin{equation}
 P_{\ell}^{-m}(x) = (-1)^{m} \frac{(\ell-m)!}{(\ell+m)!} P_{\ell}^{m}(x)
\end{equation}
the spherical harmonics
\begin{equation}
 Y_{\ell}^{m}(\theta,\phi) = \sqrt{ \frac{2\ell+1}{4\pi} \frac{(\ell-m)!}{(\ell+m)!} }
 P_{\ell}^{m}(\cos(\theta))\exp(im\phi)
\end{equation}
and the complex conjugate of the spherical harmonics
\begin{equation}
\begin{split}
 Y_{\ell}^{^{*}m}(\theta,\phi) = & (-1)^{m} Y_{\ell}^{-m}(\theta,\phi) 
 =(-1)^{m} \sqrt{ \frac{2\ell+1}{4\pi} \frac{(\ell+m)!}{(\ell-m)!} }
 P_{\ell}^{-m}(\cos(\theta))\exp(-im\phi)
\\ = & (-1)^{m} \sqrt{ \frac{2\ell+1}{4\pi} \frac{(\ell+m)!}{(\ell-m)!} }
 (-1)^{m} \frac{(\ell-m)!}{(\ell+m)!} P_{\ell}^{m}(\cos(\theta)) \exp(-im\phi)
\end{split}
\end{equation}
to plug in these into Eq.(\ref{eq:app_addrule_sphericalh0})

\begingroup
\allowdisplaybreaks
\begin{align*}
 P_{\ell}(\cos(\theta_2))
 =&  \frac{4\pi}{2\ell+1} \sum_{m=-\ell}^{\ell} Y_{\ell}^{m}(\theta,\phi)Y_{\ell}^{^{*}m}(\theta_1,\phi_1)
 =  \frac{4\pi}{2\ell+1} \sum_{m=-\ell}^{\ell} 
    \left\{
    \sqrt{ \frac{2\ell+1}{4\pi} \frac{(\ell-m)!}{(\ell+m)!} }  P_{\ell}^{m}(\cos(\theta))\exp(im\phi)
\right.
\\& 
\left.
    (-1)^{2m} \sqrt{ \frac{2\ell+1}{4\pi} \frac{(\ell+m)!}{(\ell-m)!} }
    \frac{(\ell-m)!}{(\ell+m)!} P_{\ell}^{m}(\cos(\theta_1)) \exp(-im\phi_1)
\right\}
\\=& \sum_{m=-\ell}^{\ell} \frac{(\ell-m)!}{(\ell+m)!} P_{\ell}^{m}(\cos(\theta)) P_{\ell}^{m}(\cos(\theta_1)) 
  \exp(im(\phi-\phi_1))
\\=& \underbrace{ P_{\ell}(\cos(\theta)) P_{\ell}(\cos(\theta_1)) }_{m=0\; \text{case}}
   +\sum_{m=-\ell}^{-1} \frac{(\ell-m)!}{(\ell+m)!} P_{\ell}^{m}(\cos(\theta)) P_{\ell}^{m}(\cos(\theta_1)) 
   \exp(im(\phi-\phi_1))
\\&+ \sum_{m=1}^{\ell} \frac{(\ell-m)!}{(\ell+m)!} P_{\ell}^{m}(\cos(\theta)) P_{\ell}^{m}(\cos(\theta_1)) 
   \exp(im(\phi-\phi_1))
\\=&  P_{\ell}(\cos(\theta)) P_{\ell}(\cos(\theta_1)) 
   +\sum_{m=1}^{\ell} \frac{(\ell+m)!}{(\ell-m)!} P_{\ell}^{-m}(\cos(\theta)) P_{\ell}^{-m}(\cos(\theta_1)) 
   \exp(-im(\phi-\phi_1))
\\&+ \sum_{m=1}^{\ell} \frac{(\ell-m)!}{(\ell+m)!} P_{\ell}^{m}(\cos(\theta)) P_{\ell}^{m}(\cos(\theta_1)) 
   \exp(im(\phi-\phi_1))
\\ \numberthis 
=& P_{\ell}(\cos(\theta)) P_{\ell}(\cos(\theta_1)) 
    +\sum_{m=1}^{\ell}
     \frac{(\ell+m)!}{(\ell-m)!} 
     \underbrace{(-1)^{m}\frac{(\ell-m)!}{(\ell+m)!}P_{\ell}^{m}(\cos(\theta)) }_{ P_{\ell}^{-m}(\cos(\theta))}
\\&  \underbrace{(-1)^{m}\frac{(\ell-m)!}{(\ell+m)!}P_{\ell}^{m}(\cos(\theta_1)) }_{P_{\ell}^{-m}(\cos(\theta_1))} 
     \exp(-im(\phi-\phi_1))
\\&+ \sum_{m=1}^{\ell} \frac{(\ell-m)!}{(\ell+m)!} P_{\ell}^{m}(\cos(\theta)) P_{\ell}^{m}(\cos(\theta_1)) 
   \exp(im(\phi-\phi_1))
\\=&  P_{\ell}(\cos(\theta)) P_{\ell}(\cos(\theta_1))
   +\sum_{m=1}^{\ell} \frac{(\ell-m)!}{(\ell+m)!} 
   P_{\ell}^{m}(\cos(\theta)) P_{\ell}^{m}(\cos(\theta_1)) 
   \exp(-im(\phi-\phi_1))
\\&+ \sum_{m=1}^{\ell} \frac{(\ell-m)!}{(\ell+m)!} P_{\ell}^{m}(\cos(\theta)) P_{\ell}^{m}(\cos(\theta_1)) 
   \exp(im(\phi-\phi_1))
\\=&  P_{\ell}(\cos(\theta)) P_{\ell}(\cos(\theta_1))
   +\sum_{m=1}^{\ell}
      \frac{(\ell-m)!}{(\ell+m)!} 
      P_{\ell}^{m}(\cos(\theta)) P_{\ell}^{m}(\cos(\theta_1))
 \left\{
      \underbrace{\exp[-im(\phi-\phi_1)] 
      + \exp[im(\phi-\phi_1)]}_{2\cos[m(\phi-\phi_1)]}
    \right\}
\\=&  P_{\ell}(\cos(\theta)) P_{\ell}(\cos(\theta_1)) 
   +2\sum_{m=1}^{\ell} \frac{(\ell-m)!}{(\ell+m)!} 
   P_{\ell}^{m}(\cos(\theta)) P_{\ell}^{m}(\cos(\theta_1)) 
   \cos[m(\phi-\phi_1)]   
\end{align*}
\endgroup
\end{scriptsize}
Substituting Eq.(\ref{eq:app_from_addrule_spherical}) into Eq.(\ref{eq:app_second_scattering_pdf0}) yields 
one can get 
\begin{equation}
\begin{split}
f_{n=2}(\cos(\theta)) = & \int_{-1}^{+1}\int_{0}^{2\pi}
 \left[
 \sum_{\ell=0}^{\infty} \frac{2\ell+1}{4\pi} F_{\ell} P_{\ell}(\cos(\theta_1))
 \right]
 \left[
 \sum_{\ell'=0}^{\infty} \frac{2\ell'+1}{4\pi} F_{\ell'} 
  P_{\ell'}(\cos(\theta)) P_{\ell'}(\cos(\theta_1)) 
\right.
\\&
\left.
  +2\sum_{m=1}^{\ell'} \frac{(\ell'-m)!}{(\ell'+m)!} 
   P_{\ell'}^{m}(\cos(\theta)) P_{\ell'}^{m}(\cos(\theta_1)) 
   \cos[m(\phi-\phi_1)]   
 \right]
 \mathrm{d}{\phi_1}  \mathrm{d}(\cos(\theta_1))
\end{split}
\end{equation}
Carrying out the integration over $\phi_1$ all the contributions coming from the last term will disappear. 
What remains is 
\begin{equation}
\begin{split}
f_{n=2}(\cos(\theta)) = & 2\pi \int_{-1}^{+1}
 \left[
 \sum_{\ell=0}^{\infty} \frac{2\ell+1}{4\pi} F_{\ell} P_{\ell}(\cos(\theta_1))
 \right]
 \left[
 \sum_{\ell'=0}^{\infty} \frac{2\ell'+1}{4\pi} F_{\ell'} 
  P_{\ell'}(\cos(\theta)) P_{\ell'}(\cos(\theta_1)) 
 \right]
 \mathrm{d}(\cos(\theta_1))
\end{split}
\end{equation}
Using the orthogonality of the Legendre polynomials on 
the $\{-1,+1\}$ interval (i.e. only the terms $\ell=\ell'$ will give non-zero contribution to the sum of 
integrals) 
\begin{equation}
\begin{split}
f_{n=2}(\cos(\theta)) = &  2\pi
 \sum_{\ell=0}^{\infty} \left( \frac{2\ell+1}{4\pi} \right)^{2} (F_{\ell})^{2}P_{\ell}(\cos(\theta))
  \underbrace{  
   \int_{-1}^{1} P_{\ell}(\cos(\theta_1))P_{\ell}(\cos(\theta_1)) \mathrm{d}{\cos(\theta_1)} 
  }_{2/(2\ell+1)}
\\ = &
 \sum_{\ell=0}^{\infty} \frac{2\ell+1}{4\pi} (F_{\ell})^{2}P_{\ell}(\cos(\theta))
\end{split}
\end{equation}
and in general after $n$ elastic interactions the angular distribution 
\begin{equation}
 f_{n}(\cos(\theta)) = \sum_{\ell=0}^{\infty} \frac{2\ell+1}{4\pi} (F_{\ell})^{n} P_{\ell}(\cos(\theta))
\end{equation}

\subsection{The moments of the GS angular distribution}
\label{app:GS-moments}
Having the Goudsmit-Saunderson angular distribution as given in Eq.(\ref{eq:gs_angual_distribution_final_in_cost}), which is the 
PDF of $\cos(\theta)$ after travelling an $s$ path, one can calculate $\langle P_{\ell}(\cos(\theta)) \rangle_{GS}$ 
(similarly to the single scattering case of Eq.(\ref{eq:gs_fel_def})) as
\begin{equation}
\begin{split}
\langle P_{\ell}(\cos(\theta)) \rangle_{GS}= &
 \int_{-1}^{+1} P_{\ell}(\cos(\theta)) F(s;\theta)_{GS} \mathrm{d}(\cos(\theta)) 
 =  \int_{-1}^{+1}  P_{\ell}(\cos(\theta)) 
\\& \left\{ \sum_{\ell'=0}^{\infty} \frac{2\ell'+1}{2} \exp(-s/\lambda_{\ell'})  P_{\ell'}(\cos(\theta)) \right\}
  \mathrm{d}(\cos(\theta))
\\=&  \sum_{\ell'=0}^{\infty} \frac{2\ell'+1}{2} \exp(-s/\lambda_{\ell'})
   \int_{-1}^{+1} P_{\ell}(\cos(\theta))P_{\ell'}(\cos(\theta)) \mathrm{d}(\cos(\theta))
\\=&  \frac{2\ell+1}{2} \exp(-s/\lambda_{\ell}) \frac{2}{2\ell+1}  
\\=&  \exp(-s/\lambda_{\ell}) = \exp(-s G_{\ell}/\lambda)
\end{split}
\end{equation}
where the orthogonality of the Legendre polynomials on $\{-1,+1\}$, as given by Eq.(\ref{eq:Pell-otho}), was used 
while the last equation was derived on the basis of Eq.(\ref{eq:transport_mean_free_path_with_Gl}).
This yields the first GS moment,  i.e. the mean $\cos(\theta)$ after travelling an $s$ path 
\begin{equation}
\label{eq:first-gs-mom}
\langle \cos(\theta) \rangle_{GS} = \langle P_{1}(\cos(\theta)) \rangle_{GS} = \exp(-s/\lambda_{1}) = \exp(-sG_{\ell=1}/\lambda)
\end{equation}
and the second moment (by using $P_{\ell=2}(x)=0.5(3x^{2}-1)$ and from this $x^{2}=1/3(2 P_{\ell=2}(x)+1)$)
\begin{equation}
\label{eq:second-gs-mom}
\langle \cos^{2}(\theta) \rangle_{GS}=\frac{1}{3}(2\langle P_{2}(\cos(\theta)) \rangle_{GS}+1)
 = \frac{1}{3}(2\exp(-s/\lambda_{2})+1) = \frac{1}{3}(2\exp(-sG_{\ell=2}/\lambda)+1)
\end{equation}

\subsection{The screened Rutherford DCS}
\label{app:SRDCS}
The screened Rutherford DCS (SR-DCS) for elastic scattering and the corresponding expressions for elastic scattering cross section, single scattering 
distribution, transport coefficients will be derived in this section. 
\newline\newline
The (relativistic) screened Rutherford DCS is the result of describing elastic scattering of spinless 
(unpolarised, relativistic) electrons in the exponentially \textit{screened}, point like Coulomb potential of the atom and solving the related (relativistic) Schr\"{o}dinger 
equation (that describes spinless particles) under the first Born approximation. Under these conditions, the differential elastic scattering cross section 
$\mathrm{d}\sigma/\mathrm{d}\Omega$ can be expressed 
by the scattering amplitude $f(\theta,\phi)$ as
\begin{equation}
 \frac{\mathrm{d}\sigma}{\mathrm{d}\Omega}=|f|^{2}\quad \sigma=\int \frac{\mathrm{d}\sigma}{\mathrm{d}\Omega} \mathrm{d}\Omega
\end{equation}
The scattering amplitude in the first Born approximation can be written as
\begin{equation}
 f_{B1}(\theta,\phi)=-\frac{2m}{4\pi\hbar^{2}}\int e^{i(\bar{k}_f-\bar{k}_i)\bar{r}'}V(\bar{r}')\mathrm{d}^{3}r'
\end{equation}
where $\bar{k}_i$ is the wave vector of the incident plane wave, $\bar{k}_f$ is the wave vector of the outgoing
(scattered) spherical wave. Note that in elastic scattering $k_i=k_f\equiv k$. Furthermore, 
$\hbar\bar{q}=\hbar(\bar{k}_f-\bar{k}_i)$ is the momentum transfer and 
$q^2=|\bar{k}_f-\bar{k}_i|^2=2k^2(1-\cos(\theta))=2k^2(2\sin^2(\theta/2))$ where $\theta=\angle(\bar{k}_i,\bar{k}_f)$
is the scattering angle. Let assume $V(\bar{r})\equiv V(r)$ i.e. spherically symmetric scattering potential
and substitute $\bar{q}=\bar{k}_f-\bar{k}_i$ by choosing the coordinate system for the integration such that
$\bar{q}=q\hat{\bar{z}}$
\begin{equation}
\begin{split}
 f_{B1}(\theta) &=  -\frac{2m}{4\pi\hbar^{2}} \int_{0}^{\infty} \int_{0}^{\pi} \int_{0}^{2\pi} 
           e^{iqr'\cos(\theta')}V(r')
 \mathrm{d}\phi'\sin(\theta')\mathrm{d}\theta' r'^{2}\mathrm{d}r
\\&= -\frac{2m}{4\pi\hbar^{2}} 2\pi \int_{0}^{\infty} \int_{0}^{\pi} 
           e^{iqr'\cos(\theta')}V(r')
  \sin(\theta')\mathrm{d}\theta' r'^{2}\mathrm{d}r'
\\&= -\frac{2m}{4\pi\hbar^{2}} 2\pi \int_{0}^{\infty} 
           \left[ - \frac{ e^{iqr'\cos(\theta')} }{iqr'} \right]_{0}^{\pi} V(r')
   r'^{2}\mathrm{d}r'
\\&= -\frac{2m}{4\pi\hbar^{2}} 2\pi \int_{0}^{\infty} 
           \frac{ 2\sin(qr')}{qr'} V(r')
   r'^{2}\mathrm{d}r'
\\&= -\frac{2m}{q\hbar^{2}} \int_{0}^{\infty} 
           \sin(qr')r'V(r')
   \mathrm{d}r'
\end{split}
\end{equation}
where $(\exp(ix)-\exp(-ix))/2i=\sin(x)$ was used to get $2\sin(qr')/qr'$. Using a simple exponentially screened 
Coulomb potential as a scattering potential  
\begin{equation}
\label{eq:EXP-SCREENED-COULOMB}
 V(r)=\frac{ZZ'e^{2}}{r} e^{-r/R}
\end{equation}
with a screening radius $R$ (target atomic number of $Z$ and projectile charge $Z'e$)
\begin{equation}
 \begin{split}
  f_{B1}(\theta) &=
  -\frac{2m}{q\hbar^{2}} ZZ'e^{2} \int_{0}^{\infty} 
           \sin(qr')e^{-r'/R}
   \mathrm{d}r'
\\&= -\frac{2m}{q\hbar^{2}} ZZ'e^{2} \left[
   \frac{ -Re^{-r'/R}[qr'\cos(qr')+\sin(qr')]}{q^2 R^2+1}
   \right]_{0}^{\infty}
\\&=-\frac{2m}{q\hbar^{2}} ZZ'e^{2} \left[
   \frac{ R^{2}q}{q^2 R^2+1}
   \right]=   
  -\frac{2m}{\hbar^{2}} ZZ'e^{2} \left[
   \frac{ 1}{q^2 + R^{-2}}
   \right] 
\\&=-\frac{2m}{\hbar^{2}} ZZ'e^{2} \left[
   \frac{ 1}{2k^{2}[1-\cos(\theta) + R^{-2}/(2k^{2})]}
   \right]      
 \end{split}
\end{equation}
where $q^{2}=2k^{2}(1-\cos(\theta))$ was used to obtain the last equation. The corresponding elastic DCS
\begin{equation}
\label{eq:SRF-DCS-R}
 \begin{split}
  \frac{\mathrm{d}\sigma}{\mathrm{d}\Omega} &=  |f_{B1}(\theta)|^{2} =
  (ZZ'e^{2})^{2} \frac{4m^{2}}{\hbar^{4}4k^{4}} \frac{ 1}{(1-\cos(\theta) + R^{-2}/(2k^{2}))^{2}} 
\\&=  \left( \frac{ZZ'e^{2}}{pc\beta} \right)^{2} \frac{ 1}{(1-\cos(\theta) + R^{-2}/(2k^{2}))^{2}}    
 \end{split}
\end{equation}
where $\hbar k = p$ and
\begin{equation}
  \frac{4m^{2}}{\hbar^{4}4k^{4}} 
   = \left( \frac{m}{p^{2}} \right)^{2} = \left( \frac{mc^2}{p^2c^2} \right)^{2} 
   = \left( \frac{E_t}{p^2c^2} \right)^{2} = \left( \frac{1}{pc\beta} \right)^{2}
\end{equation}
was used to obtain the last equation. One can introduce a \textit{screening parameter} $A$ such that
\begin{equation}
\label{eq:Wentzel_screening_parameter}
  A \equiv \frac{1}{4}\left(\frac{\hbar}{p}\right)^2 R^{-2} 
\end{equation}
and since
\begin{equation}
 \frac{1}{2k^2 R^2} = \frac{1}{2 (p/\hbar)^2 R^2} = \frac{1}{2 R^2}\left(\frac{\hbar}{p}\right)^2  
  = \frac{1}{2}\left(\frac{\hbar}{p}\right)^2 R^{-2} = 2 A
\end{equation}
the elastic DCS then can be written as
\begin{equation}
\label{eq:Wentzel_DCS}
  \frac{\mathrm{d}\sigma}{\mathrm{d}\Omega} = 
  \left( \frac{ZZ'e^{2}}{pc\beta} \right)^{2} \frac{ 1}{(1-\cos(\theta) + 2A)^{2}}    
\end{equation}
The total cross section then becomes
\begin{equation}
\label{eq:total_cross_section_Wentzel}
\begin{split}
 \sigma &=  \int \frac{\mathrm{d}\sigma}{\mathrm{d}\Omega} \mathrm{d}\Omega =
  \int_0^\pi \int_0^{2\pi} 
   \left( \frac{ZZ'e^{2}}{pc\beta} \right)^{2} \frac{ 1}{(1-\cos(\theta) + 2A)^{2}}    
  \mathrm{d}\phi \sin(\theta)\mathrm{d}\theta 
\\& = \left( \frac{ZZ'e^{2}}{pc\beta} \right)^{2} 2\pi \int_0^\pi 
    \frac{ 1}{(1-\cos(\theta) + 2A)^{2}}    
  \sin(\theta)\mathrm{d}\theta 
\\& = \left( \frac{ZZ'e^{2}}{pc\beta} \right)^{2} 2\pi  
    \left[ \frac{ 1}{(\cos(\theta)-1-2A)} \right]_0^\pi
\\& = \left( \frac{ZZ'e^{2}}{pc\beta} \right)^{2} 2\pi  
    \left[ \frac{ 1}{2A} -\frac{ 1}{2+2A} \right]
\\& = \left( \frac{ZZ'e^{2}}{pc\beta} \right)^{2} \frac{ \pi }{ A(1+A)}
\end{split}
\end{equation}
and the single elastic scattering angular distribution 
\begin{equation}
\label{eq:wentzel_sigle_scattering_distr}
\begin{split}
 f_1(\theta) &=\frac{1}{\sigma}\frac{\mathrm{d}\sigma}{\mathrm{d}\Omega} 
  = \left(\frac{pc\beta }{ZZ'e^{2}} \right)^{2} \frac{ A(1+A) }{\pi }
    \left( \frac{ZZ'e^{2}}{pc\beta} \right)^{2} \frac{ 1}{(1-\cos(\theta) + 2A)^{2}}
\\& = \frac{ 1 }{\pi } \frac{A(1+A)}{ (1-\cos(\theta) + 2A)^{2} }
\end{split}
\end{equation}
The corresponding $\ell$-th transport coefficient $G_\ell$~\cite{fernandez1993theory}
\begin{equation}
\label{eq:Wentzel_transport_coefs}
\begin{split}
 G_\ell &= 1-F_\ell
 = 1- 2\pi\int_{-1}^{+1} f_1(\theta)P_\ell(\cos(\theta))\mathrm{d}(\cos(\theta))
\\& = 1-2\pi \frac{ 1 }{\pi } A(1+A) \int_{-1}^{+1} 
           \frac{P_\ell(\cos(\theta))}{(1-\cos(\theta) + 2A)^{2}} 
              \mathrm{d}(\cos(\theta))
\\& =  1-\ell[Q_{\ell-1}(1+2A)-(1+2A)Q_{\ell}(1+2A)]              
\end{split}
\end{equation}
where $Q_\ell(x)$ are Legendre functions of the second kind. 

The $\ell$-th transport mean free path $\lambda_{\ell}$ with the $\ell$-th transport coefficient
$G_{\ell}$ is given by Eq.(\ref{eq:transport_mean_free_path_with_Gl}) and the elastic mean 
free path $\lambda$ is by Eq.(\ref{eq:elastic_mfp_def}) that yields
\begin{equation}
 \lambda_{\ell}^{-1} = \frac{G_{\ell}}{\lambda} = \mathcal{N}\sigma G_{\ell}
\end{equation}
Substituting the total elastic cross section Eq.(\ref{eq:total_cross_section_Wentzel}) into this expression
\begin{equation}
 \lambda_{\ell}^{-1(SR)} = \mathcal{N} G_{\ell} \left( \frac{ZZ'e^{2}}{pc\beta} \right)^{2} \frac{ \pi }{ A(1+A)}
\end{equation}
where the superscript SR indicates that the expression was derived by using a screened Rutherford DCS.
In order to get the expression for the first transport mean free path, first
one needs to derive the expression for the $G_{\ell=1}$ based on the screened Rutherford DCS 
of Eq.(\ref{eq:Wentzel_transport_coefs}) at $\ell=1$.
\begin{equation}
\label{eq:wentzel_first_transport_coef}
\begin{split}
 G_{\ell=1}^{(SR)} = & 1-[Q_{0}(1+2A)-(1+2A)Q_{1}(1+2A)]
 \\& = 2A\left[ \ln \left(\frac{1+A}{A}\right) (A+1) -1 \right]   
\end{split}
\end{equation}
by using the forms of the Legendre functions of the second kind
$Q_0(x)= 0.5\ln( (x+1)/(x-1) )$ and $Q_1(x)= 0.5x\ln( (x+1)/(x-1) )-1$.
Substituting this into the above expression for the first transport mean free path, one can get
\begin{equation}
\label{eq:simple_modified_Wentzel_with_lambda1}
\begin{split}
 \lambda_{1}^{-1(SR)} =& \mathcal{N} \left( \frac{ZZ'e^{2}}{pc\beta} \right)^{2} \frac{ \pi }{ A(1+A)}
           2A\left[ \ln \left(\frac{1+A}{A}\right) (A+1) -1 \right]
\\& = \mathcal{N} \left( \frac{ZZ'e^{2}}{pc\beta} \right)^{2}
           2\pi \left[ \ln \left(\frac{1+A}{A}\right) -\frac{1}{1+A} \right]
\end{split}
\end{equation}
Similarly
\begin{equation}
\label{eq:wentzel_second_transport_coef}
\begin{split}
 G_{\ell=2}^{(SR)} = & 1-2[Q_{1}(1+2A)-(1+2A)Q_{2}(1+2A)]
 \\& = 6A(1+A)\left[ (1+2A)\ln \left(\frac{1+A}{A}\right) -2 \right]   
\end{split}
\end{equation}
using $Q_2(x)= (3x^2-1)/4 \ln( (x+1)/(x-1) )-3x/2$ and
\begin{equation}
\label{eq:simple_modified_Wentzel_with_lambda2}
\begin{split}
 \lambda_{2}^{-1(SR)} =& \mathcal{N} \left( \frac{ZZ'e^{2}}{pc\beta} \right)^{2} \frac{ \pi }{ A(1+A)}
           6A(1+A)\left[ (1+2A)\ln \left(\frac{1+A}{A}\right) -2 \right]
\\& = \mathcal{N} \left( \frac{ZZ'e^{2}}{pc\beta} \right)^{2}
           6\pi \left[ (1+2A)\ln \left(\frac{1+A}{A}\right) -2 \right]
\end{split}
\end{equation}

\subsection{Optimal parameter of the variable transformation}
\label{app:kaw_optimal_value_of_a}
The optimal value of the parameter $a$ of the transformation Eq.(\ref{eq:kaw_transform}) can be determined by
plugging Eqs.(\ref{eq:kaw_firstderiv},\ref{eq:kaw_secondderiv}) into Eq.(\ref{eq:kaw_general_optimalaity}) and soling for $a$:
\begin{equation}
 \begin{split}
  0 = & \int_{-1}^{+1} \left[ F(s;\mu)_{GS}^{2+} \left( -\frac{[1-\mu+2a]^{2}}{2a(1+a)} \right) \right]^{2}
        \left[ -2\frac{1-\mu(1+2a)}{[1-\mu+2a]^{3}} \right]  \mathrm{d}\mu
\\=&  \int_{-1}^{+1} \left[ F(s;\mu)_{GS}^{2+} \right]^{2} 
        \left[
         \left( -\frac{[1-\mu+2a]}{4a^{2}(1+a)^{2}} \right)(-2(1-\mu(1+2a))
        \right] \mathrm{d}\mu 
\\=&   \sum_{\ell=0}^{\infty} (\ell+0.5)\xi_{\ell}(s,\lambda,A) \sum_{k=0}^{\infty} (k+0.5)\xi_{k}(s,\lambda,A)
        \int_{-1}^{+1}
        \left[ 
            - \frac{(1+2a)(1-\mu)^{2}}{2a^{2}(1+a)^{2}} + \frac{2\mu}{(1+a)^{2}}
        \right]
        \mathrm{d}\mu
\\=& -\frac{(1+2a)}{2a^{2}(1+a)^{2}} \underbrace{ \left[
        \sum_{\ell=0}^{\infty} (\ell+0.5)\xi_{\ell}(s,\lambda,A) \sum_{k=0}^{\infty} (k+0.5)\xi_{k}(s,\lambda,A)  
        \int_{-1}^{+1} P_{\ell}(\mu) P_{k}(\mu) (1-\mu)^{2} \mathrm{d}\mu
        \right] }_{\alpha}
\\& \; +\frac{2}{(1+a)^{2}} \underbrace{  \left[
        \sum_{\ell=0}^{\infty} (\ell+0.5)\xi_{\ell}(s,\lambda,A) \sum_{k=0}^{\infty} (k+0.5)\xi_{k}(s,\lambda,A)  
        \int_{-1}^{+1} P_{\ell}(\mu) P_{k}(\mu) \mu \mathrm{d}\mu
        \right] }_{\beta}
\\=&  4\beta a^{2} -2\alpha a -\alpha        
 \end{split}
\end{equation}
where 
\begin{equation}
\label{eq:app_kaw_termxi}
\xi_{i}(s,\lambda,A) = 
       \frac{ 
              \mathrm{e}^{-(s/\lambda)G_{i}(A)} -\mathrm{e}^{-(s/\lambda)}\left[1 +(s/\lambda)(1-G_{i}(A)) \right]
            }{
              1-\mathrm{e}^{-s/\lambda}-(s/\lambda)\mathrm{e}^{-s/\lambda}
            }       
\end{equation}
i.e. the last part of $F(s;\mu)_{GS}^{2+}$ given by Eq.(\ref{eq:more_scattering_base}) and the solution is
\begin{equation}
\label{eq:app_kaw_optimala}
 a=\frac{\alpha}{4\beta}+\sqrt{\left(\frac{\alpha}{4\beta}\right)^{2}+\frac{\alpha}{4\beta}}
\end{equation}
One can use the normality properties of Legendre polynomials with weight functions of 
$g(x)=1, g(x)=x, g(x)=x^{2}$ i.e.
\begin{equation}
 \int_{-1}^{+1} P_{\ell}(x)P_{k}(x)\mathrm{d}x = \frac{2}{2\ell+1}\delta_{\ell k}
\end{equation}
\begin{equation}
 \int_{-1}^{+1} P_{\ell}(x)P_{k}(x) x \mathrm{d}x = 
  \left\{
   \begin{array}{ll}
      \frac{2(\ell+1)}{(2\ell+1)(2\ell+3)} & \quad \mathrm{if}\; k=\ell+1 \\
      & \\
      \frac{2\ell}{(2\ell-1)(2\ell+1)}     & \quad \mathrm{if}\; k=\ell-1
  \end{array}
  \right.
\end{equation}
\begin{equation}
 \int_{-1}^{+1} P_{\ell}(x)P_{k}(x) x^{2} \mathrm{d}x = 
  \left\{
   \begin{array}{ll}
      \frac{2(\ell+1)(\ell+2)}{(2\ell+1)(2\ell+3)(2\ell+5)} & \quad \mathrm{if}\; k=\ell+2 \\
      & \\
      \frac{2(2\ell^{2}+2\ell-1)}{(2\ell-1)(2\ell+1)(2\ell+3)} & \quad \mathrm{if}\; k=\ell \\
      & \\
      \frac{2\ell(\ell-1)}{(2\ell-3)(2\ell-1)(2\ell+1)} & \quad \mathrm{if}\; k=\ell-2 \\
   \end{array}
  \right.
\end{equation}
to compute the integrals necessary to get the expressions for $\alpha$ and $\beta$.
Using the first property
\begin{equation}
\begin{split} 
& \sum_{\ell=0}^{\infty} (\ell+0.5)\xi_{\ell}(s,\lambda,A) \sum_{k=0}^{\infty} (k+0.5)\xi_{k}(s,\lambda,A)  
     \int_{-1}^{+1} P_{\ell}(\mu) P_{k}(\mu) \mathrm{d}\mu
\\& = \sum_{\ell=0}^{\infty} (\ell+0.5) [\xi_{\ell}(s,\lambda,A)]^{2}
\end{split}
\end{equation}
using the second property
\begin{equation}
\begin{split} 
& \sum_{\ell=0}^{\infty} (\ell+0.5)\xi_{\ell}(s,\lambda,A) \sum_{k=0}^{\infty} (k+0.5)\xi_{k}(s,\lambda,A)  
     \int_{-1}^{+1} P_{\ell}(\mu) P_{k}(\mu) \mu \mathrm{d}\mu
\\& = \sum_{\ell=0}^{\infty} (\ell+0.5) \xi_{\ell}(s,\lambda,A) \left\{ 
        (\ell-1+0.5) \frac{2\ell}{(2\ell-1)(2\ell+1)} \xi_{\ell-1}(s,\lambda,A)
        \right.
\\&\;   \left.
     +(\ell+1+0.5) \frac{2(\ell+1)}{(2\ell+1)(2\ell+3)}\xi_{\ell+1}(s,\lambda,A)
      \right\}
   = \sum_{\ell=0}^{\infty}  \xi_{\ell}(s,\lambda,A) \left\{
       \frac{\ell}{2}\xi_{\ell-1}(s,\lambda,A) 
        \right.
\\&\;   \left.
       +\frac{\ell+1}{2}\xi_{\ell+1}(s,\lambda,A)
     \right\} = \sum_{\ell=0}^{\infty} (\ell+1) \xi_{\ell}(s,\lambda,A) \xi_{\ell+1}(s,\lambda,A)
\end{split} 
\end{equation}
and the third 
\begin{equation}
\begin{split} 
& \sum_{\ell=0}^{\infty} (\ell+0.5)\xi_{\ell}(s,\lambda,A) \sum_{k=0}^{\infty} (k+0.5)\xi_{k}(s,\lambda,A)  
     \int_{-1}^{+1} P_{\ell}(\mu) P_{k}(\mu) \mu^{2} \mathrm{d}\mu
\\& = \sum_{\ell=0}^{\infty} (\ell+0.5) \xi_{\ell}(s,\lambda,A) \left\{ 
        \frac{(\ell-2+0.5)2\ell(\ell-1)}{(2\ell-3)(2\ell-1)(2\ell+1)} \xi_{\ell-2}(s,\lambda,A)
        \right.
\\&\;   \left.
     +  \frac{(\ell+0.5)2(2\ell^{2}+2\ell-1)}{(2\ell-1)(2\ell+1)(2\ell+3)} \xi_{\ell}(s,\lambda,A)
     +  \frac{(\ell+2+0.5)2(\ell+1)(\ell+2)}{(2\ell+1)(2\ell+3)(2\ell+5)} \xi_{\ell+2}(s,\lambda,A)
     \right\}
\\& = \sum_{\ell=0}^{\infty} \xi_{\ell}(s,\lambda,A) \left\{
       \frac{\ell(\ell-1)}{2(2\ell-1)} \xi_{\ell-2}(s,\lambda,A)
     + \frac{(\ell+0.5)(2\ell^{2}+2\ell-1)}{(2\ell-1)(2\ell+3)} \xi_{\ell}(s,\lambda,A)
        \right.
\\&\;   \left.
     + \frac{(\ell+1)(\ell+2)}{(2(2\ell+3)} \xi_{\ell+2}(s,\lambda,A)
      \right\}
    = \sum_{\ell=0}^{\infty} \frac{(\ell+0.5)(2\ell^{2}+2\ell-1)}{(2\ell-1)(2\ell+3)} [\xi_{\ell}(s,\lambda,A)]^{2}
\\&\;+\sum_{\ell=0}^{\infty} \xi_{\ell}(s,\lambda,A) \left\{ 
        \frac{\ell(\ell-1)}{2(2\ell-1)} \xi_{\ell-2}(s,\lambda,A)
       +\frac{(\ell+1)(\ell+2)}{(2(2\ell+3)} \xi_{\ell+2}(s,\lambda,A)  
       \right\}
\\&= \sum_{\ell=0}^{\infty} \frac{(\ell+0.5)(2\ell^{2}+2\ell-1)}{(2\ell-1)(2\ell+3)} [\xi_{\ell}(s,\lambda,A)]^{2}
   + \sum_{\ell=0}^{\infty} \frac{(\ell+1)(\ell+2)}{(2\ell+3)} \xi_{\ell}(s,\lambda,A)\xi_{\ell+2}(s,\lambda,A) 
\end{split} 
\end{equation}
Substituting these three equations into $\alpha$ and $\beta$ results in 
\begin{equation}
\begin{split}
 \alpha = & \sum_{\ell=0}^{\infty} \xi_{\ell}(s,\lambda,A) \left\{ 
   \left( 1.5\ell +\frac{0.065}{\ell+1.5}+\frac{0.065}{\ell-0.5}+0.75 \right) \xi_{\ell}(s,\lambda,A)
        \right.
\\&\;   \left.
   - 2(\ell+1)  \xi_{\ell+1}(s,\lambda,A)
   + \frac{(\ell+1)(\ell+2)}{(2\ell+3)} \xi_{\ell+2}(s,\lambda,A)  
   \right\}
\end{split}   
\end{equation}
\begin{equation}
 \beta =  \sum_{\ell=0}^{\infty} (\ell+1) \xi_{\ell}(s,\lambda,A)  \xi_{\ell+1}(s,\lambda,A)  
\end{equation}
that can be used in Eq.(\ref{eq:app_kaw_optimala}) to compute the optimal value of the parameter $a$ 
of the transformation given by Eq.(\ref{eq:kaw_transform}) 

\begin{table}
\caption{ $s/\lambda = 1$ and $\tilde{w}^{2} = 3.3675$
\\
Screening parameter $A$ and the corresponding optimal 
value of parameter $a$ of the transformation $f(a;\mu)$ with the 
corresponding $w^{2}$ values at $s/\lambda =1$ as a function of 
$G_{1}s/\lambda$. The approximate value $\tilde{w}^{2}$ value computed by Eq.(\ref{eq:kaw_aprxw2})
and the corresponding $\tilde{a}$ values are also shown.
}
\centering
  \begin{tabular}{| c | c || c | c || c |}
    \hline\hline
    $G_{1}s/\lambda$ & $A$ & $a$ & $w^{2}=a/A$ & $\tilde{a}= \tilde{w}^{2}A$\\
    \hline
1.000e-03 & 5.699162e-05 & 1.933615e-04 & 3.392806e+00  & 1.919193e-04 \\
5.090e-02 & 6.168356e-03 & 2.120923e-02 & 3.438393e+00  & 2.077194e-02 \\
1.008e-01 & 1.553012e-02 & 5.448068e-02 & 3.508066e+00  & 5.229768e-02 \\
1.507e-01 & 2.779892e-02 & 1.000408e-01 & 3.598728e+00  & 9.361288e-02 \\
2.006e-01 & 4.319982e-02 & 1.603297e-01 & 3.711352e+00  & 1.454754e-01 \\
2.505e-01 & 6.216947e-02 & 2.392574e-01 & 3.848470e+00  & 2.093557e-01 \\
3.004e-01 & 8.533942e-02 & 3.425741e-01 & 4.014254e+00  & 2.873805e-01 \\
3.503e-01 & 1.135762e-01 & 4.787333e-01 & 4.215084e+00  & 3.824679e-01 \\
4.002e-01 & 1.480622e-01 & 6.604428e-01 & 4.460576e+00  & 4.985995e-01 \\
4.501e-01 & 1.904288e-01 & 9.074555e-01 & 4.765328e+00  & 6.412689e-01 \\
5.000e-01 & 2.429743e-01 & 1.251795e+00 & 5.151963e+00  & 8.182161e-01 \\
    \hline
  \end{tabular}
\label{tb:kaw_table1}
\end{table}

\begin{table}
\caption{ $s/\lambda = 10$ and $\tilde{w}^{2} =2.289938e+01$
\\Same as Table\ref{tb:kaw_table1}
}
\centering
  \begin{tabular}{| c | c || c | c || c |}
    \hline\hline
    $G_{1}s/\lambda$ & $A$ & $a$ & $w^{2}=a/A$ & $\tilde{a}= \tilde{w}^{2}A$\\
    \hline
1.000e-03 & 4.412635e-06 & 1.011501e-04 & 2.292283e+01 & 1.010466e-04 \\
5.090e-02 & 3.683146e-04 & 8.502289e-03 & 2.308431e+01 & 8.434178e-03 \\
1.008e-01 & 8.253681e-04 & 1.922263e-02 & 2.328976e+01 & 1.890042e-02 \\
1.507e-01 & 1.339376e-03 & 3.150770e-02 & 2.352415e+01 & 3.067090e-02 \\
2.006e-01 & 1.899663e-03 & 4.518052e-02 & 2.378344e+01 & 4.350111e-02 \\
2.505e-01 & 2.500655e-03 & 6.018001e-02 & 2.406570e+01 & 5.726345e-02 \\
3.004e-01 & 3.138897e-03 & 7.649451e-02 & 2.436987e+01 & 7.187882e-02 \\
3.503e-01 & 3.812052e-03 & 9.413963e-02 & 2.469526e+01 & 8.729364e-02 \\
4.002e-01 & 4.518446e-03 & 1.131484e-01 & 2.504145e+01 & 1.034696e-01 \\
4.501e-01 & 5.256847e-03 & 1.335664e-01 & 2.540809e+01 & 1.203785e-01 \\
5.000e-01 & 6.026322e-03 & 1.554486e-01 & 2.579493e+01 & 1.379991e-01 \\
    \hline
  \end{tabular}
\label{tb:kaw_table2}
\end{table}

\begin{table}
\caption{ $s/\lambda = 10^{3}$ and $\tilde{w}^{2} =6.070075e+03$
\\Same as Table\ref{tb:kaw_table1}
}
\centering
  \begin{tabular}{| c | c || c | c || c |}
    \hline\hline
    $G_{1}s/\lambda$ & $A$ & $a$ & $w^{2}=a/A$ & $\tilde{a}= \tilde{w}^{2}A$\\
    \hline
1.000e-03 & 3.067520e-08 & 1.859172e-04 & 6.060830e+03 & 1.862008e-04 \\
5.090e-02 & 2.108620e-06 & 1.294233e-02 & 6.137819e+03 & 1.279948e-02 \\
1.008e-01 & 4.451370e-06 & 2.771704e-02 & 6.226632e+03 & 2.702015e-02 \\
1.507e-01 & 6.925284e-06 & 4.378737e-02 & 6.322827e+03 & 4.203699e-02 \\
2.006e-01 & 9.493600e-06 & 6.100034e-02 & 6.425417e+03 & 5.762686e-02 \\
2.505e-01 & 1.213737e-05 & 7.930613e-02 & 6.534047e+03 & 7.367473e-02 \\
3.004e-01 & 1.484477e-05 & 9.869717e-02 & 6.648617e+03 & 9.010885e-02 \\
3.503e-01 & 1.760764e-05 & 1.191890e-01 & 6.769165e+03 & 1.068797e-01 \\
4.002e-01 & 2.041995e-05 & 1.408122e-01 & 6.895816e+03 & 1.239506e-01 \\
4.501e-01 & 2.327707e-05 & 1.636088e-01 & 7.028755e+03 & 1.412935e-01 \\
5.000e-01 & 2.617528e-05 & 1.876300e-01 & 7.168216e+03 & 1.588859e-01 \\
    \hline
  \end{tabular}
\label{tb:kaw_table3}
\end{table}

\begin{table}
\caption{ $s/\lambda = 10^{5}$ and $\tilde{w}^{2} =9.629423e+05$
\\Same as Table\ref{tb:kaw_table1}
}
\centering
  \begin{tabular}{| c | c || c | c || c |}
    \hline\hline
    $G_{1}s/\lambda$ & $A$ & $a$ & $w^{2}=a/A$ & $\tilde{a}= \tilde{w}^{2}A$\\
    \hline
1.000e-03 & 2.362253e-10 & 2.269574e-04 & 9.607668e+05 & 2.274713e-04 \\
5.090e-02 & 1.495453e-08 & 1.457531e-02 & 9.746419e+05 & 1.440035e-02 \\
1.008e-01 & 3.093671e-08 & 3.063026e-02 & 9.900942e+05 & 2.979027e-02 \\
1.507e-01 & 4.750198e-08 & 4.781334e-02 & 1.006554e+06 & 4.574167e-02 \\
2.006e-01 & 6.447247e-08 & 6.601397e-02 & 1.023909e+06 & 6.208327e-02 \\
2.505e-01 & 8.175856e-08 & 8.520270e-02 & 1.042126e+06 & 7.872878e-02 \\
3.004e-01 & 9.930534e-08 & 1.053831e-01 & 1.061203e+06 & 9.562531e-02 \\
3.503e-01 & 1.170754e-07 & 1.265770e-01 & 1.081158e+06 & 1.127368e-01 \\
4.002e-01 & 1.350413e-07 & 1.488185e-01 & 1.102022e+06 & 1.300370e-01 \\
4.501e-01 & 1.531824e-07 & 1.721514e-01 & 1.123833e+06 & 1.475058e-01 \\
5.000e-01 & 1.714820e-07 & 1.966278e-01 & 1.146638e+06 & 1.651273e-01 \\
  \hline
  \end{tabular}
\label{tb:kaw_table4}
\end{table}

\subsection{On the run time interpolation of the pre-computed PDF-s}
\label{app:pdf-interpol}
Let $x\in \mathbb{R}\;\vert\; x \in [x_{\text{min}}, x_{\text{max}}], x_{\text{max}}>x_{\text{min}}$ a continus variable. Dividing 
the $[x_{\text{min}}, x_{\text{max}}]$ interval into $N$ equal sub-intervals (bins) corresponds to introduce the $\{x_0,x_1,...,x_N\}\; N+1$ 
discrete values such that
\begin{equation}
  x_0 = x_{\text{min}},\; x_i = x_0 + i\Delta, i=0,...,N\; \text{with } \Delta\equiv (x_{\text{max}} - x_{\text{min}})/N
\end{equation}
Then the index $k$, such that $x_k \leq x < x_{k+1}$ for any $x\in \mathbb{R}\;\vert\; x \in [x_{\text{min}}, x_{\text{max}})$ continuous value, 
is the $k = [(x-x_0)/\Delta]$ where $[\cdot]$ is used for the \textit{integer part} (or \texttt{floor}) of the expression. Furthermore, let $y(x)$ a single value 
continuous function of $x$, i.e. $y:[x_{\text{min}}, x_{\text{max}}]\to\mathbb{R}$. Linear interpolation of $y(x)$ at any $x \in [x_{\text{min}}, x_{\text{max}})$
such that $x_k \leq x < x_{k+1}$ corresponds to
\begin{equation}
  y(x)\approx\tilde{y}(x) = y_k + \frac{y_{k+1}-y_k}{x_{k+1}-x_k} (x-x_k) 
\end{equation}
with the $y_i\equiv y(x_i)$ for brevity. This can also be written in the alternative form of 
\begin{equation}
  y(x)\approx\tilde{y}(x) = \left[1-\frac{x-x_k}{x_{k+1}-x_k}\right] y_k + \left[\frac{x-x_k}{x_{k+1}-x_k}\right] y_{k+1}
\end{equation}
that shows more explicitly that the result of the linear interpolation $\tilde{y}(x)$ is nothing but the linear combination of the two $y_k,\; y_{k+1}$ exact 
function values taken at $x_k$ and $x_{k+1}$ with the $1-w$ and $w\equiv (x-x_k)/(x_{k+1}-x_k)$ weights respectively.  

Replacing the single valued $y$ function of $x$ now with a pdf of a random variable $u$ that depends on $x$ as a parameter, the goal now is to interpolate 
the $p(u; x)$ distribution at the $x \in [x_{\text{min}}, x_{\text{max}})$ such that $x_k \leq x < x_{k+1}$. The above expressions for linear interpolation 
can also be used to interpolate the pdf leading to
\begin{equation}
  p(u; x)\approx\tilde{p}(u; x) = \left[1-\frac{x-x_k}{x_{k+1}-x_k}\right] p(u; x_k) + \left[\frac{x-x_k}{x_{k+1}-x_k}\right] p(u;x_{k+1})
\end{equation}
This shows clearly that the result of the linear interpolation of the pdf $\tilde{p}(u; x)$ is nothing but a \textit{mixture distribution} of the exact $p(u;x_k)$ and 
$p(u;x_{k+1})$ pdf-s with the $1-w$ and $w$ weights respectively. This is because the weights $1-w$ and $w$ are both $\geq 0$ while sum up to unity.
Therefore, providing $u$ samples distributed according to the $\tilde{p}(u; x)$ interpolated pdf can be done by generating samples form $p(u;x_k)$ 
with a probability of $1-w$ and from $p(u; x_{k+1})$ with a probability of $w\equiv (x-x_k)/(x_{k+1}-x_k)$. 

The most important is to notice 
that computing the $\tilde{p}(u; x)$ interpolated pdf is not required for producing samples from it as being able to generate $u$ samples according to the 
$p(u; x_i)$ pdf-s at the discrete ${x_0,x_1,...,x_i,...,x_N}$ points is sufficient. Therefore, for any $x \in [x_{\text{min}}, x_{\text{max}})$:
\begin{enumerate}
 \item determine $k$ such that $x_k \leq x < x_{k+1}$:
     \begin{equation}
       k = [(x-x_0)/\Delta],\; \text{ with }\Delta\equiv (x_{\text{max}}-x_{\text{min}})/N
     \end{equation}
 \item sample either from the $p(u; x_{k})$ or $p(u; x_{k+1})$ pdf: 
 \begin{equation}
 \begin{array}{ll}
         \text{generate a random number:} & \eta \in \mathcal{U}(0,1) \\
         \text{sample }u\text{ from } p(u; x_{k+1}) &  \text {if }  \eta < (x-x_k)/(x_{k+1}-x_k): \\
         \text{sample }u\text{ from }  p(u; x_{k})    & \text{otherwise}\\
 \end{array}    
 \end{equation} 
\end{enumerate}

\subsection{The Mott DCS}
\label{app:mott-dcs}
The simplest DCS, that already includes the electron/positron projectile spin, can be obtained by solving the Dirac equation of elastic scattering 
in a point like, \textit{unscreened} Coulomb potential, i.e. elastic scattering of Dirac particles on the point-like, bare nucleus. 
The resulted DCS is know as the Mott DCS (M-DCS) \cite{mott1929scattering}. The computation flow followed 
in this work will be summarised below while more details can be found in \cite{salvat2005elsepa, jicru-7-2007} or \cite{jicru-7-2007-Apdix-C}.

Using a simple, point like Coulomb potential (i.e. taking $R \to \infty$ in Eq.(\ref{eq:EXP-SCREENED-COULOMB}))  
\begin{equation}
 V(r)=\frac{ZZ'e^{2}}{r}
\end{equation}
(target atomic number of Z and projectile charge Z'e) as a scattering potential in the Dirac equation leads to the Mott DCS 
\begin{equation}
\label{eq:mott-dcs-scat-amps}
\frac{\mathrm{d}\sigma^{(R)}}{\mathrm{d}\Omega} = |f^{(C)}(\theta)|^2 + |g^{(C)}(\theta)|^2
\end{equation}  
with the corresponding direct $f^{(C)}(\theta)$ and spin-flip $g^{(C)}(\theta)$ scattering amplitudes
\begin{equation}
\label{eq:Dirac-Coulomb-scat-amps}
\begin{split}
 f^{(C)}(\theta)  & = \frac{1}{2ik} \sum_{l=0}^{\infty}\left\{ 
   (l+1)\left[ 
     \exp\left( 
       2i\Delta_{-l-1}
     \right) - 1
   \right]    +
   l\left[ 
     \exp\left( 
       2i\Delta_{l} 
     \right) -1
   \right]
 \right\} P_{l}(\cos(\theta)) \\
  g^{(C)}(\theta)  & = \frac{1}{2ik} \sum_{l=0}^{\infty}\left\{  
       \exp\left( 2i\Delta_{l} \right)       -  \exp\left(  2i\Delta_{-l-1}  \right)
 \right\} P^{1}_{l}(\cos(\theta))
\end{split}
\end{equation}
where $k=p/\hbar$ is the relativistic wave number of the projectile with $(cp)^2=(c\hbar k)^2=E_{\text{kin}}(E_{\text{kin}}+2m_ec^2)$, $P_{l}(\cos(\theta))$ 
are Legendre polynomials, $P^{1}_{l}(\cos(\theta))$ are associated Legendre functions and 
\begin{equation}
\Delta_\kappa = \arg[\zeta E_t - i(\kappa+\lambda)c\hbar k] - \frac{\pi}{2}(\lambda -l -1) + \arg\Gamma(\lambda+i\eta) - \mathcal{S}(\zeta,\kappa)\pi
\end{equation}
are the corresponding Dirac-Coulomb phase shifts \cite{salvat1995accurate,salvat2005elsepa,jicru-7-2007-Apdix-C}
with $\zeta=ZZ'e^2/(\hbar c)=ZZ'/\alpha$,  $\lambda=\sqrt{\kappa^2+\zeta^2}$,  $\eta=ZZ'e^2 m_e/(\hbar k)$ is the Sommerfeld parameter, 
$\mathcal{S}(\zeta<0,\kappa<0) = 1$ and zero otherwise while $E_t = E_{\text{kin}}+2m_ec^2$ is the total energy of the projectile.

At polar scattering angles $\theta > 1 ^{\circ}$, the Mott DCS can be calculated directly from the corresponding Dirac-Coulomb scattering 
amplitudes as given by Eq.(\ref{eq:mott-dcs-scat-amps}) relying on the direct summation of their partial wave expansions Eq.(\ref{eq:Dirac-Coulomb-scat-amps}) using the 
reduced series method \cite{yennie1954phase} as suggested in \cite{salvat1993elastic,salvat2005elsepa,jicru-7-2007-Section4}.
At lower $\theta < 1 ^{\circ}$ scattering angles, where the series in Eq.(\ref{eq:Dirac-Coulomb-scat-amps}) converge very slowly, the asymptotic 
formula (see e.g. in \cite{salvat2005elsepa,jicru-7-2007-Section4})
\begin{equation}
R_{MR}(\theta) = 1 + \pi\beta^2\zeta\text{Re} 
  \left[ 
   \frac{\Gamma(0.5-i\zeta)}{\Gamma(0.5+i\zeta)}\frac{\Gamma(1+i\zeta)}{\Gamma(1-i\zeta)}
 \right]\sin(\theta/2)
\end{equation}
for $R_{MR}(\theta) \equiv \mathrm{d}\sigma^{(M)}(\theta)  / \mathrm{d}\sigma^{(R)}(\theta)$, i.e. the Mott-to-Rutherford DCS ratio, can be utilised for the indirect 
calculation of the Mott DCS. As discussed in Section \ref{sec::spin-effect}, the Rutherford DCS is the result of solving the same scattering problem as above but 
neglecting the projectile spin, i.e. considering elastic scattering of spinless electrons in a simple (\textit{unscreened}) point-like Coulomb potential and solving the 
corresponding Schr\"{o}dinger equation under the first born approximation. Therefore, $R_{MR}(\theta)$ captures the spin effect to the DCS and by its definition 
the Mott DCS can be expressed as \cite{fernandez1993cross,salvat2005elsepa} 
\begin{equation}
  \frac{\mathrm{d}\sigma^{(M)}}{\mathrm{d}\Omega} =  \frac{\mathrm{d}\sigma^{(R)}}{\mathrm{d}\Omega} R_{MR} (\theta)
\end{equation}
with $\mathrm{d}\sigma^{(R)}/\mathrm{d}\Omega$ as given by Eq.(\ref{eq:RF-DCS}). Some results of the computation are shown in Section \ref{sec::spin-effect}. 

\end{appendices}

\bibliographystyle{ieeetr}
\bibliography{bib}

\begin{thebibliography}{10}

\bibitem{kawrakow1998representation}
I.~Kawrakow and A.~F. Bielajew, ``On the representation of electron multiple
  elastic-scattering distributions for monte carlo calculations,'' {\em Nuclear
  Instruments and Methods in Physics Research Section B: Beam Interactions with
  Materials and Atoms}, vol.~134, no.~3-4, pp.~325--336, 1998.

\bibitem{kawrakow2000accurate}
I.~Kawrakow, ``Accurate condensed history monte carlo simulation of electron
  transport: {I}. {EGS}nrc, the new {EGS}4 version,'' {\em Medical physics},
  vol.~27, no.~3, pp.~485--498, 2000.

\bibitem{kawrakow2000egsnrc}
I.~Kawrakow and D.~Rogers, ``The {EGS}nrc code system: Monte carlo simulation
  of electron and photon transport,'' Tech. Rep. PIRS-701, National Research
  Council of Canada, Ottawa, Canada, 2000.

\bibitem{kawrakow1997improved}
I.~Kawrakow, ``Improved modeling of multiple scattering in the voxel monte
  carlo model,'' {\em Medical physics}, vol.~24, no.~4, pp.~505--517, 1997.

\bibitem{kawrakow1996electron}
I.~Kawrakow, ``Electron transport: lateral and longitudinal correlation
  algorithm,'' {\em Nuclear Instruments and Methods in Physics Research Section
  B: Beam Interactions with Materials and Atoms}, vol.~114, no.~3,
  pp.~307--326, 1996.

\bibitem{kawrakow1998condensed}
I.~Kawrakow and A.~F. Bielajew, ``On the condensed history technique for
  electron transport,'' {\em Nuclear Instruments and Methods in Physics
  Research Section B: Beam Interactions with Materials and Atoms}, vol.~142,
  no.~3, pp.~253--280, 1998.

\bibitem{kadri2009incorporation}
O.~Kadri, V.~Ivanchenko, F.~Gharbi, and A.~Trabelsi, ``Incorporation of the
  goudsmit--saunderson electron transport theory in the geant4 monte carlo
  code,'' {\em Nuclear Instruments and Methods in Physics Research Section B:
  Beam Interactions with Materials and Atoms}, vol.~267, no.~23-24,
  pp.~3624--3632, 2009.

\bibitem{goudsmit1940multiple}
S.~Goudsmit and J.~Saunderson, ``Multiple scattering of electrons,'' {\em
  Physical Review}, vol.~57, no.~1, p.~24, 1940.

\bibitem{fernandez1993theory}
J.~Fern{\'a}ndez-Varea, R.~Mayol, J.~Bar{\'o}, and F.~Salvat, ``On the theory
  and simulation of multiple elastic scattering of electrons,'' {\em Nuclear
  Instruments and Methods in Physics Research Section B: Beam Interactions with
  Materials and Atoms}, vol.~73, no.~4, pp.~447--473, 1993.

\bibitem{berger1988multiple}
M.~J. Berger and R.~Wang, ``Multiple-scattering angular deflections and
  energy-loss straggling,'' in {\em Monte Carlo transport of electrons and
  photons}, pp.~21--56, Springer, 1988.

\bibitem{negreanu2005calculation}
C.~Negreanu, X.~Llovet, R.~Chawla, and F.~Salvat, ``Calculation of
  multiple-scattering angular distributions of electrons and positrons,'' {\em
  Radiation Physics and Chemistry}, vol.~74, no.~5, pp.~264--281, 2005.

\bibitem{bielajew1996hybrid}
A.~F. Bielajew, ``A hybrid multiple-scattering theory for electron-transport
  monte carlo calculations,'' {\em Nuclear Instruments and Methods in Physics
  Research Section B: Beam Interactions with Materials and Atoms}, vol.~111,
  no.~3, pp.~195--208, 1996.

\bibitem{bielajew1994plural}
A.~F. Bielajew, ``Plural and multiple small-angle scattering from a screened
  rutherford cross section,'' {\em Nuclear Instruments and Methods in Physics
  Research Section B: Beam Interactions with Materials and Atoms}, vol.~86,
  no.~3, pp.~257--269, 1994.

\bibitem{moliere1947theorieI}
G.~Moliere, ``Theorie der streuung schneller geladener teilchen i.
  einzelstreuung am abgeschirmten coulomb-feld,'' {\em Zeitschrift f{\"u}r
  Naturforschung A}, vol.~2, no.~3, pp.~133--145, 1947.

\bibitem{bethe1953moliere}
H.~Bethe, ``Moliere's theory of multiple scattering,'' {\em Physical Review},
  vol.~89, no.~6, p.~1256, 1953.

\bibitem{nelson1985egs4}
W.~R. Nelson, H.~Hirayama, and D.~W. Rogers, ``Egs4 code system,'' tech. rep.,
  Stanford Linear Accelerator Center, Menlo Park, CA (USA), 1985.

\bibitem{scott1963theory}
W.~T. Scott, ``The theory of small-angle multiple scattering of fast charged
  particles,'' {\em Reviews of modern physics}, vol.~35, no.~2, p.~231, 1963.

\bibitem{walker1977efficient}
A.~J. Walker, ``An efficient method for generating discrete random variables
  with general distributions,'' {\em ACM Transactions on Mathematical Software
  (TOMS)}, vol.~3, no.~3, pp.~253--256, 1977.

\bibitem{salvat2006penelope}
F.~Salvat, J.~M. Fern{\'a}ndez-Varea, J.~Sempau, {\em et~al.}, ``Penelope-2006:
  A code system for monte carlo simulation of electron and photon transport,''
  in {\em Workshop proceedings}, vol.~4, p.~7, Citeseer, 2006.

\bibitem{salvat2005elsepa}
F.~Salvat, A.~Jablonski, and C.~J. Powell, ``{ELSEPA}—{D}irac partial-wave
  calculation of elastic scattering of electrons and positrons by atoms,
  positive ions and molecules,'' {\em Computer physics communications},
  vol.~165, no.~2, pp.~157--190, 2005.

\bibitem{fernandez1993cross}
J.~Fern{\'a}ndez-Varea, R.~Mayol, and F.~Salvat, ``Cross sections for elastic
  scattering of fast electrons and positrons by atoms,'' {\em Nuclear
  Instruments and Methods in Physics Research Section B: Beam Interactions with
  Materials and Atoms}, vol.~82, no.~1, pp.~39--45, 1993.

\bibitem{zeitler1964screening}
E.~Zeitler and H.~Olsen, ``Screening effects in elastic electron scattering,''
  {\em Physical Review}, vol.~136, no.~6A, p.~A1546, 1964.

\bibitem{kawrakow2000cross}
I.~Kawrakow, ``Cross section improvements for egsnrc,'' in {\em Proceedings of
  the 22nd Annual International Conference of the IEEE Engineering in Medicine
  and Biology Society (Cat. No. 00CH37143)}, vol.~3, pp.~1668--1671, IEEE,
  2000.

\bibitem{mott1929scattering}
N.~F. Mott, ``The scattering of fast electrons by atomic nuclei,'' {\em
  Proceedings of the Royal Society of London. Series A, Containing Papers of a
  Mathematical and Physical Character}, vol.~124, no.~794, pp.~425--442, 1929.

\bibitem{jicru-7-2007}
``Elastic scattering of electrons and positrons,'' {\em Journal of the ICRU},
  vol.~7, no.~1, pp.~iii--iii, 2007.
\newblock PMID: 24170956.

\bibitem{jicru-7-2007-Apdix-C}
``Elastic scattering of electrons and positrons: Appendix c. quantum theory of
  scattering by a central potential,'' {\em Journal of the ICRU}, vol.~7,
  no.~1, pp.~131--149, 2007.

\bibitem{salvat1995accurate}
F.~Salvat, J.~Fern{\'a}ndez-Varea, and W.~Williamson~Jr, ``Accurate numerical
  solution of the radial schr{\"o}dinger and dirac wave equations,'' {\em
  Computer physics communications}, vol.~90, no.~1, pp.~151--168, 1995.

\bibitem{yennie1954phase}
D.~Yennie, D.~G. Ravenhall, and R.~Wilson, ``Phase-shift calculation of
  high-energy electron scattering,'' {\em Physical Review}, vol.~95, no.~2,
  p.~500, 1954.

\bibitem{salvat1993elastic}
F.~Salvat and R.~Mayol, ``Elastic scattering of electrons and positrons by
  atoms. schr{\"o}dinger and dirac partial wave analysis,'' {\em Computer
  physics communications}, vol.~74, no.~3, pp.~358--374, 1993.

\bibitem{jicru-7-2007-Section4}
``Elastic scattering of electrons and positrons: 4. calculations for atoms used
  for the data generation,'' {\em Journal of the ICRU}, vol.~7, no.~1,
  pp.~57--76, 2007.
\newblock PMID: 24170947.

\end{thebibliography}

\end{document}